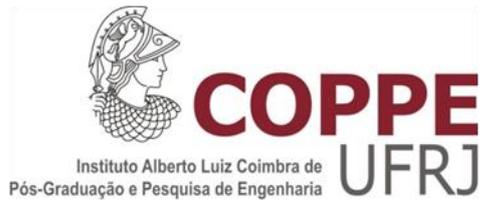

# TOWARDS A STRATEGY FOR SUPPORTING THE ENGINEERING OF CONTEMPORARY SOFTWARE SYSTEMS

Rebeca Campos Motta

Exame de Qualificação apresentado ao Programa de Pós-Graduação em Engenharia de Sistemas e Computação, COPPE, da Universidade Federal do Rio de Janeiro, como parte dos requisitos necessários à obtenção do título de Doutor em Engenharia de Sistemas e Computação.

Orientadores: Guilherme Horta Travassos, UFRJ
                 Káthia Marçal de Oliveira, UPHF

Rio de Janeiro
Novembro de 2018

# TOWARDS A STRATEGY FOR SUPPORTING THE ENGINEERING OF CONTEMPORARY SOFTWARE SYSTEMS

Rebeca Campos Motta

Novembro/2018

Orientadores:
Guilherme Horta Travassos
Káthia Marçal de Oliveira

Programa:
Engenharia de Sistemas e Computação

Sistemas de software contemporâneos (CSS), como a Internet das Coisas (IoT), Indústria 4.0 e Cidades Inteligentes, oferecem desafios para a sua construção, uma vez que questionam nossa forma tradicional de desenvolvimento de software. Eles representam um paradigma promissor para a integração de dispositivos e tecnologias de comunicação. Isso está levando a uma mudança da visão monolítica clássica do desenvolvimento, onde as partes interessadas recebem um produto de software no final, para sistemas de software materializados através de objetos físicos interconectados por redes e com inteligência embarcada para apoiar atividades. É necessário, portanto, revisitar nosso modo de desenvolver sistemas de software e começar a considerar as particularidades exigidas por esses novos tipos de aplicativos. Esta tese tem como objetivo investigar as particularidades destes novos tipos de aplicações para apoiar a definição de um framework para sustentar a tomada de decisões na engenharia desse tipo de aplicações e sistemas. Para alcançar esse objetivo, utilizamos sistemas IoT para representar CSS, uma vez que apresentam a contemporaneidade e da multidisciplinaridade que buscamos investigar.

TOWARDS A STRATEGY FOR SUPPORTING THE ENGINEERING OF CONTEMPORARY SOFTWARE SYSTEMS

Rebeca Campos Motta

November/2018

Advisors:
Guilherme Horta Travassos
Káthia Marçal de Oliveira

Program:
Computing and Systems Engineering

Contemporary software systems, such as the Internet of Things, Industry 4.0 and Intelligent Cities, present challenges for their engineering, since they question our traditional form of software development. They represent a promising paradigm for the integration of communication devices and technologies. It is leading to a shift from the classical monolithic view of development where stakeholders receive a software product at the end, to materialized software systems through physical objects interconnected by networks and with embedded smartness to support activities. Therefore, it is necessary to revisit our way of developing software systems and begin to consider the particularities required by these new types of applications. This thesis aims to investigate the particularities of these new types of applications to support the definition of a framework to support decision-making in the engineering of this kind of applications and systems. To this end, we use IoT systems as surrogate for CSS, since they present the contemporaneity and multidisciplinarity that we aim to investigate.

Résumé de l'examen de qualification présenté à COPPE / UFRJ comme un accomplissement partiel des exigences pour le diplôme de docteur en sciences (D.Sc.).

# TOWARDS A STRATEGY FOR SUPPORTING THE ENGINEERING OF CONTEMPORARY SOFTWARE SYSTEMS

Rebeca Campos Motta

Novembre/2018

Directeurs de thèse:
Guilherme Horta Travassos
Káthia Marçal de Oliveira

Programme:
Génie Informatique et Systems

Les systèmes logiciels contemporains, tels que l'Internet of Things, Industry 4.0 et Intelligent Cities, présentent des défis pour leur ingénierie, car ils remettent en question notre forme traditionnelle de développement de logiciels. Ils représentent un paradigme prometteur pour l'intégration des dispositifs et des technologies de communication. Cela conduit à un changement de la vision monolithique classique du développement où les parties prenantes reçoivent un produit logiciel à la fin, à des systèmes logiciels matérialisés à travers des objets physiques interconnectés par des réseaux et avec une intelligence intégrée pour soutenir les activités. Il est donc nécessaire de revoir notre façon de développer des systèmes logiciels et de commencer à considérer les particularités requises par ces nouveaux types d'applications. Cette thèse vise à étudier les particularités de ces nouveaux types d'applications pour soutenir la définition d'un cadre pour soutenir la prise de décision dans l'ingénierie de ce genre d'applications et de systèmes. À cette fin, nous utilisons les systèmes IoT comme substituts de CSS, car ils présentent la contemporanéité et la multidisciplinarité que nous souhaitons étudier.

# INDEX

# 1  Introduction

*This chapter presents the context of the work, along with motivation and the research questions. This section also presents the objectives, the research methodology adopted and the organization of the text.*

## 1.1  Motivation

In the past, software applications were not integrated in the daily activities, leading to a barrier to scalability, with much effort to customize and maintain since they were independent of each other. A solution was to go for the "big software," more extensive and more extended projects with standardized processes that end up with a more expensive one-size-fits-all approach to technology that failed due overwhelming technical complexity and inflexibility (Andriole, 2017). However, even with such limitations, the advancement of technologies made the society rely on software-based systems increasingly. This scenario brings to the discussion the software development for current paradigms such as the Internet of Things (IoT), Industry 4.0, Intelligent Cities, Context Sensitive Systems, among others that make up a set of Contemporary Software System (CSS). These systems rely on software solutions for their complete operation and offer the opportunity for a reality where "things" can act, products can "command" production lines and other features different from conventional computing solutions.

Contemporary means *to be marked by characteristics of the present period, or belonging to or occurring in the present* (Oxford Dictionary) and *existing or happening now* (Cambridge Dictionary), in contrast to the previous initiatives.

New challenges are emerging as a result of these new possibilities such as the higher need for the software to be embedded in the product (Miranda *et al.*, 2015; Lu, 2017) and technology diversity and multidisciplinarity to deliver the variety of possible solutions (Chapline and Sullivan, 2010; Gubbi *et al.*, 2013) considering communication and interoperability, essential for the paradigm materialization (Gyrard, Serrano and Atemezing, 2015; Lin *et al.*, 2017). Thus, attention to the development of software with a holistic vision integrated with different disciplines can represent an excellent differential for the development of such systems, since complex systems require systems engineering that integrates across each part to meet requirements (Chapline and Sullivan, 2010). It is necessary given the high uncertainties about the system and its problem domain, the multidisciplinarity among the solution and the business needs regarding CSS. This scenario promotes a high degree of innovation where software engineers need to build new software technologies to solve new problems, many of which are still unknown (Atzori, Iera and Morabito, 2010; Haller, 2011).

Our interest as a research group in the area of emerging and contemporary software technologies started with an investigation concerned with **Pervasive** and **Ubiquitous Systems** (Spínola, Massollar, and Travassos, 2007; Spínola and Travassos, 2012; Mota, 2013). These two terms are intimately connected, and some authors have addressed them interchangeably (Satyanarayanan, 2001; Baldauf, Dustdar and Rosenberg, 2007; Spínola, Pinto and Travassos, 2008). The working on these topics is usually motivated by the idea that "the most profound technologies are those that disappear" as stated in (Weiser, 1991). In his seminal work, Weiser defines ubiquitous computing as being the use of the computer through its availability in the physical environment, making it effectively invisible to the user, and proposes that in the future computers will be embedded in the environment, invisible to the users, becoming ubiquitous and creating a new paradigm to access information and to interact with devices. A software system can be considered ubiquitous according to their adherence to ubiquity characteristics (Spínola and Travassos, 2012): context sensitivity, adaptable behavior, service omnipresence, heterogeneity of devices, experience capture, spontaneous, interoperability, scalability, privacy and trust, fault tolerance, quality of service, and universal usability. Ubiquity becomes a transversal property of a software system as they fulfill these ubiquity characteristics.

In our interpretation, Ubiquitous Computing is one of the perspectives that motivates CSS. It was the beginning of a new paradigm that changes the style and form of interacting with software beyond the traditional desktop, bringing it to the larger real world. It is a challenge since it is not a well-understood or well-controlled environment, it is, however, complex and dynamic, with an ever-changing context of use.

Following our research motivations, we delve deeper into the context awareness feature investigating **Context-Aware Software Systems**. Several works have been developed in this direction, primarily concerned with testing and interoperability in this type of system (Motta, Oliveira, and Travassos, 2016; Matalonga, Rodrigues and Travassos, 2017; Santos *et al.*, 2017). "Context" is defined as any piece of information that may be used to characterize the situation of an entity (physical objects present in the systems environment) and "context-awareness" as a dynamic property of the system that can affect the overall software system behavior when realizing interaction between an actor and the system (Abowd *et al.*, 1999). Context-Aware systems usually are equipped with identification and sensing capabilities, bridging the physical to the virtual world. It leaves systems closer to what is proposed for ubiquitous systems since it becomes more embedded in the environment. Different technologies can be used, such as RFID tags and smartphones, engaging devices and things with enhanced capabilities such as identification and sensing. With this possibility of expanding the initial capacities of an object and physical objects having a presence in the virtual world, new concepts emerged based on these ideas. From the research done and the knowledge acquired in these areas, we decided to look at other related areas in order to observe the similarities and differences between them.

Another related area is the **Machine-to-Machine (M2M)** domain, where devices are communicating end-to-end with no human intervention (Madakam, Ramaswamy and Tripathi, 2015). M2M refers to technologies allowing both wireless and wired systems to communicate with the capability of acting (Wan *et al.*, 2013). M2M systems are meant to operate in a specific application, which means that M2M solutions do not allow the broad sharing of data or opened connection of devices into the Internet (Holler *et al.*, 2014). We see M2M as a leading paradigm towards the idea of IoT (Atzori, Iera, and Morabito, 2010).

One more area is represented by **Cyber-Physical Systems** (CPS) that aims "to bring the cyber-world of computing and communications together with the physical world" (Rajkumar *et al.*, 2010; Madakam, Ramaswamy and Tripathi, 2015). According to (Miorandi *et al.*, 2012), "a Cyber-physical infrastructure is the result of the embedding of electronics into everyday physical objects, making them "smart" and letting them integrate seamlessly within the global." CPS is the evolution of M2M systems, contributing to the bridge between the physical and virtual world, in the same manner, but introducing more intelligent and interactive operations (Chen, 2012).

Moreover, a more recent concept is the **Internet of Things (IoT)** that allows composing systems from uniquely addressable objects (things) equipped with identifying, sensing or acting behaviors and processing capabilities that can communicate and cooperate to reach a goal (Motta, de Oliveira, and Travassos, 2018). IoT has emerged as a new paradigm where software systems are no longer limited to computers but can materialize into a great variety of different objects, or to specific users' goals and closed environments. The interaction between humans and the cyber-physical world is changing since software can be deployed everywhere and in everything, such as cars, smartphones, refrigerators, watches, and clothes (Giusto, 2010; Weber, 2010; Zorzi *et al.*, 2010; Haller, 2011). This perspective enables pervasively connecting things (like what is proposed in ubiquitous systems) with identification, sensing or actuation capabilities, which makes possible to interact with our environment (similar to what is expected in CPS).

The IoT paradigm is closely related to the context of **Industry 4.0.** In this case, IoT is deployed in factories and production environments, turning them more intelligent, leading toward the fourth industrial revolution (Wortmann, Combemale and Barais, 2017). Other examples are **smart cities**, **smart homes** and other **smart environments** (Aziz, Sheikh, and Felemban, 2016; Cicirelli *et al.*, 2018), where the *smartness* is directly related to IoT proposal of enhancements in the things, extending their original behaviors or purpose.

Therefore, interactive systems, ubiquitous systems, and IoT compose a new generation of software systems. From our point of view, such contemporary software systems require the integration of different engineering domains (e.g., software engineering, human-machine interaction) and not only software solutions but must make use of various technologies (communications, mobile devices, sensors and actuators, big data, cloud computing, artificial intelligence) (Miorandi *et al.*, 2012; Whitmore, Agarwal and Da Xu, 2015). Currently, it is possible to develop small devices, with embedded intelligence, seamless communication, thing-thing interaction, wireless connections, using different technologies. We can observe that the evolution of several technological areas and the collaboration among them is enabling the realization of what we are calling CSS.

All of these engineering issues justify the need for evolving knowledge, skills, and technologies distinct from those offered to support the traditional engineering of software (Skiba, 2013; Zambonelli, 2016; Larrucea *et al.*, 2017). Therefore, new software engineering research and development challenges emerge in this paradigm, without prejudice to the original software life-cycle concerns with deadlines, costs and quality levels of products and processes (Pfleeger and Atlee, 1998; Fitzgerald and Stol, 2017), but involving the intensive internalization of software into the products, high distribution of solutions, diversity and technological multidisciplinarity, communication and systemic interoperability.

Due to the broadness and importance of IoT, we get motivated to work with it as a surrogate for CSS. This resolution was reached since from the many paradigms composing CSS, IoT brings characteristics that represent well the contemporaneity and multidisciplinarity that we aim to investigate. Besides, it is considered an enabler for other areas (Abuarqoub *et al.*, 2017; Trappey *et al.*, 2017). Also, there is a great effort and interest in academia and industry for the advancement of software technologies regarding this paradigm (Lee and Lee, 2015; Caron *et al.*, 2016). Therefore, it is going to be useful for our proposal to have more data points, inputs, and resources to investigate.

Our motivation to investigate and contribute to the evolution of CSS engineering, focusing at the IoT paradigm, is therefore supported by:
1. The relevance of CSS (and IoT) in the national and international environments (Borgia, 2014; CNI, 2016; Lu, 2017; BNDES, 2018);

2. The need for a holistic approach and multidisciplinary view for the development of new software solutions (Higgins, 1966; Chapline and Sullivan, 2010; de Lemos *et al.*, 2013; Gubbi *et al.*, 2013; Bauer and Dey, 2016; Aniculaesei *et al.*, 2018).
3. The demand for technical competencies and skills detained by different practitioners to engineer such software systems (Yan Yu, Jianhua Wang and Guohui Zhou, 2010; Movahedi *et al.*, 2015; de Farias *et al.*, 2017; Desolda, Ardito and Matera, 2017).
4. The lack of proper software engineering methodologies to support the engineering of CSS (and IoT) (Zambonelli, 2016; de Farias *et al.*, 2017; Jacobson, Spence and Ng, 2017; Larrucea *et al.*, 2017).

## 1.2 Research Problem and Questions

Software Engineering, as a discipline, has undergone constant changes since its conception, with the Internet being an evolution that strongly influenced changes in the area. With each change, it becomes clear the need to evolve the software technologies previously proposed and used for the development of computational solutions (Jacobson, Spence and Ng, 2017). Some concepts, methods, tools, and standards have been proposed to support the development of contemporary software systems (Zorzi *et al.*, 2010) but we do not yet have established solutions (Zambonelli, 2016; Larrucea *et al.*, 2017). The engineering of IoT and others CSS is not limited only to computing or software, but it covers several problem domains and can be applied in different solution domains. Using the IoT paradigm as a surrogate for CSS, its applications present a set of characteristics that differ from conventional ones. Therefore, CSS drives us to "engineer" multidisciplinary solutions involving, in addition to Software Engineering, the integration of different disciplines for the accomplishment of successful systems and by its purposes, emphatically including the presence of software, essential for the materialization of systemic solutions.

During our investigation, observing the challenges in engineering such new software systems, we argue that Software Engineering for CSS and IoT needs more than a single perspective since these solutions usually cover other disciplines (network, hardware, and others) alongside software. Thus, our concerns are configured in a multidisciplinary way. The notion of IoT departs somewhat from the notion of a pure and straightforward software system, demanding approaches more closely to the comprehensive view of Systems Engineering. The purpose of Systems Engineering is to embrace multidisciplinarity, uniting the areas necessary for the realization of successful systems according to its purposes, including the part of Software (BKCASE Governing

Board,2014). Therefore, we conjecture that the principles of Software Engineering should intertwine with those of other disciplines in order to deliver contemporary and adequate engineered solutions with a strong software emphasis, composing a comprehensive view of Software Systems Engineering.

The IoT characterization (presented in Chapter 3) allows observing that its characteristics are orthogonal to the areas associated with the applications and bring challenges to its development and quality (Atzori, Iera, and Morabito, 2010). As far as we could observe, we cannot confirm that existing software engineering technologies are adherent to the construction of CSS and IoT applications (Zambonelli, 2016; Aniculaesei *et al.*, 2018). This scenario impacts their development, revealing risks to project decisions and the development process itself. Therefore, the problem to be addressed in this thesis is to support decision-making on engineering CSS (and IoT) considering its characteristics and multidisciplinarity.

The main research question of this thesis proposal is formulated as follows:

**How to support the decision-making on engineering CSS (focusing on IoT applications) considering its characteristics and multidisciplinarity?**

This research question can be broken down into the following secondary research:
**RQ1.1)** What characterizes IoT applications?
**RQ1.2)** What are the challenges for engineering IoT applications?
**RQ1.3)** What areas are involved in engineering IoT applications?
**RQ1.4)** What software technologies support the engineering of IoT applications?

## 1.3 Research Goals

Because of CSS immense potential, in addition to presenting a characterization of the area and organizing the existing challenges (Chapter 3 and 4), with this work we want to propose a framework that:

- Is generic enough, at a higher level of abstraction, to represent all the particularities and characteristics of the CSS paradigm, starting with IoT applications.
- Is flexible enough to be extended and evolved so that it continues to represent contemporaneity. The proposed update procedure is based on Rapid Reviews protocols (Chapter 4).
- Is adaptable enough so that it can be instantiated more concretely in the different paradigms that compose CSS. The proposed instantiation in this work will be in the context of IoT, working as a surrogate to CSS (Chapter 5).

The primary objective of this work is to propose a framework to support the decision-making regarding the engineering of CSS (focusing on IoT applications) as depicted in the RQ´s. This primary objective can be broken down and better detailed in the following sub-objectives:

- Investigate what characteristics define IoT applications and differentiate them from conventional ones;
- Investigate what the challenges on engineering IoT applications are;
- Investigate what disciplines are involved in the development of IoT applications;
- Identify software technologies supporting the engineering of IoT applications;
- Organize a body of knowledge regarding the engineering of CSS (focusing on IoT);
- Define a framework to use such a body of knowledge to support the decision-making on engineering IoT applications, considering their characteristics, challenges and involved technologies and disciplines;
- Develop a computational infrastructure to enable and ease the framework use;
- Evaluate the proposed framework and its computational infrastructure through a family of experimental studies, in order to assess its feasibility, applicability, and validity.

## 1.4 Research Methodology

According to (Kitchenham, Dyba and Jorgensen, 2004) the use of quality processes for software engineers is not a sufficient condition for the improvement of quality in the development. It is recommended the characterization of technology before its adoption in a way to determine its feasibility, contributing to Evidence-based Software Engineering.

The research methodology to be used in this work is adapted from (Spínola, Dias-neto and Travassos, 2008). This methodology relies on primary and secondary studies to support the conception of new software technologies. We selected this methodology because it is adequate for the research purpose since it is an evidence-based approach to technology proposal. The methodology is composed of two main phases. The first stage, the conceptual phase, involves the execution of a secondary study with the objective of obtaining an initial proposal for the technology. In the second phase, named Development and Evaluation phase, different studies are planned and executed to give us evidence of the feasibility, applicability, and validity of the proposal, contributing to an incremental development with its improvement (Figure 1).

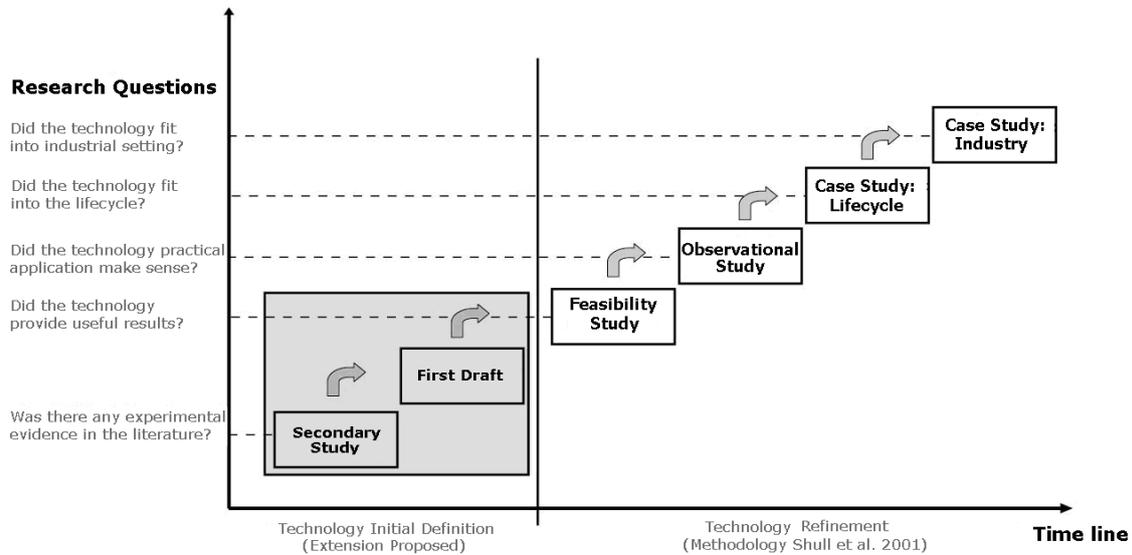

Figure 1. The methodology proposed by (Spínola, Dias-neto and Travassos, 2008).

We adapted this methodology as detailed in Figure 2, the colored part was executed so far, and the gray part represents the expected next steps. For the conception phase, four activities were executed:

1. Secondary Study - to characterize IoT about its definition, attributes, and current applications. We followed adequate procedures focusing on secondary studies. This activity is presented in Chapter 3.

2. Investigate IoT concerns - to recover issues based on technical literature, field professionals and public initiative. In this way, it is possible to find research gaps and the main problems that need an effort for IoT development. This activity is also presented in Chapter 3.

3. Investigate IoT facets - to extend our vision beyond software to systems engineering; capture other disciplines involves in the development. We also present the challenges for IoT development, mapping the concerns in each facet recovered. This activity is presented in Chapter 4.

4. Proposed framework - the initial proposal considers a conceptual organization of a body of knowledge to support the decision-making on engineering IoT applications with tools, methods, and techniques that can contemplate the different facets and perspectives involved. This activity is presented in Chapter 5.

Our next research investigation is for Development and Evaluation phases. The Development phase aims to operationalize the framework. An initial activity is to propose a Problem Characterization Template, which aims to define the problem domain from a conceptual perspective. We want to propose the Problem Characterization Template based on the findings from the technical literature and from real projects. In the development phase, we also aim to fill the conceptual body of knowledge proposed by for each facet with inputs from technical literature; from real projects and from interviews with practitioners. And propose a decision-making strategy based on the problem domain characterization and the content of the body of knowledge. A computational infrastructure is proposed to assist the framework usage, also to be developed in this phase.

The Evaluation part involves the application of primary studies to assess the technology as proposed by Shull *et al.* (Shull, Carver and Travassos, 2001). The proposed solution needs to be evaluated regarding its feasibility, applicability, and validity, different studies are planned for this purpose. This phase also involves improving the proposal from the study's results, according to the study cycle presented. Initially, we established a generic cycle that begins with the **definition** of the type of study and the purpose within the research and will guide the **planning** phase, in which we must follow established guidelines in the technical literature (Wohlin *et al.*, 2012). The study will then be **executed** and **analyzed** according to the appropriate scope. After the analysis the results could contribute to **refining** the proposal (that means the proposed framework) as improvements and new studies could be executed in the same way if necessary.

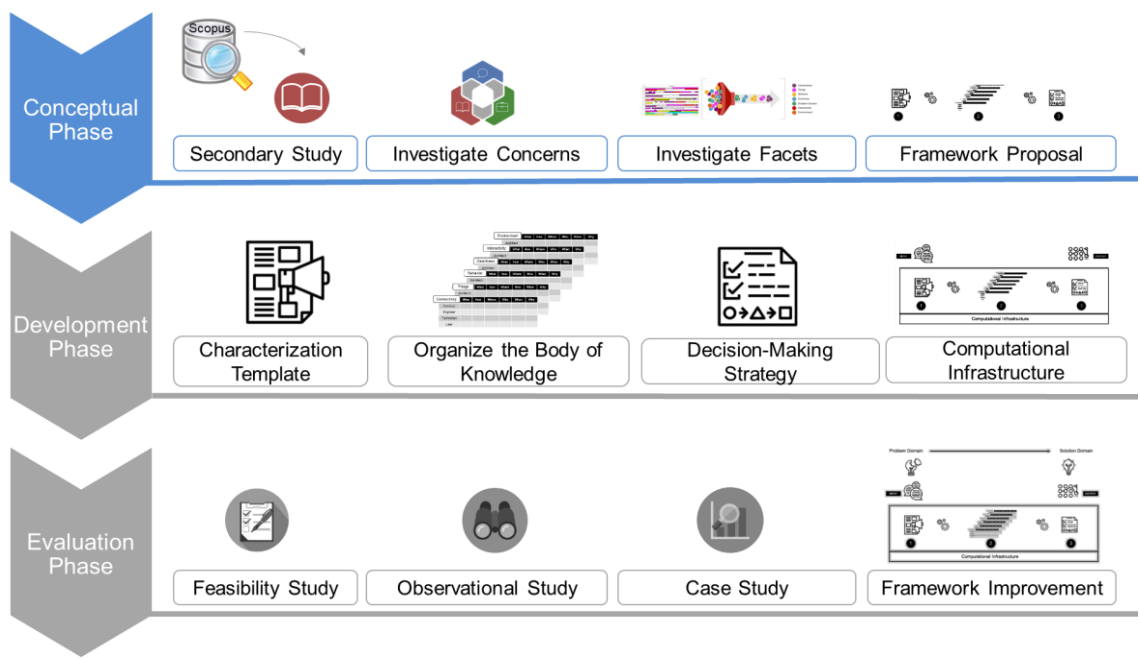

Figure 2. Details of the Research Methodology performed so far.

As presented in the previous section (1.1 Motivation) our study on this topic emerged while working with ubiquitous computing, context-aware applications and other software systems that lead us to investigate IoT as well. IoT is now one of the CSS that had gained considerable attention in the community being chosen to the applicability of this work. As consequence, we have performed the Conceptual phase focusing on IoT applications. For the framework proposal (Chapter 5), we consider other aspects of CSS to extend the vision of our study, but the central point of our investigation remains on IoT.

Therefore, we consider outside the scope of this proposal to implement the proposal in all the different CSS paradigms and to cover all the traditional phases in the development process defined in Software Engineering. We plan to implement the proposal for the IoT context and cover the decision-making in the conception phase of the engineering process.

## 1.5 Organization

This thesis proposal is organized into six chapters. In this first one, we have presented the motivations that led to work on this topic, the research problem and questions as well as the research methodology to be followed.

Chapter 2 presents a theoretical background of this work, presenting concepts of Systems Engineering, the Zachman Framework used as the basis for our research, related work and other relevant content for the development of this work.

Chapter 3 presents the studies conducted in the conception phase to characterize and support the present research. It corresponds to the IoT characterization, recovered from a secondary study and the concerns for its development that comprise sources of technical literature, practitioners and a government report.

Chapter 4 discuss the multidisciplinarity in IoT. First, the concept of facets is introduced, and then each of the facets proposed is discussed in detailed. It involved procedures of qualitative analysis and the execution of seven rapid reviews to support the discussion presented and is an evolution of the IoT characterization first presented, to some of the CSS set.

Chapter 5 presents the framework proposal to support the development of IoT in the context of CSS. The proposal was derived from the results obtained in the studies and based on research in the area.

Finally, Chapter 6 presents the final considerations, the current state of the research, objectives achieved as well as the activities in progress and future activities.

# 2 Theoretical Background

*This chapter presents the theoretical foundations of this work. Through the sections, we present concepts of Contemporary Software Systems, the Zachman Framework, used as the basis for our research, and other relevant content for the development of this work.*

## 2.1 Introduction

This proposal presents a conceptual framework to support the decision-making of engineering CSS. For this, we are based on existing concepts in the software area and the research performed. In the proposal, we are inserted in the CSS context, using a **System Engineering** approach, and the Zachman Framework inspires the f**ramework**. In this chapter, we also present **Related Works** to our theme. These concepts are detailed in the present chapter and are the theoretical foundation necessary for the understanding and realization of our proposal.

## 2.2 Internet of Things Overview

For this overview, we present some of the IoT definitions found throughout our literature review, organized in chronological order to observe how the concept has evolved through the years.

*"An intelligent infrastructure linking objects, information and people through the computer networks, and where the RFID technology found the basis for realization."*
*Defined in 2001 by* (Brock, 2001), *cited by* (Borgia, 2014).

In this definition, we can observe that the idea is to connect objects, information, and people, being those the actors in this system. It makes clear the network necessity as a way to connect the actors, and the realization was limited by the RFID identification technology, which represents the starting point of IoT discussions.

*"Internet of Things as a paradigm in which computing and networking capabilities are embedded in any conceivable object. We use these capabilities to query the state of the object and to change its state if possible." Defined in 2005 by* (Li, Xu and Zhao, 2015), *cited by* (Bandyopadhyay and Sen, 2011).

This other definition does not propose the use of any technology, like RFID, but includes the idea of expanding the original capabilities of an object through technology and brings attention to objects' behaviors. However, to perceive changes in the objects' state, it is only possible by identifying the object first. It leads to an effort to make the things identifiable.

Once they are identifiable, it is possible to make things to communicate automatically (Dunkels and Vasseur, 2008). We consider this as a concept an evolution since this kind of autonomy was not previously discussed. This next definition is also introducing the purpose-idea and reinforce it, even vaguely:

*"A world where things can automatically communicate to computers and each other providing services to the benefit of the human kind." Defined in 2008 by (Dunkels and Vasseur, 2008), cited by (Atzori, Iera and Morabito, 2010; Gil et al., 2016).*

Another definition is:

*"A dynamic global network infrastructure with self-capabilities based on standard and interoperable communication protocols where physical and virtual "things" have identities, physical attributes, virtual personalities and use intelligent interfaces, and are seamlessly integrated into the information network" Defined by in 2009 (Gusmeroli, Sundmaeker and Bassi, 2015), cited by (Borgia, 2014; Whitmore, Agarwal and Da Xu, 2015).*

In this definition, we can see that the central concept of communication and integration remains, but we noticed the introduction of requirements such as interoperability and integration in a seamlessly way. This definition also details what are the things in IoT, as things being virtual or physical, that can have different personalities and may use different communication protocols.

*"The basic idea of this concept is the pervasive presence around us of a variety of things or objects such as Radio-Frequency Identification (RFID) tags, sensors, actuators, mobile phones, etc. which, through unique addressing schemes, are able to interact with each other and cooperate with their neighbors to reach common goals." Defined in 2010 by (Atzori, Iera and Morabito, 2010), cited by (Miorandi et al., 2012; Gubbi et al., 2013; Singh, Tripathi and Jara, 2014).*

It is one of the most used IoT definitions, and we consider it complete regarding a rationale involving actors, relations among actors, requirements and what it enables. It presents the vast amount and heterogeneity of actors that can engage an interaction, and a requirement to achieve that through unique addressing schemes. In this case, new actors are included, and we can observe that sensing and acting are other possible behaviors that a system can possess, differing from previous definitions. Therefore, these actors can cooperate to reach some goals.

*"Interconnection of sensing and actuating devices providing the ability to share information across platforms through a unified framework, developing a common operating picture for enabling innovative applications. This is achieved by seamless large-scale sensing, data analytics and information representation using cutting-edge ubiquitous sensing and cloud computing." Defined in 2012 by (Gubbi et al., 2013).*

Once more, sensing and acting have essential roles in IoT, as presented in this definition. The vast amount of data collection and sharing among actors can be a source to compose diversified, innovative applications. This definition also makes it clear the multidisciplinary nature of IoT as there are areas that support or leverages it, such as data analytics, ubiquitous and cloud computing.

*"Everyday objects can be equipped with identifying, sensing, networking and processing capabilities that will allow them to communicate with one another and with other devices and services over the Internet to achieve some useful objective (…). Every day "things" will be equipped with tracking and sensing capabilities. When this vision is fully actualized, "things" will also contain more sophisticated processing and networking capabilities that will enable these smart objects to understand their environments and interact with people." Defined in 2015 by (Whitmore, Agarwal and Da Xu, 2015).*

Once the everyday "things" can sense the environment, they become more aware of what is around them, which characterizes context-awareness. In this definition, we see again that the main concern in IoT is to leverage the connection among different things to achieve a system objective. Also, the authors explain that "things" in the IoT context are those objects equipped with identifying, sensing, networking, and processing capabilities, whereas other definitions exemplify things as being the providers of such capabilities, that is, tags, sensors, and actuators.

The goal of this section is to present an overview of IoT from its definitions. We can observe the evolution of the paradigm over the years and what it currently represents, clarifying points of multidisciplinarity, heterogeneity and other characteristics that motivate the proposal of this work. In a broader sense, we can see that other concepts such as context-awareness and ubiquity are present in IoT definitions. By the popularization of the IoT term and for presenting the characteristics that we aim to investigate, it was chosen to be the representative of CSS in the investigations of this work.

Further details of the IoT characterization and the literature review conducted are presented in Section 3.2. The definitions presented are also used for the initial definition of facets, as discussed in Section 4.2.

## 2.3 Systems Engineering

From our research, we observed that new challenges are emerging as a result of these new possibilities provided by contemporary systems as a consequence of the recent advances in technology. Some of the main challenges reported are:

- Software embedded in the product (Miranda *et al.*, 2015; Lu, 2017);
- High distribution of solutions (de Lemos *et al.*, 2013; Jacobson, Spence and Ng, 2017);
- Technology diversity and multidisciplinarity of the solutions (Chapline and Sullivan, 2010; Gubbi *et al.*, 2013);
- Communication and interoperability (Gyrard, Serrano and Atemezing, 2015; Lin *et al.*, 2017).

Considering this context, we argue that the solutions should consider Software Engineering intertwined with other disciplines to deliver engineered solutions, configuring a broader Systems Engineering vision.

Systems engineering presents "an interdisciplinary approach and means to enable the realization of successful systems. Successful systems must satisfy the needs of their customers, users, and other stakeholders" (BKCASE Governing Board, 2014). Usually, software development emphasized the following activities. However, the authors depart from tradition to emphasize the inevitable intertwining of system requirements definition and design, for example.

The scope of Systems Engineering is composed of three complementary areas that contribute to the realization of a successful system (BKCASE Governing Board, 2014):

- **Systems Engineering:** it concerns activities to discover, create and describe in detail a system to satisfy an identified need. The activities are grouped and described as general processes that cover build artifacts, decisions for concept definition, the needs and requirements of stakeholders, and preliminary operational concepts.
- **Systems Implementation:** it uses the structure created during the architectural design and results of system analysis to construct the system elements that meet the stakeholder requirements and system requirements developed in the early phases. These elements are then integrated to form intermediate aggregates and finally the complete system-of-interest.
- **Project / System management:** this area is about managing the resources and assets allocated to perform systems engineering, often in the context of a project or a service. Implementing systems engineering requires the coordination of technical and managerial endeavors. Management provides the planning, organizational structure, collaborative environment, and program controls to ensure that stakeholder needs are met.

Alongside with the areas, System Engineering presents a generic life cycle since "no single "one-size-fits-all" system life cycle model can provide specific guidance for all project situations" (BKCASE Governing Board, 2014). Their proposal covers the following general phases:

- **Definition:** this phase includes Concept Definition, with the need to build or change an engineered system where activities include developing the main concepts of operations and business, and System Definition, where requirements are sufficiently well defined to define a solution
- **Realization:** it begins with the commitment to deliver operational capability and activities include the construction of the developmental elements as well as the integration of them with each other. System Production (improvements), Support (maintenance), and Utilization (operation) stages follow the System Realization.
- **Retirement:** this stage is often executed incrementally as the systems become obsolete or are no longer economical to support and therefore undergo disposal or recycling of their content.

From the problem statement - the solutions in CSS involves more than Software Engineering - the System Engineering in this vision (BKCASE Governing Board, 2014) motivated us to have a multidisciplinary view of the problem. In this section, we briefly present the proposed areas and phases. With our work, we seek to contribute in the **System Engineering** area and in the **Definition** phase proposing a framework that embraces multidisciplinarity and presents a guideline for action based on initial requirements and project characterization reflecting principles of Systems Engineering.

## 2.4 The Zachman Framework

From the data recovered in our research, we realize that concepts and properties related to CSS change according to the context and actors involved. This multifaceted view of CSS shows once again that it is a multidisciplinary paradigm. For this reason, a representation of the concepts should be as comprehensive as possible to represent all aspects involved.

In the latest technologies, software is one of the components since further development is necessary for requirements representation, data infrastructure, network configuration and others (Tang, Jun Han and Pin Chen, 2004). Our aim is regarding the conceptual organization of the data we recovered that should take into account the requirements of different stakeholders and the activities in the different facets related to CSS. Having such a structure, we aim to organize the concepts more explicitly and support the decision-making in engineering CSS systems.

With this goal in mind, we have identified a structure that could support the organization of the concepts: the Zachman Framework (Zachman, 1987). It was introduced in 1987 to comprehend the scope of control within an enterprise and to provide a holistic view of the enterprise architecture that may be used as a base for its management. It still is an essential reference for enterprise architecture, and it is still supported by many types of modeling tools and languages (Goethals *et al.*, 2006).

Zachman's motivation to develop the framework was that "*with increasing size and complexity of the implementation of information systems, it is necessary to use some logical construct for defining and controlling the interfaces and the integration of all of the components of the system*" (Tang, Jun Han and Pin Chen, 2004).

This framework is primarily defined considering a table, crossing perspectives and interrogative questions as presented in Table 1 (Zachman, 1987; Sowa and Zachman, 1992).

Table 1. Zachman Framework.

|  | **Interrogative questions** | | | | | |
|---|---|---|---|---|---|---|
| **PERSPECTIVES** | | **What** | **How** | **Where** | **When** | **Who** | **Why** |
| | **Planner** | | | | | | |
| | **Owner** | | | | | | |
| | **Designer** | | | | | | |
| | **Builder** | | | | | | |
| | **Implementer** | | | | | | |
| | **User** | | | | | | |

The framework formalization its conception was presented as a metaphor from the building architecture to system architecture. The perspectives are therefore described as (Sowa and Zachman, 1992):

- **Planner** - The first architectural sketch depicts in gross terms the size, shape, spatial relationships, and primary purpose of the final structure. In the framework, it corresponds to an executive summary for a planner or investor who wants an estimate of the scope of the system, what it would cost, and how it would perform.
- **Owner** - Next is the architect's drawings that depict the final building from the perspective of the owner, who will have to live in it. They correspond to the enterprise business model, which constitutes the design of the business and shows the business entities and processes and how they interact.
- **Designer** - The architect's plans are the translation of the drawings into detailed specifications from the designer's perspective. They correspond to the system model designed by a systems analyst who must determine the data elements and functions that represent business entities and processes.
- **Builder** - The contractor must redraw the architect's plans to represent the builder's perspective, which must consider the constraints of tools, technology, and materials. The builder's plans correspond to the technology model, which must tailor the information system model to the details of the programming languages, I/O devices, or other technology.
- **Implementer** - Subcontractors work from shop plans that specify the details of parts or subsections. These correspond to the detailed specifications that are given to programmers who code individual modules without being concerned with the overall context or structure of the system.
- **User** - The user perspective was added in a later version and represents the view of the functioning building, or system, in its operational environment.

The framework presents six fundamental questions in the columns to outline each perspective:
- The answer to the question **what** is some type of entity. For Rows 1 and 2 (Planner's and Owner's perspectives), the entities are real-world objects. For Row 3 (Designer's perspective), they are logical information types in the model. For Row 4 (Builders's perspective), they are physical data types in the technology model. For Row 5 (Implementer's perspective), they are more specialized data types for each component.
- The answer to the question **how** is some type of process. For Rows 1 and 2, they are real-world processes. For the lower rows, they are computational functions that model the process.

- The answer to the question **where** is some type of location. For the top two rows, they are locations in the world. For the lower rows, they are logical or physical nodes in a computer network.
- The answer to the question **who** is some type of role played by a person or a computational agent. For Rows 1 and 2, they are persons who play some role in the enterprise. For the lower rows, they may be programs that act for the user at a higher level.
- The answer to the question **when** is time, a subtype such as a date, or time that is coincident with some event.
- The answer to the question **why** is some goal or subgoal that provides the reason that motivates the model for that row

The framework does not prescribe a process, notation, tool or method. The primary purpose is to represent an organization holistically, keeping it simple but comprehensive as a classification scheme. To remain straightforward, Zachman defines seven rules for using the framework:

Rule 1: Do not add rows or columns to the framework

Rule 2: Each column has a simple generic model

Rule 3: Each cell model specializes its column's generic model

Rule 4: No Meta concept can be classified into more than one cell

Rule 5: Do not create diagonal relationships between cells

Rule 6: Do not change the names of the rows or columns

Rule 7: There is no column order. However, the rows should be fulfilled from top to bottom.

The definitions presented here are related to the formalization of the original framework. Since its proposal (1987) and formalization (1992), the framework evolved was implemented for different uses and was the base for several adaptations. In the evolution, the initial name of perspectives were updated for new names: Planner is named Executive, Owner is named Business, Designer is named Architect, Builder is named Engineer and Implementer is named Technician.

Together with its extensive use for enterprise architecture, the framework is suitable for working with complex systems as well as in other areas.

The Zachman Framework has been used to assess the Rational Unified Process - RUP (de Villiers, 2001). The RUP is defined regarding roles, artifacts, activities, and workflows, presenting the lifecycle in temporal terms, using phases and iterations. The four phases are Inception, Elaboration, Construction, and Transition. The idea of the study was to observe RUPs effectiveness regarding its coverage of software development deliverables, using the Zachman Framework. In the paper, the authors tailor the perspectives and questions initially proposed by the framework to fit their purposes (de Villiers, 2001). In conclusion, the authors claim that the Zachman Framework cannot assess the full capabilities of RUP because despite its adequate cover of the static part (addressing the artifacts and their relationships to one another, plus roles and activities and their relationship to artifacts) the framework does not capture the dynamic point of view (how the static aspects relate to each other across the lifecycle).

The Zachman Framework was also used to support a method to infer business activities to support business processes modeling, in order to facilitate the consistent representation of business process (Sousa *et al.*, 2007). It proposed rules to identify business process activities by analyzing the framework dimensions with the questions.

Another work aims to support product traceability along the product lifecycle and presents the Zachman framework as a guideline for applying the IEC 62264[1] standards balancing conceptual and implementation information (Panetto, Baïna, and Morel, 2007). The authors affirm that the framework could define different models at different abstract levels, for different purposes with different views.

The framework has also been used for requirements engineering (Chen and Pooley, 2009; Lee, Ann, and Lee, 2014). In both studies, they use the Zachman Framework for requirements engineering and to provide alternatives for a meta-model to fill each cell in the framework and recommendations for a modeling method.

Zhang et al. used this framework for safety analysis in Avionics Systems (Zhang, Shi, and Chen, 2014). They justified its use by their difficulty in describing "a system composed of the interconnected physical and functional elements. The difficulty is the mixture of the physical and functional layers while no structure is defining the relation instantiation", which was achieved through the framework.

It was also applied to Systems of Systems - SoS (Bondar *et al.*, 2017) where the framework guided the development of SoS architecture, including emergent behavior. In the paper, the essential features of the framework, no specific models, no methodology and no notation, are considered advantages, since it enables a certain level of freedom to the architects and developers to incorporate different modeling techniques.

More evidence on using the framework can be observed in different case studies (Panetto, Baïna and Morel, 2007; Nogueira *et al.*, 2013; Aginsa, Matheus Edward and Shalannanda, 2016), the latter claiming that "*Zachman framework continues to represent*

---

[1] From the International Organization for Standard - IEC 62264-1:2013 for Enterprise-control system integration.

*a modeling tool of great utility and value since it can integrate and align the IT infrastructure and business goals.*"

The flexibility of the framework observed in the works presented and others accessed during the research was one of the factors that motivated us to use it as inspiration and initial conceptual structure for our proposal.

## 2.5 Related Work

For this thesis, we propose a holistic view, based on the principles of System Engineering, for the construction of applications that matches the CSS category defined. Therefore, we searched for related works that fit into the intercession of some of the concepts that we present in this section of theoretical background.

With more practical work, Patel and Cassou (Patel and Cassou, 2015) propose a development methodology and framework to support the implementation of IoT applications. Their approach is designed to address essential challenges (lack of division of roles, heterogeneity, scale, different lifecycle phases) that differentiate IoT applications from others (Patel and Cassou, 2015).

In the methodology, the proposal is based on the separation of concerns: domain, functional, deployment, and platform. Each concern has specific steps to guide the development, implemented in a defined process.

We can see some similarities with the work of Patel and Cassou (Patel and Cassou, 2015) to our proposal. We highlight their strategy to attack multidisciplinarity by using four concerns with a varied set of skills performed by five different roles. However, our proposal is different from that because it has a broader view of the concerns and is more focused on supporting the development team moving out of the problem domain with an action plan stepping into the solution domain. Moreover, since we are worried about different kinds of CSS and not only IoT, our proposal does not define a specific process, but a general framework that may be specialized in each situation. It is better detailed in Chapter 4.

Two interesting works (Alegre, Augusto, and Clark, 2016) and (Sánchez Guinea, Nain, and Le Traon, 2016) are literature reviews, focusing on engineering strategies to develop Context-Aware Software Systems (CASS) and Ubiquitous Systems, respectively.

In (Alegre, Augusto, and Clark, 2016) the results are based on a literature review, and the results of a questionnaire carried out with specialists in CASS. It presents an extensive work in the CASS area, analyzing and characterizing the concept of context and as well as their interaction types and main features. The most exciting part for the perspective of our work is that they search the literature for developing techniques and methods that have been adapted from conventional systems to CASS throughout the most common stages of a development process: Requirements Elicitation, Analysis & Design, Implementation, and Deployment & Maintenance.

In the paper, they present a brief analysis of the different techniques found and conclude that, usually, the proposals are focused on addressing a specific issue in the development independent of each other. Several aspects were presented to justify a lack of a unified vision such as diversity (many alternatives require many developments type in different possible scenarios) and a lack of a shared understanding.

None of the techniques presented fully meets the CASS requirements, and the authors conclude the work by recommending a more holistic and unified approach for the development of CASS and arguing that it should be different from the conventional software engineering approach for creating these systems (Alegre, Augusto, and Clark, 2016).

With a similar motivation, Guinea *et al.* (Sánchez Guinea, Nain, and Le Traon, 2016) performed a systematic review also to investigate development strategies but focusing on Ubiquitous Systems. The review aimed to answer the following questions, with the subsequent main findings:

- **RQ1** - What are the stages of the development life cycle of software for ubiquitous systems that have been mainly considered? They identified 8 phases: requirements, design, implementation, verification and validation, testing, deployment, evolution/maintenance, and feedback.
- **RQ2** - What are the main approaches that have been proposed for each of these stages? They recovered 134 approaches, and 5 of them claimed to address the whole development process.
- **RQ3** - What are the current limitations of such approaches? The authors detail some of the approaches distributed in each phase, showing their strengths and limitations.
- **RQ4** - What are the open issues to be further investigated regarding the development of software for ubiquitous systems? They present 31 open issues, also distributed by the developments phases and highlight nine challenges to develop ubiquitous systems.

The authors conclude the review by indicating that one of the main challenges is the lack of support for developers, since the lack of techniques and methodologies that help developers design and deploy their applications to different ubiquitous systems and there is no support the entire development cycle (Sánchez Guinea, Nain, and Le Traon, 2016). Some of the other challenges presented shows the need for a multidisciplinary strategy to deal with software alongside connectivity, security and other concerns in a unified way.

These two works fit into the context of CSS addressing concepts of context-aware and ubiquitous systems. Although they do not propose solutions, they present an overview of the area that corroborates the motivation of our work regarding multidisciplinarity and the need for a holistic vision.

An interesting work is from Costa *et al.* (Costa, Pires and Delicato, 2017) that more than just presenting the requirements and needs of an IoT application, focus on this challenges and proposes an approach to support the requirements specification for IoT systems named the IoT Requirements Modeling Language (IoT-RML). We share some of the motivations with this work since it states that different perspectives and the heterogeneous nature of IoT should be taken into account in the development. The Domain Model composes their proposal for the abstraction and a SysML profile for the specification. In their model, a stakeholder expresses a requirement as a proposition, and the requirement may influence or conflict with other requirements. Their approach supports both functional and nonfunctional requirements, which is crucial in this scenario.

Through their solution, four requirements specification activities are supported: the elicitation of system's requirements from the stakeholders that will generate an initial model in their tool, the analysis to identify influences and conflicts among requirements updating the model to represent them, then conflict resolution and the last activity is to decide on a candidate solution containing the requirements to be addressed. A proof of concept is presented to illustrate the use of the approach in the context of a smart building, focusing on employees' safety and energy efficiency.

Our proposal can somehow be related to the IoT-RML approach (Costa, Pires and Delicato, 2017). However, we aim to address the problem understanding in the conceptual phase, which focuses on a step before the specification requirements considering a multi-perspective and multidisciplinary strategy.

Another related work is from Aniculaesei *et al.* where they argue that conventional engineering methods are not adequate for providing guarantees some of the challenges specific of autonomous systems, such as the dependability focus of their work (Aniculaesei *et al.*, 2018). Some of the main points discussed is the possibility of adaptive behavior present in CSS, as they adapt their behavior to better interact with other systems and people or to solve problems more effectively, and variations in the context, the formerly closed and valid development artifacts may not capture the changes and be inadequate since the environment and the behavior of the system can no longer be fully predicted or described in advance (Aniculaesei *et al.*, 2018).

In response to these challenges and gaps, the authors propose an approach based on the notion of Dependability Cages. Their approach is to deal with external risks (uncertainties in the environment) and internal risks (system changing behavior), both at the development and at the operation.

Although it is an initial proposal, it is interesting and focused on observing the boundary conditions defined and tested at development time and intervene if necessary to maintain the configuration of the system while promoting an iterative development process with new development artifacts (Aniculaesei *et al.*, 2018). One of the limitations observed in the proposal is related to the multidisciplinarity. The authors identify this aspect of the systems, but the proposed approach does not present a mechanism to deal with it. Another point that is missing is a breakdown about the necessary initial content (say the requirements) to use the approach.

In the current moment of our research, we found a lack of more concrete proposals for the materialization of this paradigm. We aim to address the challenges presented in (Alegre, Augusto, and Clark, 2016) and (Sánchez Guinea, Nain, and Le Traon, 2016), filling the gaps from (Patel and Cassou, 2015), (Costa, Pires and Delicato, 2017) and (Aniculaesei et al., 2018) and in this proposal we intend to focus on the issue of multidisciplinarity. Besides, to support decision-making in the initial phase of understanding during development.

## 2.6  Conclusion

This chapter presented a proposal the theoretical foundation necessary for the understanding and realization of our proposal. We aim to address CSS, with its particularities, with a multidisciplinary approach based on **System Engineering** in a framework by the **Zachman Framework**. In this chapter, we also present **Related Works** to our theme. These concepts are detailed in the present chapter, being. Until the moment of this proposal, the intercession of these concepts and their application of innovative shows in the sense that the literature has not found proposals that fit in this direction

.

# 3 Characterizing Internet of Things

*In this chapter, we present three studies conducted to characterize and support the Internet of Things, the starting point for the present research. The studies comprise sources of technical literature, practitioners and a government report. The results of the studies provided the necessary knowledge as a starting point to the thesis proposal.*

## 3.1 Introduction

Before any decision to direct the proposal, the objective was to characterize the IoT paradigm so that we could observe research opportunities to investigate. Thus, it was defined as one of the specific objectives to identify the characteristics presented by IoT and give an overview of the area, aiming to promote a better perception of current development needs. For this, a literature review was performed, that is partially described in this chapter and with a technical report[2] available. The results of this review were accepted in an article for the *Journal of Software Engineering Research and Development* and are currently under review.

In addition to the characterization, we also wanted to investigate the current status and concerns in the development of IoT applications. At this point, the academy's perspective, achieved through the literature review, was complemented by visions of academia and government, achieved from qualitative studies, to give us a broader view of the development concerns.

This chapter details the literature review to characterize IoT and the other studies to capture IoT concerns.

## 3.2 Internet of Things Characterization

As presented in the introduction, the focus of the research is on contemporary software systems represented by IoT applications, which are quickly emerging and have been widely discussed in our community (Lin *et al.*, 2017). After reviewing the literature and in other researches, we noticed that the characteristics presented in IoT are also part of other paradigms and concepts, especially regarding multidisciplinarity and heterogeneity. IoT fits together CSS since it also involves a multidisciplinary domain, it encompasses a large number of topics, many disciplines, and different technologies requiring professionals with different skills.

---

[2] https://goo.gl/cZVVDc

We performed a literature review as a starting point for the investigations in this work. The review reported in this section is focused on secondary studies and was conducted to investigate IoT. The purpose is to characterize this research area for the further development of this work, as proposed in the Research methodology and highlighted in Figure 3.

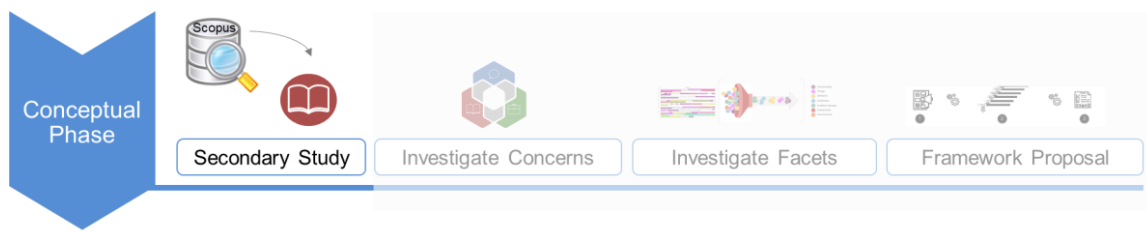

Figure 3. Secondary Study - methodology step.

We followed a defined methodology and guidelines (Budgen and Brereton, 2006; de Almeida Biolchini *et al.*, 2007) in order to provide a formal and well-defined process. With the review, we aim to summarize the technical literature related to IoT and identify possible research gaps as well as expand the conceptual background of this investigation, focused mainly on the characterization of IoT. The following sections of this chapter expose segments of the used protocol as well as the data retrieved and the review contribution. The complete protocol of this review was documented as a technical report[3].

**Planning**: prepare the protocol based on the research questions. The search string should be formulated considering possible terms and synonyms. Studies selection and inclusion criteria are also decided. It is crucial the protocol approval to proceed to the execution step. The summary of the protocol is presented in Table 2.

**Execution**: this step is carried in trials where the search string iteratively evolves aiming to improve precision and recall. Each trial involves reading and consensus from the readers' part in the studies retrieved. Decisions encompass whether to continue and include papers considering the criteria established in the planning step or refine search string and perform a new trial. It is important the reader's consensus to proceed with the analysis step. After applying the search string to Scopus, 76 articles were returned, from which 24 remained after applying the inclusion and exclusion criteria defined in the protocol. After a detailed reading of them, seven were kept for analysis. From these seven we performed Snowballing procedures. It refers to using the reference list of an article or its citations to identify additional material (Wohlin, 2014). In this step, we performed Backward and Forward Snowballing Sampling, tracking down references in the seven articles selected in the previous step and their citations. This step resulted in the inclusion of five new articles.

---

[3] https://goo.gl/cZVVDc

Table 2. Protocol Summary.

| | | |
|---|---|---|
| **Goal** | **Analyze**<br>**With the purpose of**<br>**Regarding**<br>**From the point of view of**<br>**In the context of** | Internet of Things<br>Characterizing<br>its definitions, characteristics and application areas<br>software engineering researchers<br>knowledge available in the technical literature |
| **Research questions** | (RQ1) What is Internet of Things?<br>(RQ2) Which characteristics define IoT applications?<br>(RQ3) Which are the applications for IoT? | |
| **Search string** | **Population** | ("*systematic literature review" OR "systematic* review*" OR "mapping study"<br>OR "systematic mapping" OR "structured review" OR "secondary study" OR "literature survey" OR "survey of technologies" OR "driver technologies" OR "review of survey*" OR "technolog* review*" OR "state of research") **AND** |
| | **Intervention** | ("internet of things" OR "iot") |
| **Search Strategy** | SCOPUS (www.scopus.com) + Backward and Forward Snowballing (Wohlin, 2014) | |
| **Inclusion Criteria** | - To provide an IoT definition; OR to provide IoT properties; OR to provide applications for IoT. | |
| **Exclusion Criteria** | - Not provides an IoT definition; AND not provides IoT properties; AND not provides applications for IoT; AND studies in duplicity; AND register of proceedings. | |
| **Study type** | Secondary Studies | |
| **Acceptance Criteria** | Three distinct readers:<br>- all readers accept => paper is accepted<br>- all readers exclude => paper is excluded<br>- the majority of accept, others in doubt => paper is accepted<br>- else => discuss and consensus | |
| **Technical Report** | Detailed information about the planning and execution - https://goo.gl/cZVVDc | |

In total 12 articles compose our final set for the review (Atzori, Iera and Morabito, 2010; Bandyopadhyay and Sen, 2011; Miorandi *et al.*, 2012; Gubbi *et al.*, 2013; Xu, He and Li, 2014; Borgia, 2014; Singh, Tripathi and Jara, 2014; Whitmore, Agarwal and Da Xu, 2015; Madakam, Ramaswamy and Tripathi, 2015; Gil *et al.*, 2016; Sethi and Sarangi, 2017; Trappey *et al.*, 2017). We used an extraction form to retrieve the following information from the secondary sources: Reference information, Abstract, IoT definition, IoT related terms, IoT application features, IoT application domain, Development Strategies for IoT, Study Type, Study Properties, Challenges and Article focus.

**Analysis**: the readers agree upon a set of candidate papers, considering the inclusion/exclusion criteria. After full reading, the candidate papers, extract relevant data based on the extraction form. In this step, based on the results, we performed a qualitative analysis based on Grounded Theory (GT) procedures (Strauss and Corbin, 1990) in part of the findings. In the analysis phase, we found 28 IoT definitions, 28 characteristics and several applications domains to answer our research questions.

**Packaging**: this step is performed through all the review process, aiming to document every decision in each activity, as well as the information collected and analyzed.

The complete discussion is detailed in a technical report, and we briefly present here a summary of the answers to the research questions.

- **RQ1: What is "Internet of Things"?** The 12 selected papers supported the extraction of 28 different IoT definitions. From the analysis of these 28 definitions, we noticed that the existing definitions followed a specific pattern in their structure, in the concern of explaining the actors involved, the requirements and the consequences of relations among actors as part of a system - not necessarily presented in all definitions. We considered this structure not to limit our interpretation, but to support a more thorough IoT concept understanding and thus finding an appropriate and updated definition for this work. In the report, we organized some of the definitions found in chronological order to observe how the concept has evolved. In our understanding, the "things" in the IoT context exist in the physical realm, such as sensors, actuators and anything that is equipped with identification (tag reading), sensing or actuation capabilities, which excludes entities in the Internet domain (hosts, terminals, routers, among others). The things should also have communication, networking and processing functionalities varying according to the systems requirements. In the beginning, the things in IoT systems were objects attached with electronic tags, so these systems present the behavior of Identification. Subsequently, with the evolution of the concept, sensors, and actuators begun to be part of the paradigm and enabled the Sensing and Actuation behaviors respectively. It means that an IoT system may have Identification, Sensing or Actuation behaviors, or a combination of them. To answer RQ1, from the understanding of all the definitions found, IoT is a paradigm- that allows composing systems from uniquely addressable objects (things) equipped with identifying, sensing or acting behaviors and processing capabilities that can communicate and cooperate to reach a goal. This definition helped us to unify our understanding of the research regarding

IoT and motivated us to follow in the direction taken, and it is from this definition that other activities were carried out.

- **RQ2: Which characteristics define an IoT domain?** The 12 papers provided 211 excerpts, which were coded following the principles of open coding, as described in Grounded Theory (Strauss and Corbin, 1990), from what we identified 28 characteristics (Table 3). One point of discussion is that the authors do not define all the characteristics presented in the articles or referred to the original work defining them. The lack of definitions hinders the research and understanding of the area since we cannot know the characteristic's meaning or what the author meant by that. It makes it challenging to characterize IoT and to develop more suitable solutions that meet all the desired characteristics, since they were not defined, only listed. For the same reason, it is not possible to infer that the authors are discussing the same issues, such as efficiency for instance, which from the sources can be regarding cost, size, resources or energy. We list the characteristics without definition and detail the defined characteristics in Table 4.

Table 3. IoT Characteristics.

| Characteristics | # |
|---|---|
| Characteristics not defined | 19 |
| Characteristics defined | 9 |
| Total | 28 |

List of characteristics **not defined** by the papers in the set: Accuracy, Adaptability, Availability, Connectivity, Efficiency, Extensibility, Flexibility, Manageability, Modularity, Performance, Privacy, Reliability, Robustness, Scalability, Smartness, Sustainability, Traceability, Trust and Visibility. Even with the lack of definition, these characteristics are relevant for the characterization scenario of IoT systems.

List of characteristics **defined** by the papers in the set:

Table 4. IoT Defined Characteristics.

| Characteristic | Definition | Reference |
|---|---|---|
| Addressability | The ability to distinguish objects using unique IDs | (Atzori, Iera, and Morabito, 2010; Bandyopadhyay and Sen, 2011; Miorandi *et al.*, 2012; Borgia, 2014). |
| Unique ID | It is necessary for unique identification for every physical object. Once the object is identified, it is possible to enhance it with personalities and other information and enable the control over it | (Atzori, Iera, and Morabito 2010; Bandyopadhyay and Sen 2011; Borgia 2014; Gubbi *et al.* 2013; Li, Xu, and Zhao 2015a; Miorandi *et al.* 2012) |
| Object Autonomy | Smart objects can have individual autonomy, not needing direct human interaction to perform established actions, while reacting or being influenced by real/physical world events | (Atzori, Iera, and Morabito, 2010; Gubbi *et al.*, 2013; Madakam, Ramaswamy and Tripathi, 2015) |
| Mobility | Object availability of across different locations | (Atzori, Iera, and Morabito, 2010; Bandyopadhyay and Sen, 2011; Borgia, 2014; Sethi and Sarangi, 2017) |
| Autonomy | Refers to systems not needing direct human intervention to perform established actions such as data capture, autonomous behavior, and reaction | (Atzori, Iera and Morabito, 2010; Miorandi *et al.*, 2012; Gubbi *et al.*, 2013; Borgia, 2014; Whitmore, Agarwal and Da Xu, 2015; Sethi and Sarangi, 2017) |
| Context-awareness | The use of context to provide task-relevant information and/or services to a user | (Atzori, Iera, and Morabito 2010; Bandyopadhyay and Sen 2011; Borgia 2014; Gubbi *et al.* 2013; Li, Xu, and Zhao 2015a; Miorandi *et al.* 2012) |
| Heterogeneity | Several services taking part in the system, which present very different capabilities from the computational and communication standpoints | (Atzori, Iera, and Morabito 2010; Bandyopadhyay and Sen 2011; Borgia 2014; Gubbi *et al.* 2013; Li, Xu, and Zhao 2015a; Madakam, Ramaswamy, and Tripathi 2015; Miorandi *et al.* 2012; Sethi and Sarangi 2017) |
| Interoperability | Interoperability is of three types: Network interoperability that deals with communication protocols. Syntactic interoperability ensures conversion of different formats and structures. Semantic interoperability deals with abstracting the meaning of data within a particular domain | (Atzori, Iera, and Morabito 2010; Bandyopadhyay and Sen 2011; Borgia 2014; Gubbi *et al.* 2013; Li, Xu, and Zhao 2015a; Madakam, Ramaswamy, and Tripathi 2015; Miorandi *et al.* 2012; Sethi and Sarangi 2017) |
| Security | To ensure the security of data, services, and entire IoT system, a series of properties, such as confidentiality, integrity, authentication, authorization, non-repudiation, availability, and privacy, must be guaranteed | (Atzori, Iera, and Morabito 2010; Bandyopadhyay and Sen 2011; Borgia 2014; Gubbi *et al.* 2013; Li, Xu, and Zhao 2015a; Madakam, Ramaswamy, and Tripathi 2015; Miorandi *et al.* 2012; Sethi and Sarangi 2017; Whitmore, Agarwal, and Da Xu 2015) |

- **RQ3: Which areas of application do use IoT?** Several application domains will leverage the Internet of Things advantages. All the application domains are only examples of areas that benefit from IoT or are supposed to do it in the future. As declared in Whitmore *et al.* "the domain of the application areas for the IoT is limited only by imagination at this point" (Whitmore, Agarwal and Da Xu, 2015). Atzori *et al.* (Atzori, Iera, and Morabito, 2010) [1] describe five domains: (A) Transportation and logistics, (B) Healthcare, (C) Smart environment (home, office, plant), (D) Personal/social and (E) Futuristic domain (whose implementation of such applications is still too complicated). Gubbi *et al.* (Gubbi *et al.*, 2013) describe (A) Personal and Home, (B) Enterprise, (C) Utilities, and (D) Mobile domain. Also, there is also a classification of the applications for Consumer (Home, Lifestyle, Healthcare, Transport) and Business (manufacturing, retail, public services, energy, transportation, agriculture, cities, and others) (Trappey *et al.*, 2017). Those domain categorizations can be a subpart of a categorization, which grouped the applications in three major domains (Borgia, 2014): (A) Industrial domain, (B) Smart city domain, and (C) Health well-being domain. They are not isolated from each other, but there is a partial overlapping since some applications are shared across the contexts. For example, tracking of products can be a demand for both Industrial and Health well-being domains.

With this review, we addressed the purpose of a general IoT characterization, presenting a definition and identified characteristics. It was an initial step in the Conceptual phase of the proposal, and one of the first contributions in this work is the knowledge organized and presented in the Technical Report[4].

After the IoT characterization, we performed different studies to complement this characterization and to identify the main issues and concerns when dealing with IoT. This set of studies was one of the initial activities of the research and focused on characterizing only IoT, and its challenges served as the basis for a later, more complete adaptation for CSS (Figure 4).

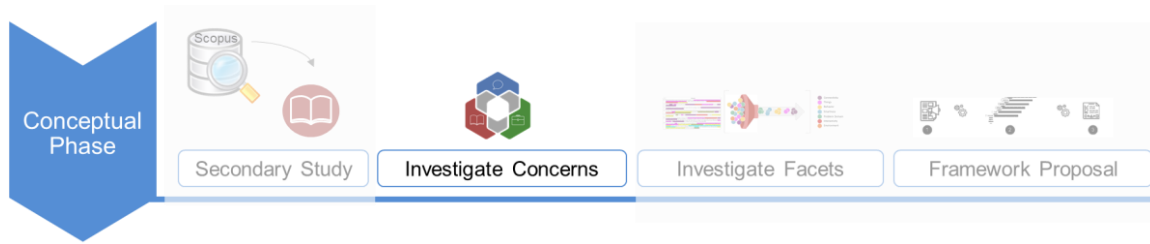

Figure 4. Investigate Concerns - methodology step.

Each study was planned considering a specific perspective on the subject. Initially, we contemplate the perspective of the academy, recovered through a literature review previously presented. Then we decided to broaden the range to represent two other perspectives collected from industry, with practitioners and a government report, contributing to a more comprehensive representation, retrieved through discussions at technical events and a government report, respectively (Figure 5).

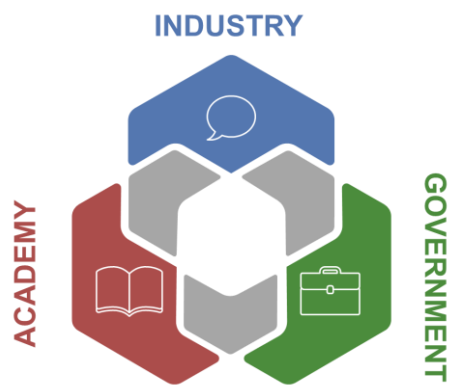

Figure 5. The three characterization Perspectives.

---

[4] https://goo.gl/cZVVDc

Although they are different visions, they discuss the same topic. Thus they become complementary giving us a more comprehensive view of the area. Throughout the next sections, we show the impressions of each perspective and each study is detailed with planning, execution, and results.

For the analysis of the data resulting from each study, we rely on the procedures of qualitative analysis, based on Grounded Theory (Strauss and Corbin, 1990). The idea is that the analysis arises from and is grounded in research data, through constant comparison and have been extensively used and adequate to Software Engineering research (Seaman, 1999; Carver, 2007; Badreddin, 2013). This approach was selected since GT provides reference support for the procedures and is adequate to work with a large amount of information, such as the data extracted from a literature review and other sources, and to interpret data. Considering that some concepts have different meanings, this methodology is suitable to establish the similarities and differences among them. The same analysis strategy was used throughout the study and is based on *coding* - the process of breaking down, examining, comparing, conceptualizing and categorizing data (STRAUSS; CORBIN, 1998).

## 3.3 Concerns from Academy

The review presented in the previous section we followed a structured process, divided into different steps (Figure 6).

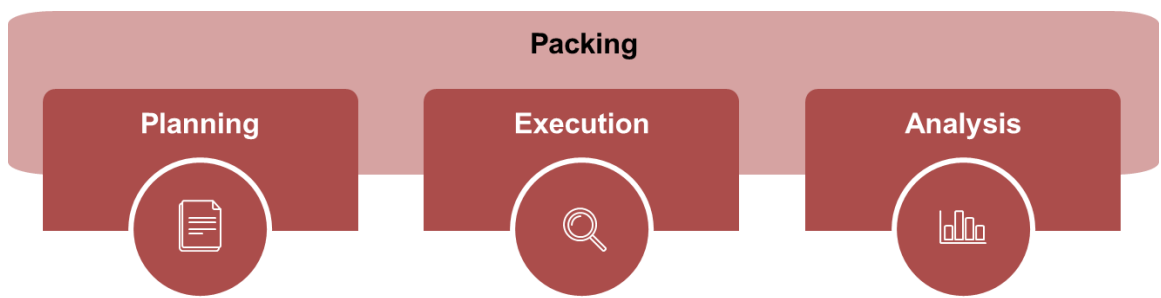

Figure 6. Literature Review Process (de Almeida Biolchini *et al.*, 2007).

Alongside with the analysis to answer the proposed research questions, we also recovered information from issues, challenges, gaps and open questions regarding IoT development, that we are calling here as concerns. The 12 papers provided 38 excerpts regarding IoT concerns. Then we used codes to assign concepts to a portion of data, with constant comparative analysis to identify patterns from similarities and differences emergent from the data. This procedure was based on GT practices (Strauss and Corbin, 1990). This textual analysis was conducted by two researchers, with crosschecking to achieve consensus. The 38 excerpts were organized into seven main concerns:

- **Architecture** -Issues and concerns regarding design decisions, styles and the structure of IoT systems.
- **Data** -It refers to the management of a large amount of data, and how to recover, represent, store, interconnect, search, and organize data generated by IoT from so many different users and devices.
- **Interoperability** - Related to the challenge of making different systems, software and things to interact for a purpose. Standards and protocols are also included as issues.
- **Management** - The application of management activities, such as planning, monitoring and controlling, in the IoT system will raise the interaction of different things.
- **Network** - Technical challenges related to communication technologies, routing, access and addressing schemes considering the different characteristics of the devices.
- **Security** - Issues related to several aspects to ensure data security in the IoT system. For that, a series of properties, such as confidentiality, integrity, authentication, authorization, non-repudiation, availability, and privacy should be investigated.
- **Social** - Concerns related to the human end-user to understand the situation of its users and their appliances.

It is interesting to notice that some concerns can be interrelated, indicating the multidisciplinary nature of IoT. For example: "For technology to disappear from the consciousness of the user, the Internet of Things demands software architectures and pervasive communication networks to process and convey the contextual information to where it is relevant" (Gubbi *et al.*, 2013)., this excerpt is coded for an architectural issue and network as well. Another example is "Central issues are making full interoperability of interconnected devices possible, providing them with an always higher degree of smartness by enabling their adaptation and autonomous behavior, while guaranteeing trust, privacy, and security." (IEEE, 2004), which was coded both for interoperability and for security issues. Provided solutions to the issues presented in the technical literature can be tricky to achieve due to the diversity of concerns, variety of devices and uncertainties in the area.

## 3.4 Concerns from Industry

Another perspective used to recover IoT concerns was the practitioners' opinion. From the characterization obtained with the literature review, we had the opportunity to hear people from industry and academia, who are interested or already work with IoT.

The intent of capturing the information from this source was to increase our observation dataset and triangulate the concerns found in the literature with the ones reported by practice. With this new vision, we deal with other aspects that are relevant and put the research closer to the people who are working in the area.

We performed qualitative studies during two scientific events from which all the participants were working on the IoT domain. Therefore we considered the participants representative, insightful and experienced in the topic. We organized the discussions at the events inspired by the focus group process and experiences from previous studies. The general process with some details is presented in Figure 7.

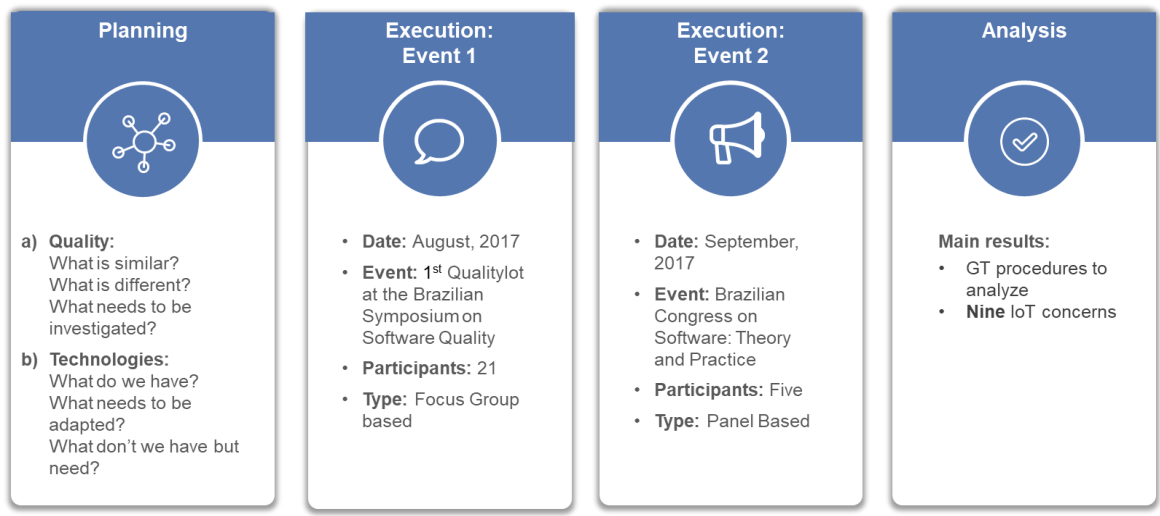

Figure 7. Research Questions.

The questions seek to capture participants' perceptions regarding IoT and to parallel the differences between the conventional and these new applications: a) Regarding product **quality** between conventional software and IoT: What is similar? What is different? What needs to be investigated? b) Regarding the software **technologies** between conventional software and IoT: What can be used directly? What needs adaptation? What don't we have?

For the discussions, we were mainly focused on the quality of the product, the technologies used, and the necessary knowledge of software engineering used in conventional software systems projects and contemporary software systems projects. Both in planning and in execution a researcher assumed the role of moderator accompanying the whole process. The questions aimed to foster discussions and participants were free to express their perceptions.

Based on the outlined questions we had the opportunity to execute the study in two events. In the first event, the 21 participants were divided by their interests into three discussions groups to deal with the mentioned questions in the following perspectives:

- **People:** Discussion focused on human end-user. Challenges and impact of this technology in our daily lives, such as social, legal and ethical. A group composed of five (5) participants.
- **Product:** Discussion focused on IoT products that can be generated, considering the inclusion of software and "smartness" in general objects and the possibilities of new products in this scenario. A group composed of nine (9) participants.
- **Process:** Discussion focused on the software development process that should be included in the *things* and consider the big picture of organizing the *things* together. A group composed of seven (7) participants.

The groups had 1h for discussion. A representative of each group wrote down the main points identified and later presented the ideas for all the participants.

The second event was a panel in the Brazilian Congress on Software: Theory and Practice (CBSOFT) conducted by the same moderator of the first event. In this panel, five (5) practitioners (experts from academy and industry) and audience were motivated to discuss the same previous study questions for 1h30. The moderator acted as the reporter in the panel discussion, gathering the central issues, and producing a document reporting the notes.

At the end of this round of studies, all the notes from both events were collected and analyzed leading to the findings and results discussed here. Discussions were reported through text, and the analysis was based on coding procedures based on GT (Strauss and Corbin, 1990) were used that allowed the identification of nine categories of IoT concerns:

- **Architecture** - More attention is required to the software system architecture since the boundaries between hardware and software are no longer well defined. Also, the architecture should reflect in its conception the concerns on portability and interoperability including a form of orchestrating the connected devices, which is not trivial.
- **Interoperability** - Aside from the primary concern with the interaction of so many different devices, an important issue is how to address the programming for multi-devices. Thus, interoperability can be considered for the development as well.
- **Professional** - The current developers are not entirely prepared to develop for IoT. For the practitioners, the professionals should evolve together with the technologies, so it is necessary an educational evolution and the training of software system engineers.

- **Quality Properties** - Although some specific properties such as interoperability, privacy, and security are primarily discussed, several other quality attributes are considered different in the IoT domain such as capacity (device and network), installation difficulty, responsiveness, context awareness. Contemplate non-functional requirements by considering what the individual sees, feels and how the *things* can contribute to that.
- **Requirements** - Considering the IoT nature, with a tendency for more innovation mainly based on ideas, the requirements can be presented in a less structured form. Another concern is that the user can also be a developer since the solutions reach different types of individuals and devices and new features can be attached.
- **Scale** - To develop, manage and maintain a large-scale software system is a concern. As the number of devices in the software system increases along with the number of relationships, new technologies are needed to maintain a software system with the quality level required.
- **Social** - Aligning the technical with the social, Human-Computer Interaction and User Experience is of great importance in the IoT development and should provide new methods and tools for the IoT scenario.
- **Security** - In the center of many discussions, security-related issues such as privacy and confidentiality are significant concerns, such as the software system scale, mobility, and performance. To balance several dimensions in a secure software system is required to turn IoT into reality, but the current software technologies do not support it yet.
- **Test** - IoT will provide unprecedented universal access to connected devices. Testbed and acceptance tests are sophisticated, and there is a greater need for other types of tests, for example, usability, integrity, security, performance, and context-awareness.

## 3.5 Concerns from the Brazilian Government

In 2016, the Brazilian Federal Government, together with the National Bank for Economic and Social Development (BNDES), began a series of surveys with a prospective vision and with the objective of conducting diagnosis and proposing public policies for IoT. The motivation for this call is based on the tendency of IoT to spread across virtually all sectors of the economy since it is positioned as one of the major technological trends in the Information.

The purpose of the Technical Study performed is to assess the stage and perspectives of implementation of IoT in the world and Brazil, to proposing public policies that potential economic, technological and productive impacts, as well as those linked to the well-being of Brazilian society. In addition to a general diagnosis, the Technical Study should go more in-depth into mapping possible application segments as well as structural and technical issues, which present the greatest balanced potential between the densification visions of the chain productive and impact on the economy and well-being. Based on this in-depth diagnosis, public policies should be developed together with the competent bodies.

The study was planned and executed by the McKinsey / Fundação CPqD / Pereira Neto Macedo consortium selected through the Public Call BNDES / FEP Prospecção nº 01/2016 - Internet of Things (IoT) and all results are public domain and can be accessed for detailed information[5]. In this section, we will present some information about the conducted study (Study Background and Execution). The purpose of our research is to analyze the results obtained (Using the Results) to look for IoT concerns in this perspective.

### 3.5.1 Study Background and Execution

The consortium conducted the planning and execution, we only add here for contextualization, we only based our part in their results that are discussed in the next section. With the objective of conducting diagnosis and proposing public policies in the theme Internet of Things for Brazil and was organized in 4 phases (BNDES, 2017).

The study aimed to have a benchmark with successful international experiences (public policies and projects) that could serve as inspiration and to answer the main questions:

- Which are the primary application segments and structural issues that will be approached?
- What are the technologies to be developed and which are the leading global players?
- What are the challenges/opportunities in the country that IoT can address?
- What are the skills and opportunities for the industry?

These central questions were further detailed and developed throughout the study. The study was performed between January 2017 and March 2018. Their study recovery data from several public sources, among them a Public Consultation, they also conducted interviews with experts from various sectors relevant to the deployment of the

---

[5] https://www.bndes.gov.br/wps/portal/site/home/conhecimento/pesquisaedados/estudos/estudo-internet-das-coisas-iot/estudo-internet-das-coisas-um-plano-de-acao-para-o-brasil

Internet of the Things in Brazil, as well as those obtained five workshops executed during the period. This collaborative effort involving several actors thus constitutes the foundation upon which the results rest.

Both planning and execution were performed by the McKinsey / Fundação CPqD / Pereira Neto Macedo consortium. From our side, in the context of this research, we relied on their results to conduct an analysis based on GT, separated in Figure 8 by the dotted line and detailed in the next section.

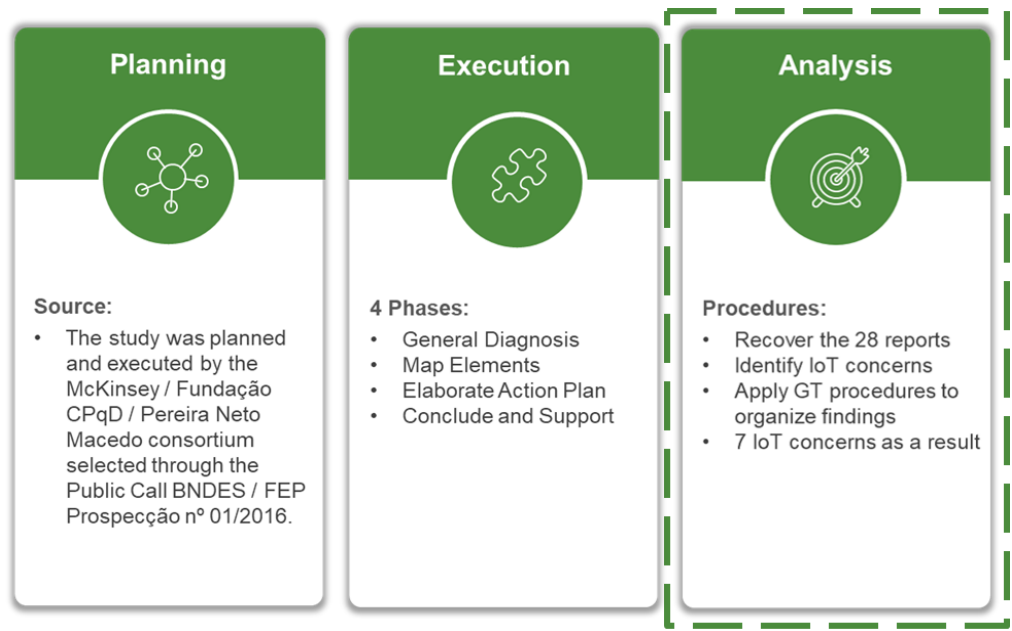

Figure 8. Work division the IoT Technical Study. Only the dotted line was executed in the context of this work.

### 3.5.2 Using the Results

The result of the official study comprises a vast amount of information distributed in 28 documents that serve to serve the strategic purposes that led to the conduction of the study in the first place.

Our interest relies on the set of materials available in a textual format and conducts an analysis. The author conducted the procedure from the complete reading of the content and extraction of portions of data that are mainly associated with challenges, opportunities, gaps, concerns and issues related to IoT from a government perspective, in performed similarly as in the previous studies. The extraction of concerns from the data from the reports of the government study occurred after the execution of the other two studies, contributing to the identification and faster assimilation of the content.

Reading the material allowed extracting information focusing on the presented concerns, analyzing and similarly organizing them as the two previous information sources (the literature review and practitioners). From this material, seven categories of IoT concerns emerged:

- **Regulation** - Governments are working on crucial issues that require significant investment and coordination between the public and private sectors. Within regulatory issues, standardization is one of the most critical, and there is no single strategy to follow. In some cases, it is necessary for the creation of specific laws and institutions regulate privacy and security issues, a topic that is debated today by all the countries mentioned in the report.
- **Interoperability** - To allow devices to communicate with each other, regardless of model, manufacturer or industry. There is a concern that, if left free to the market, the standards developed by technology giants may result in monopolies, leading to the exclusion (or cost-intensive inclusion) of technologies in the global IoT ecosystem.
- **Security** - The vast amount of data generated results in numerous challenges regarding security in IoT, such as increasing the network attack, restricting the devices to support robust security techniques and mechanisms, misuse by the user and even some product design flaws. Thus, security can be considered one of the leading technological concerns of IoT, comprising components of any solution.
- **Professionals** - To invest resources in the training of engineers and other professionals can result in the creation of a strategic differential. However, the scenario is different, so more than proficiency in programming languages of lower level; the professional who develops software for IoT should be able to carry out the customization of solutions already developed for specific demands.
- **Things** - For the devices, which includes their access and gateways there are several non-functional restrictions inherent to IoT that should be present in the products. These restrictions increase the total cost of the objects, such as an energy consumption alternative when it is not possible to connect to the power grid.
- **Network** - There are quite heterogeneous concerns, since IoT covers a number of use cases for which the network requirements are specific, such as: (i) for real-time applications, such as autonomous vehicles, communication latency as well as response time are crucial factors directly related to the network; (ii) applications requiring low data traffic and coexisting with a broad geographic dispersion (e.g., precision agriculture) impose a new paradigm for the evolution of technologies, contrary to what has been developed in the last decade, where the higher bandwidth capacity was predominant. In summary, the IoT access to the network should be heterogeneous, with different technologies composing a vast ecosystem.

- **Data** - The concentration of the data generated and transmitted by smart objects should be processed and analyzed, generating the expected use cases value. Thus, there is the concern of storing and handling a vast amount of data, especially when there are strict low latency and greater agility in response requirements to be met.

## 3.6 Putting all Together

Extracting the perception and concerns of IoT from different points of view was essential for the strengthening and direction of our research. For instance, it is possible to observe that, although there are different perspectives, they become complementary to represent the concerns to produce quality software for this kind of system. Together, the three sources provided 14 different concerns, which must be met in favor of a higher quality IoT software system (Figure 9). We can see that each source has its particularities, and some are consistent with its origin. It is expected that practitioners have a more technical and in-depth view presenting more individual and software-oriented issues regarding IoT software systems. The concerns with **management** and **quality** are transversal to the implementation of such software systems and can be observed in any point of view, but the practitioners have specific concerns of quality, such as meeting non-functional requirements, which bring more specificity and definition to this issue. Also, **requirements** and **testing** issues are still somewhat open on how to represent, describe and integrate software systems. These three aspects must be met in the software systems regardless of their scale, which in IoT software systems can reach ultra-large scale, bringing their associated problems. These three concerns are affected by one aspect that we observed in the literature review. From the characteristics extracted we could observe that properties and characterization are not explicit, neither the characteristics that can affect the development process of such applications. Unclear characteristics can impair requirements, which in turn affects the testing, hindering the overall system quality. We consider that this difficulty is partially due to conceptual aspects, since IoT and the related concepts are not yet established and not enclosed by a single definition, being the concept still under discussion (Shang *et al.*, 2016).

Considering the increasing number of interconnected devices, the size or **scale** of IoT can grow consistently. The systems can achieve a more extensive scale coupled with complicated structure-controlling techniques, which brings new challenges to the design and deployment (Huang *et al.*, 2017). New solutions for architectural foundations, orchestration, and management are essential for dealing with scale issues, especially for Ultra Large Scale Systems such as Smart Cities and autonomous vehicles (Roca *et al.*, 2018).

Concerning **regulation**, some actions are being made, from governments[6] and other institutions[7], to form an adequate legal framework. It is necessary to prompt action to provide guidance and decisions regarding governance and how to operate IoT applications in a lawful, ethical, socially and politically acceptable way, respecting the right to privacy and ensuring the protection of personal data (Caron *et al.*, 2016; Almeida, Doneda and Moreira da Costa, 2018).

For the devices, sensors, actuators, tags, smart objects and all the *things* in the Internet of Things, or Everything, these are some of the aspects that should be taken into consideration: a) resources and energy consumption, since intelligent devices should be designed to minimize required resources as well as costs; b) Deployment since they can be deployed one-time, or incrementally, or randomly depending on the requirements of applications; c) Heterogeneity and Communication: different things interacting with others, they must be available, able to communicate and accessible (Li, Xu and Zhao, 2015; Madakam, Ramaswamy and Tripathi, 2015).

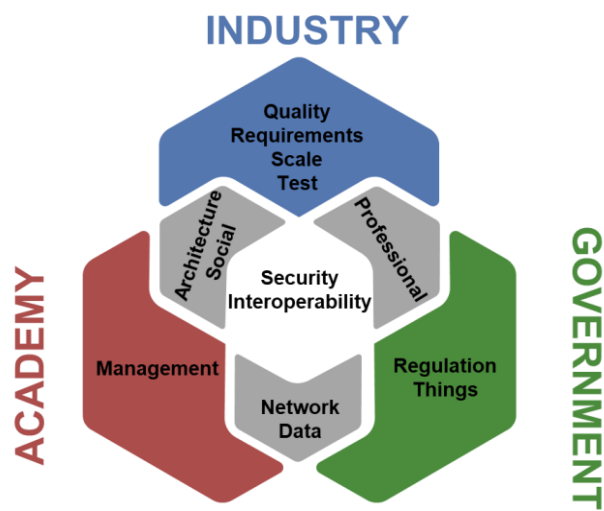

Figure 9. CSS Concerns.

At the intersection between Industry and Literature, we have **architectural** and **social** issues. Both concerns are open due to the area novelty in which there is still an uncovering of how to deal and what to expect. **Architecture** is a recurrent issue in the literature being point out by (Liao *et al.*, 2017) as one of the priority areas for action and reported by (Trappey *et al.*, 2017) to be one of the official objectives of ISO/IEC JTC1. In general, the status is that there still no consolidated standard nor well-established terminologies to uniform advancements for architecture in IoT.

---

[6] https://aioti.eu/ and https://ec.europa.eu/commission/priorities/digital-single-market_en
[7] https://www.kiot.or.kr/main/index.nx and https://www.digicatapult.org.uk/

Regarding **social** concerns, given that the objects, devices and a myriad of *things* are likely to be connected to many others, being people one of the actors as well (Matalonga, Rodrigues and Travassos, 2017), it is necessary to explore the potential sociotechnical impacts of these technologies (Whitmore, Agarwal and Da Xu, 2015). Using such devices to provide information *about* and *for* people are one of the applications. A number of challenges and concerns should be addressed to achieve the benefits aimed with IoT. In facilitating the development is required the design of data dissemination protocols, and evolve the solutions for privacy, security, trust maintenance, and effective economic models (Guo *et al.*, 2012). As affirmed by Dutton 2014, if not designed, implemented and governed in appropriate ways these new IoT could undermine such core values as equality and individual choice.

At the intersection between Industry and Government, we have the concern of **professionals**, with is represented by the preparation of their skills and knowledge as for the teams that should be multidisciplinary to meet IoT premises. If requirements, test and other technical activities are under discussion, we need to think about the professional who will satisfy and perform such activities (Yan Yu, Jianhua Wang, and Guohui Zhou, 2010). With the development of IoT, different people, systems, and parties will have a variety of requirements, one of the abilities required is how to translate these requirements into new technologies and products. Other skills are related to manage the frequency of information generated, manage the ubiquity and actors involved in interactions, develop and maintain privacy and security policies (Tian *et al.*, 2018). As the area is new and is defining the professionals and teams that will work on it too, so it is essential to discuss the professional, develop skills and knowledge necessary for this new generation of innovators, decision-makers and engineers (Kusmin *et al.*, 2017).

Connectivity, Communication, **Network** and the multiple related concepts that enable the evolution of interconnected objects is a critical point for the materialization of IoT (Gubbi *et al.*, 2013). One of the main challenges of this scenario is a vast amount of information identified, sensed and act upon that must be processed mostly in real- or near-real time with an unobtrusive delivery of personalized manner, ensuring availability and reliability of the data and the channel between devices and between the human and devices (Mihovska and Sarkar, 2018). There are many open challenges that require new approaches to a quality network in this scenario. Therefore research should progress into practice to ensure the benefits for the users. Together with Network concerns, we have **Data** issues. In a world with "anytime, anyplace connectivity for anyone and connectivity for anything" (Conti, 2006), we can see how quickly the data can be generated and how vast amounts of information are created. Some of the challenges are related to the continuous and unstructured creation of connection points (devices, things); the persistence of data objects, unknown scale, and data quality (Uncertainty, Redundancy, Ambiguity, Inconsistency, Incompleteness) (Gil *et al.*, 2016).

However, above these, **security** and **interoperability** concerns are at the center of all IoT related discussions. For IoT, for example, it enables computing capabilities in *things* around us and interoperability is the attribute that enables the interaction among heterogeneous devices, with varied requirements of different applications. Interoperability can range in different levels like technical, syntactical, semantic and organizational, which varies according to the software system needs. Complete interoperability is an open question for current software and essential for IoT due to its comprehensive nature. Issues like encryption, trust, privacy, and any security-related concerns are of utmost importance since IoT are inserted in someone's personal life or into the industry. High coverage procedures should guarantee the software system security and trustworthiness.

## 3.7 Validity Threats

Like any empirical study, different threats to the validity of our results can be identified.

- The **literature review** used only Scopus as a search engine, so it may be missing some relevant studies. However, from our experience, it can give a reasonable coverage when performing together snowballing procedures (backward and forward) (Matalonga, Rodrigues and Travassos, 2015; Motta, Oliveira and Travassos, 2016). Data extraction and interpretation biases were mitigated with crosschecking between two researchers and by having a third researcher to revise the results. All phases of this review were peer-revised, any doubt was discussed among the readers, to reduce selection bias. We have not performed a Quality Assessment regarding the research methodology of the selected studies due to the lack of information in the secondary reports. Therefore, it is a threat to this study validity. However, the triangulation with data acquired of practitioners and information extracted from the government report strengthened the representativeness of data and reduced the researchers' bias, powering the results.
- From both the data collected from **industry** and the **government** the interpretation of data was supported by the practices of Grounded Theory, which allowed to get consistency among researchers and shared understanding of the central concepts. However, other perspectives could be used for data interpretation imposing a risk of changing the results. It represents a threat to any qualitative study and constitutes a menace that we cannot completely mitigate.

## 3.8 Conclusion

Based on the results of the studies it is possible to note that, in the context of Software Engineering, although the topic is widely discussed, there seems to be a set of concerns that are still open, with research and development opportunities that should be investigated for the development of these new applications. These concerns represent a wide range of gaps and issues that are not limited to software technologies, but as a set, that should evolve together with the field if we aim to develop quality solutions. As one of the early contributions of this work, it is possible to use these results as a starting point for future research in the community.

From these results, it is possible to see that even in the concerns there is a multidisciplinarity where the difficulties of one area impact on another. The solutions are no longer punctual or individual; it requires aligned and ingrained actions in the system as a whole. For this, presented in the next chapter, is the study conducted to investigate the facets that should work together to develop the solutions in this scenario.

# 4 Characterizing Contemporary Software Systems under the lens of IoT

*In this chapter, we present two activities performed to investigate the areas involved in engineering CSS extrapolating from IoT. One is the qualitative analysis conducted to extract the facets from the results of the Secondary Study for IoT. The other is the Rapid Reviews performed to deepen and detail the vision in each facet. A conceptual organization is proposed, and some challenges discussed.*

## 4.1 Introduction

Our vision in this research seeks to be more comprehensive, in the sense of Systems Engineering, and during the activities, we seek to see the possible disciplines and areas of knowledge involved in CSS, what we are calling facets (Figure 10). We understand facets as "one side of something many-sided" (Oxford Dictionary), "one part of a subject, a situation that has many parts" (Cambridge Dictionary), representing the multidisciplinarity required in such systems.

To support this vision, we have analyzed the material extracted from the IoT literature review in order to arrive at the facets that represent this multidisciplinarity. This is presented in Section 4.2. Also, in this chapter, we also present the rapid reviews conducted for each of the six proposed facets (Section 4.3). At the end of the chapter is presented the union of concerns with the facets, which defined the challenges to engineer CSS (Section 4.4). The six proposed facets are the core that makes up the body of knowledge of CSS (Section 5.2.1). A preliminary version of this discussion was accepted in the XXXII Brazilian Symposium on Software Engineering (Motta, de Oliveira and Travassos, 2018).

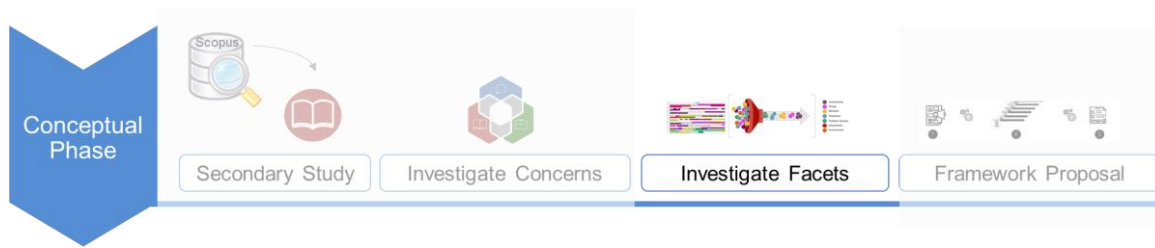

Figure 10. Investigate Facets - methodology step.

## 4.2 From definitions to facets

Aiming at identifying those different facets that characterize this multidisciplinarity, we performed an analysis of the IoT definitions identified in the literature review (Section 3.3) and the analysis was based on GT procedures (Strauss and Corbin, 1990) in the same way as previously defined. The coding procedure leads us to the six facets proposed (Figure 11)

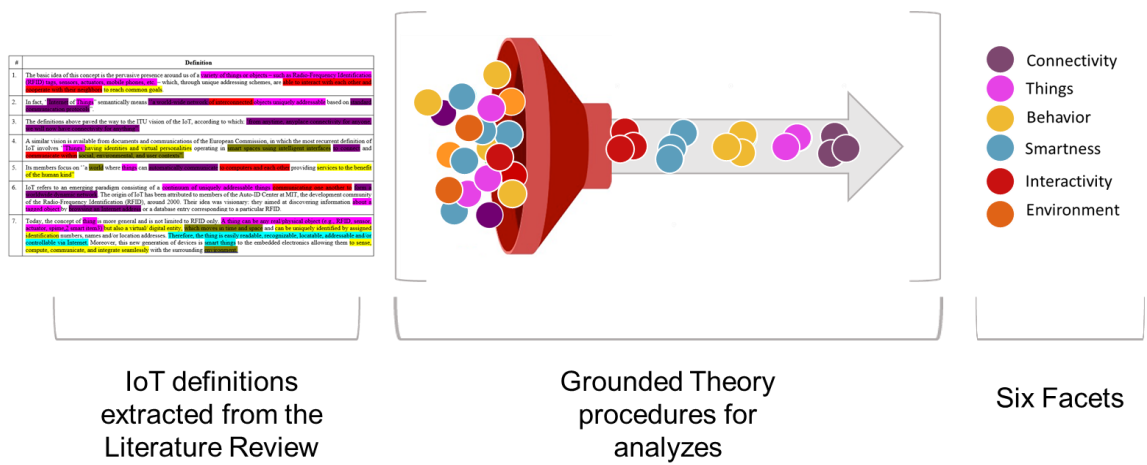

Figure 11. Analysis Procedure performed.

The 28 extracted IoT definitions were organized in a table with one field of "code" to assign an area, topic, discipline (named here as a facet) related to a definition excerpt. This coding process was executed by three researchers separately, using separate and independent documents. An example of the document is presented in Figure 12. It is composed of three columns: a) Index: with the definition number; b) Definition: where each definition is presented as extracted from the paper; c) Code: with the codes associated with portions of the definition, with a color scheme to help their identification.

There were three rounds of discussions, first with two then with all of the three researchers. It was done to discuss the similarity and differences in the coding, support the concepts and reduce bias, until reaching a consensus. From this analysis, we would like to have a set of facets, based on the data we had so far, and be able to sort among the most used to present a set of areas that must be considered. After the documents merge, meetings for discussions were held, some of the discussion was regarding the coding granularity level. For example, network and telecommunication can all be part of a single facet called connectivity, aiming to encompass several concepts and keep the same level of abstraction.

| # | Definition | Code |
|---|---|---|
| 1. | The basic idea of this concept is the pervasive presence around us of a variety of things or objects – such as Radio-Frequency Identification (RFID) tags, sensors, actuators, mobile phones, etc. – which, through unique addressing schemes, are able to interact with each other and cooperate with their neighbors to reach common goals. | Things (Hw), Interoperation, Behavior (Sw), Communication |
| 2. | In fact, "Internet of Things" semantically means "a world-wide network of interconnected objects uniquely addressable, based on standard communication protocols". | Communication (connectivity), Things (Hw), Interoperation |
| 3. | The definitions above paved the way to the ITU vision of the IoT, according to which: "from anytime, anyplace connectivity for anyone, we will now have connectivity for anything". | Communication (Connectivity) |
| 4. | A similar vision is available from documents and communications of the European Commission, in which the most recurrent definition of IoT involves "Things having identities and virtual personalities operating in smart spaces using intelligent interfaces to connect and communicate within social, environmental, and user contexts". | Things (HW), Environment, Smartness (behavior), Communication (connectivity), Interoperation, Socio-Technical |
| 5. | Its members focus on "a world where things can automatically communicate to computers and each other providing services to the benefit of the human kind" | Environment, Thinks (Hw), Communication (connectivity), Interoperation, Behavior (Sw), Socio-Technical (peopleware) |
| 6. | IoT refers to an emerging paradigm consisting of a continuum of uniquely addressable things communicating one another to form a worldwide dynamic network. The origin of IoT has been attributed to members of the Auto-ID Center at MIT, the development community of the Radio-Frequency Identification (RFID), around 2000. Their idea was visionary: they aimed at discovering information about a tagged object by browsing an Internet address or a database entry corresponding to a particular RFID. To address the above idea, they worked on the development of the Electronic Product Code (EPC), i.e., a universal identifier that provides a unique identity for every physical object, with the aim of spreading the use of RFID in worldwide networks. | Thinks (Hw), Communication (connectivity), Interoperation, Environment |

Figure 12. Example of document filled with the definitions and marked with coding.

For the identified excerpts we discussed and organized the understanding in the same level of abstraction for all of them, in order to represent the different needs for the development and construction of IoT software systems. As a result of this process, we came to the consensus (based on the definitions) that for IoT we should take into account six different facets:

**1. Connectivity**

Connectivity is one of the main aspects of contemporary systems. We argue that it is necessary to have available a medium by which things can connect to materialize the CSS paradigm. It is essential some form of connection, a network for the development of solutions, and our idea is not to limit Internet-only connectivity, but to be able to cover other media such as Intranet, Bluetooth, among others, means **the manner by which objects are connected**. This facet was mentioned in 26 definitions.

It is important to note that there is no one-fit-for all solution (Luzuriaga *et al.*, 2015) since it englobes many domains and each one of them will have particular characteristics and requirements. However, we can observe in the literature that specific requirements more related to the devices' nature or the application needs, that influence communication directly - such as low latency, bandwidth, and robustness (Poluru and Naseera, 2017). Even though some of the requirements are not directly related to connectivity, but they show aspects that will profoundly influence communication, thus, they are requirements that need to be well understood and addressed to make CSS work. This facet was mentioned in 26 definitions.

**2. Things**

In this sense, it means the things by themselves in IoT. Tags, sensors, actuators, mobile phones, **all hardware that can traditionally replace the computer** expanding the connectivity reach.

In our understanding, *things* exist in the physical realm, such as sensors, actuators or any objects equipped with identifying, sensing or acting behaviors and processing capabilities that can communicate and cooperate to reach a goal, varying according to the systems requirements (Whitmore, Agarwal and Da Xu, 2015). When an object has enhanced capabilities and uses connectivity to interact with others, it can be considered a *thing* in our context. This facet was mentioned in 25 definitions.

**3. Behavior**

The existence of things is not new, nor their natural capacities. What IoT provides is the chance of **enhancements in the things, extending their original behaviors**. In the beginning, the things in IoT systems were objects attached to electronic tags, so these systems present the behavior of Identification. Subsequently, sensors and actuators composing the software systems enabled the Sensing and Actuation behaviors respectively. It can be necessary the use of software solutions, semantic technologies, data analytics, and other areas to enhance the behavior of things.

The idea of the system behavior results from its constituent parts, that is, the behavior is generated by the interaction and collaboration of two or more devices and a more complex behavior can be managed by the combination of simpler behaviors. The behavior of a CSS can be aggregative and emergent being capable of performing different actions (Jackson, 2015). This facet was mentioned in 13 definitions.

**4. Smartness**

Smartness or Intelligence is related to Behavior but as to managing or organizing it. It is more referring to **orchestration associated with things and to what level of intelligence technology can evolve their initial behavior**.

Artificial intelligence and machine learning techniques can be applied to enhance the intelligence and effective interactions between things to manage smartness. About the development of smart applications, it is critical to highlight that having only sensors collecting data, does not make it smart. For a system to be *smart*, it needs a set of actions, for example treating data, making decisions and acting. The level of the smartness depends on the application domain and user need. This facet was mentioned in 14 definitions.

**5. Interactivity**

It refers to the **involvement of actors in the interaction** to exchange of information with things and the degree to which it happens. The actors engaged with IoT applications are not limited to humans. Therefore, beyond the sociotechnical concerns surrounding the human actors, we also have concerns with other actors like animals and the interactions thing-thing. The degree to which it happens works together with the medium through which things can connect (connectivity) so that in addition to being connected they can understand (interoperability). This facet was mentioned in 4 definitions.

**6. Environment**

The problem and the solution are embedded in a domain, an environment, or a context. This facet seeks to represent such an environment and how the context information can influence its use. The environment is **the place where things are, actions happen, events occur, and people are**. Smart Environments or Smart Spaces provide intelligent services by acquiring knowledge about itself and its inhabitants to adapt to users' needs and behavior (Aziz, Sheikh, and Felemban, 2016). These systems have a set of *things* which are capable of sensing, reason, collaborate and act upon ambient. An essential characteristic of this ambient is the user-centric thinking approach in which all of the systems have to be developed to attend the users in first place. This facet was mentioned in 4 definitions.

**Problem domain**

In addition to the facets with the vision of construction and development, we also perceive the concept of Problem Domain, as usually is perceived in conventional software. A problem domain is the area of expertise or application that needs to be examined to solve a problem. IoT software systems are developed to reach a goal, for a specific purpose. At this point, we are starting from a goal (problem domain) to reach a solution (software system). Focusing on a problem domain is merely looking at only the topics of interest and excluding everything else. It, in general, directs the objective of that solution. We do not see this concept as a facet since it is presented in any software solution. However, it is important to consider since it will direct and contextualize how the other facets will be derived, implemented and managed. This concept was mentioned in five definitions.

From the IoT definition proposed from the findings of the secondary study (Section 3.2) we did the exercise to fit it in the facets proposed to exemplify the following demands to develop an IoT software system:

- **A paradigm that allows composing systems:** IoT is not just the things by themselves. It represents a more substantial aggregate consisting of several parts. It implies that there is not a single IoT solution, but a myriad of options that can derive from the things and other systems available. It will require some domain and business-specific strategies.

- **From uniquely addressable objects (things):** Things should be able to be distinguished using unique IDs, a unique identification for every physical object. It concerns the network solutions and hardware technologies required to devise the composing parts of the IoT paradigm, representing the things facet.
- **Equipped with identifying, sensing or acting behaviors and processing capabilities:** Once the object is identified, it is possible to enhance their original behaviors it with personalities and other information and enable it to connect, monitor, manage and control things. This understanding implies that depending on the "behavior" and "smartness" degree required for a setting. A software solution can be naturally more robust and involve other technical arrangements, such as artificial intelligence.
- **That can communicate and cooperate:** The other part of the paradigm, alongside with the things, is the connection channel of the available things. Together with this network solution, the things should be able to communicate, but not only that. Also cooperate, interchange, interact, and share, with one another and also with other actors and humans, therefore the connectivity and interactivity facets.
- **To reach a goal:** This whole scenario is set for a purpose, for a reason, motivated by something. This primary goal is what will guide the development that is to address the problem, inserted in the problem domain.

## 4.3 From IoT to CSS

The initial conceptual basis for the proposal of the facets was based on the research focused on the IoT (Motta, de Oliveira, and Travassos, 2018). From the inputs and findings in IoT and with the progress of the discussions and research we decided to investigate the facets in the Contemporary Software Systems vision. To do so, we performed a study to confirm the findings and see their feasibility of application in the broader context of CSS. The strategy used was to review the technical literature looking for the facets in the context of CSS and not only IoT.

We conducted the review in the Rapid Reviews (RR) format, which are adaptations of regular, systematic literature reviews made to fit practitioners constraints (Tricco *et al.*, 2015) and begin to be used in the context of Software Engineering (Cartaxo, Pinto and Soares, 2018). The procedure to be performed is the same of regular, systematic literature reviews, and for this, we formatted a generic meta-protocol that was instantiated for each of the six facets presented (**Connectivity**, **Things**, **Behavior**, **Smartness**, **Interactivity,** and **Environment**). Our goal was to observe if these facets are also relevant in CSS. Thus, the reviews sought to answer if each facet represented a concern in the engineering of contemporary software systems. This central question was broken into minor questions regarding *what, how, where, when and why* (5W1H) a facet can be used, verifying the existence of published studies supporting our previous results.

The 5W1H aims to give the observational perspective on a general understanding and characterization of which information is required to the understanding and management of the facet in a system (what); to the software technologies (techniques, technologies, methods and solutions) defining their operationalization (how); the activities location being geographically distributed or something external to the software system (where); the roles involved to deal with the facet development (who); the effects of time over the facet, describing its transformations and states (when); and to translate the motivation, goals, and strategies going to what is implemented in the facet (why), in respect of CSS projects. This was a first step to understand these questions in relation to CSS development, but not as detailed as required by the Zackman framework considering the different perspectives (as presented in Section 2.4).

The revision was carried out in the context of a postgraduate discipline of the program of Systems Engineering and Computing of the Federal University of Rio de Janeiro. The discipline was Special Topics in Software Engineering, and the revisions were carried out by six students at the master's level, being accompanied by one doctoral student and the professor. Follow-up was carried out weekly, and the discussions and doubts handled individually. The facets were distributed randomly through a lottery, and each student was responsible for instantiating the protocol for their respective facet. All protocols share the same PICOC structure - Population, Intervention, Comparison, Outcome, Context (Petticrew and Roberts, 2006), altering only the intervention for the respective facet. The discussion of the strings and the trials were done together with all participants. The execution occurred in the second half of 2018.

All the results are presented in an individual protocol, and a summary of the findings are also presented in the format of Evidence Briefings (EB), one-page documents used as mediums to transfer knowledge acquired from systematic reviews to practitioners that reports the main findings of empirical research (Cartaxo *et al.*, 2016). The meta-protocol and the results of each review were compiled in a single Technical Report with detailed information. Some highlights of the results this study are presented in this section and Table 5 presents a summary of the meta-protocol.

Table 5. Meta-Protocol Summary.

| Goal | **Analyze** | each facet |
|---|---|---|
| | **With the purpose** | of characterizing it |
| | **Regarding** | what, how, where, when and why is used in CSS projects |
| | **From the point of view of** | software engineering researchers |

|  | **In the context of** | knowledge available in the technical literature |
|---|---|---|
| **Research questions** | colspan="2" | Main RQ: "Does <<facet>> represent a concern in the engineering of contemporary software systems?"<br>(RQ1) What is the understanding and management of <<facet>> in CSS projects?<br>(RQ2) How do CSS projects deal with software technologies (techniques, technologies, methods, and solutions) and their operationalization regarding <<facet>>?<br>(RQ3) Where do CSS projects locate the activities regarding <<facet>>?<br>(RQ4) Whom do CSS projects allocate to deal with <<facet>>?<br>(RQ5) When do the effects of time, transformations, and states of <<facet>> affect CSS projects?<br>(RQ6) Why do CSS projects implement <<facet>>? |
| **Search string** | **Population** | ("ambient intelligence" OR "assisted living" OR "multiagent systems" OR "systems of systems" OR "internet of things" OR "Cyber Physical Systems" OR "Industr 4" OR "fourth industrial revolution" OR "web of things" OR "Internet of Everything" OR "contemporary software systems" OR "smart manufacturing" OR digitalization OR digitization OR "digital transformation" OR "smart cit*" OR "smart building" OR "smart health" OR "smart environment") **AND** |
|  | **Intervention** | <<facet>> **AND** |
|  | **Comparison** | No |
|  | **Outcome** | (understanding OR management OR technique OR "technolog*" OR method OR location OR place OR setting OR actor OR role OR team OR time OR transformation OR state OR reason OR motivation OR aim OR objective) **AND** |
|  | **Context** | (engineering or development or project or planning OR management OR building OR construction OR maintenance) |
| **Search Strategy** | colspan="2" | SCOPUS (www.scopus.com) + Backward and Forward Snowballing (Wohlin, 2014) |
| **Inclusion Criteria** | colspan="2" | • The paper must be in the context of software engineering; and<br>• The paper must be in the context of contemporary software systems; and<br>• The paper must report a primary or a secondary study; and<br>• The paper must report an evidence-based study grounded in empirical methods (e.g., interviews, surveys, case studies, formal experiment, etc.); and<br>• The paper must provide data to answering at least one of the RR research question; and<br>• The paper must be written in the English language. |
| **Technical Report** | colspan="2" | Detailed information about the planning and execution<br>https://drive.google.com/file/d/1PyUC4U0p-AjGxC0pmtqrhP0_TNv6SOJX/view?usp=sharing |

In this activity, we seek to expand the view beyond IoT and consider other concepts such as Cyber-Physical Systems, Industry 4.0 and others presented in the string population. With the results, we can have a broader characterization of the facets and the area in the context of this research. The idea is not to present an exhaustive and complete report but to see if the IoT facets can be captured in CSS and also an initial characterization of how they are presented (5W1H questions).

**1. Connectivity**

The search resulted in 781 articles, with 752 remaining after removing duplicates and proceedings. Later we applied Title and Abstract selection with 27 remaining for a full reading. We also applied backward and forward snowballing procedures. After the final selection, data set selected in this review is composed of 13 papers, focusing on communication technologies, inserted mainly in the domains of IoT, Smart Cities, Cyber-Physical Systems and Health Care. The evidence showed that, although it has some characteristics well-defined which help to understand the connectivity in the contemporary software systems scenarios, there are lots of open questions and specific solutions according to the domains.

| Research Questions | Summary of the Answers |
|---|---|
| What | Some information and requirements need to be understood to understand and manage connectivity:<br>• CSS is highly scalable, highly available and robust systems with a large number of devices, geographically distributed through an extended area;<br>• It requires a seamless connection as well as network traffic control and management, providing low latency even with limited bandwidth available;<br>• Is deeply influenced by devices limitations and domain requirements, such as low power and high mobility devices;<br>• Deal with limited resources (low memory capacity and low processing power), thus, require efficient operations.<br>• From the 13 articles in the final set, 13 present some input to characterize what. |
| How | • It uses specific solutions according to the application domain<br>• It tries to re-use legacy cellular infrastructure and invest in novel communication solutions<br>• It is mostly based on wireless communication technologies that could be divided into Short-Range, Long-Range, and Cellular-based.<br>• From the 13 articles in the final set, 13 present some input to characterize how. |
| Where | • Through the Network Architecture and the Network layers.<br>• From the 13 articles in the final set, nine present some input to characterize where. |
| Who | No evidence found in the current set. |
| When | No evidence found in the current set. |
| Why | • Some reasons to implement connectivity is to provide communication among the devices, to enable a connection among huge number or applications.<br>• From the 13 articles in the final set, four present some input to characterize why. |

**2. Things**

The search resulted in 830 articles, with 782 remaining after removing duplicates and proceedings. Later we applied Title and Abstract selection with 29 remaining for a full reading. We also applied backward and forward snowballing procedures. After the final selection, the data set selected in this review is composed by 30 papers, focusing in objects and devices, inserted mainly in the domains of IoT, Smart Cities, Smart Buildings, Smart Agriculture, Water Management and Health Care. CSS employ smart devices (things) with the capacity to sense, actuate and interact with users or even the own environment where are embedded. The evidence showed that are different demands and concerns that demonstrates that building *things* is not limited to hardware but involves an intertwining of different areas that need to work together to deliver quality and secure solutions.

| Research Questions | Summary of the Answers |
|---|---|
| What | • Things in the context of contemporary software systems are every device that can sense, actuate or interact with the user or environment; |
| | • In other words, these devices are all hardware that can traditionally replace the computer expanding the connectivity reach;<br>• Tags, home controller devices, mobile phones, wearables, vehicles and transports like buses, cars and trucks, health devices, farm devices, indoor environment devices, water devices, indoor location solutions, and tracking devices are examples of things;<br>• From the 30 articles in the final set, 30 present some input to characterize what. |
| How | • Regarding technologies there are many solutions that were combined to build devices like sensors, actuators, smartphones, microcontrollers, interactables, cameras, communication and network enablers, and others;<br>• Some systems treat Things giving a virtual representation of these devices enabling remote access, and control of them;<br>• To achieve this is necessary to connect the device with the internet. Some technologies were applied to provide communications services to these devices like WSN, Wi-Fi APs (Access Point), ZigBee, 4G Network, Bluetooth, Bluetooth Low Energy (BLE), Wi-Fi, SMS Gateway, GSM/GPRS, Cellular IoT, and iBeacons;<br>• From the 30 articles in the final set, 28 present some input to characterize how. |
| Where | There is no general response to this question. The activities' location is the own environment and depends on the domain that is employed. Based on the literature found the authors built systems in places like houses, shopping places, transport, smart cities, factory, road/streets, military, industry, farm, lignite coal mines, hospital, office, water, airport, and buildings. Some solutions were generics like outdoor and indoor locations.<br>• From the 30 articles in the final set, 17 present some input to characterize where. |
| Who | In software engineering, there is no evidence about who correctly deal with these devices. Some solutions presented the own user construct and program the Thing, in a "do it yourself" approach.<br>• From the 30 articles in the final set, 4 present some input to characterize where. |
| When | No evidence found in the current set. |
| Why | Things with part of solutions in contemporary software systems provide a series of benefits for users: comfort, reduce costs, security, increase the quality of life efficiency, decrease energy consumption, support in the decision-making process, automate a manual process, remote control and monitoring, and indoor environmental quality.<br>• From the 30 articles in the final set, 30 present some input to characterize where. |

## 3. Behavior

The search resulted in 592 articles, with 563 remaining after removing duplicates and proceedings. Later we applied Title and Abstract selection with 27 remaining for a full reading. We also applied backward and forward snowballing procedures. After the final selection, the data set selected in this review is composed by 19 papers, focusing mainly in emergent behaviors, inserted mainly in the domains of Cyber-Physical Systems, Systems of Systems, IoT and Ultra-Large-Scale Systems. The behavior of a system is its central point, and therefore, it needs to have a good understanding. All the actions of CSS are triggered by some event, which can be a stimulus or a reaction to another event. It has a very delicate feature called emergency, making it possible to emerge at unexpected moments. When it comes to contemporary systems, it is often difficult to predict how correctly the system will behave in advance; however, for primarily all practical applications, there must be specific assurances about the behavior of the system, since it would not be safe to implement it otherwise.

| Research Questions | Summary of the Answers |
|---|---|
| What | • All behavior exerted by the CSS system is triggered by some event. Therefore, it is necessary to know when this event will happen and what this event will be. |
| | • The behavior of the whole CSS is more than the sum of the behaviors of its constituent systems. Therefore, it is necessary to know how this greater behavior is generated and when it will arise.<br>• From the 19 articles in the final set, 18 present some input to characterize what. |
| How | • The first and most common way to treat behavior is in stages, where the greater behaviors are constituted by, the smaller ones, with this it is possible to reduce the complexity of taking care of the behaviors.<br>• Another way to manage behavior is through the use of a state machine (Jackson, 2015; Giammarco, 2017).<br>• SosADL and Monterey Phoenix are behavioral modeling frameworks for SoS architecture which describes these systems regarding abstract specifications of possible constituent systems, mediators, and behaviors (Giammarco, Giles and Whitcomb, 2017; Oquendo, 2017).<br>• From the 19 articles in the final set, 13 present some input to characterize how. |
| Where | No evidence found in the current set. |
| Who | • The leading roles for managing the CSS found were: software engineers, programmers, software architects, and systems architects. The other roles encountered were the system users, who are the people who get involved with the system and the role of each of the objects within a system.<br>• From the 19 articles in the final set, ten present some input to characterize who. |
| When | • Frequent updates are expected on projects involving CSS over the lifetime of the project.<br>• The main phases of the life cycle that were identified were initialization, development, validation, implementation and change verification.<br>• To have a good understanding of the behaviors of a system the primary emphasis is assigned to the initial phase of requirements engineering.<br>• From the 19 articles in the final set, ten present some input to characterize when. |
| Why | • The behavior of the system is regarded as the central object of software development and is proposed as the core object of software development. Early and clear identification of behaviors contributes to a reduction of cost schedule risk.<br>• From the 19 articles in the final set, 15 present some input to characterize why. |

## 4. Smartness

The search resulted in 2070 articles, with 2035 remaining after removing duplicates and proceedings. Later we applied Title and Abstract selection with 91 remaining for a full reading. We also applied backward and forward snowballing procedures. After the final selection, the data set selected in this review is composed by 24 papers, focusing in communication technologies, inserted mainly in the domains of IoT, Smart Environments in general, Resource Management, Ambient Intelligence, Context-Aware Systems, and Health Care. One of the reasons for this smartness concern in CSS may be the lack of standardization or understanding of what a "smart system" is. According to the research, to attend smartness, the system should have some level of autonomy and a set of operations such as sensing, data collection, data processing, decision-making, actuation and orchestration in the environment that it is immersed. However, to the system be "smart," it is not necessary to have all these capabilities.

| Research Questions | Summary of the Answers |
|---|---|
| What | • The system should have some level of autonomy and is a set of operations such as sensing, data collection, data processing, decision-making and acting to orchestrate things in the environment that are immersed and to understand smartness in the context of CSS.<br>• In the scenario of CSS projects, smartness deals with data collected, data analyses, treatment and transmission of data to manage and make a decision. All these data |

| | collected from the ambient help the CSS to be aware of what is occurring in the environment. |
| --- | --- |
| | • From the 24 articles in the final set, 21 present some input to characterize what. |
| How | CSS projects use technologies such as sensors or wearables to collect data from the environment: |
| | • It uses actuators, maker decision, and acting according to the data collected and treated to perform some activity in the environment autonomously. |
| | • It uses techniques from artificial intelligence, machine learning, neural networking, fuzzy logic to deal with the data. Hence make a decision and act. |
| | • From the 24 articles in the final set, 22 present some input to characterize how. |
| Where | • Smartness is handled in software architecture, such as Client-server architecture, Representational State Transfer (REST), Service Oriented Architecture (SOA). It is also treated in the process of system implementation or system design. |
| | • From the 24 articles in the final set, 21 present some input to characterize where. |
| Who | No evidence found in the current set. |
| When | • When CSS needs to decide according to the data collected in real-time. CSS project needs to deal with real-time information. In real-time monitoring and visualization to manage the data obtained. |
| | • From the 24 articles in the final set, five present some input to characterize when. |
| Why | • To make the system more autonomy without user interaction; |
| | • To improve the quality of life of end users; |
| | • Management of ambient, such as: save energy, sustainable building, healthcare and so on. |
| | • From the 24 articles in the final set, 20 present some input to characterize why. |

## 5. Interactivity

The search resulted in 955 articles, with 936 remaining after removing duplicates and proceedings. Later we applied Title and Abstract selection with 20 remaining for a full reading. We also applied backward and forward snowballing procedures. After the final selection, the data set selected in this review is composed of 21 papers, focusing in communication technologies, inserted mainly in the domains of IoT, SoS, Cyber-Physical Systems, and Health Care. The evidence showed that, although it has some characteristics well-defined which help to understand the connectivity in the contemporary software systems scenarios, there are lots of open questions and specific solutions according to the domains.

| Research Questions | Summary of the Answers |
| --- | --- |
| What | • In CSS projects, interactivity is characterized by the interaction involving things, systems, and humans where interaction is characterized by the ability to communicate, exchange information and control actions. |
| | • Data must be collected (sensing the environment), processed (generally in some cloud), stored (using databases) and transmitted. To transmit and receive the information, as well as interact with humans, they utilize networks a medium of communication. |
| | • From the 21 articles in the final set, 21 present some input to characterize what. |
| How | • To guarantee connectivity: Zig-Bee, Bluetooth, Radio Frequency, RFID, 6LowPAN, WSN, WiFi, IPv6 and others. |
| | • To guarantee communication: HTTP, XMPP, TCP, UDP, CoAP, MQTT and others. |
| | • To guarantee understanding: JSON, XML, OWL, SSN Ontology, COCI and others. |
| | • Also, real-world objects are virtualized and represented as Web Resources and accessed through Web Interfaces based on REST principles and Producer and Consumers methods. |
| | • From the 21 articles in the final set, 21 present some input to characterize how. |
| Where | No evidence found in the current set. |

| | | |
|---|---|---|
| Who | | • Designers, architects, developers, managers, and engineers deal with interactivity in different phases of CSS projects.<br>• Changing the scenario: "Engineering is no more a set of vertical activities developed by different engineers but a collaborative process in which people and technology is completely involved in the engineering process".<br>• From the 21 articles in the final set, 6 present some input to characterize who. |
| When | | No evidence found in the current set. |
| Why | | • To bridge the gap between the massive heterogeneity present in CSS in order to create an interoperable systems, that can overcome different standards, protocols and technologies to perform more efficiently than isolated ones.<br>• Interactivity is one of the main characteristics of CSS projects, making new types of application possible (such as smart environments), facilitating everyday life, enhancing products competitivity, and sustainability.<br>• From the 21 articles in the final set, 15 present some input to characterize why. |

## 6. Environment

The search resulted in 925 articles, with 827 remaining after removing duplicates and proceedings. Later we applied Title and Abstract selection with 57 remaining for a full reading. We also applied backward and forward snowballing procedures. After the final selection, the data set selected in this review is composed by 22 papers, focusing on the requirements for such environments, inserted mainly in the domains of IoT, Ambient Assisted Living, Smart Cities, Cyber-Physical Systems, Industry 4.0, Smart House, Smart Campus, and Health Care. The environment can involve many devices composed by sensors, actuators and other objects generating a significant amount of data leading to issues related with connectivity and interoperability, data processing and storage, to be efficient and reliable. This high complex ambient requires system integration, and there is a necessity for trustful and legal regulation. Also, sustainability is a crucial concept for these environments.

| Research Questions | Summary of the Answers |
|---|---|
| What | • The environment is the place where things are, actions happen, events occur, and people are. Smart Environments (SE) or Smart Spaces provide intelligent services by acquiring knowledge about itself and its inhabitants to adapt to users' needs and behavior. Contemporary Software Systems apply various technological solutions to attend specific requirements that differ according to the project.<br>• From the 22 articles in the final set, 21 present some input to characterize what. |
| How | • In general, the environments are composed of sensors and actuators to sense and change a state of the ambient. Technologies like IoT, cloud, smart objects, middleware's, Wireless Sensor Networks, Vehicular Ad-hoc Networks, edge computing, artificial intelligence, machine learning, data mining can be employed on these systems.<br>• Techniques for designing smart systems using Use Cases and Smart Environment Metamodels can be applied. User interaction, autonomy, and easy management are essential requirements on Smart Environment.<br>• From the 22 articles in the final set, 19 present some input to characterize how. |
| Where | • The activities' location is the own environment and depends on the domain that is employed. Based on the literature found environment can be places like city, home, ambient assisted living, campus, office, industry, building, transportation, street, road, bike station, parking space, and others<br>• From the 22 articles in the final set, 22 present some input to characterize where. |
| Who | • In software engineering, the phases that allocate environment activities allocate developers, system designers, domain experts, technical professionals, end-users and stakeholders to build an ambient solution. |
| | • From the 22 articles in the final set, 22 present some input to characterize who. |
| When | • Concerning the solutions presented, the majority deals with software activities related to analysis, design, and implementation phases on activities like system architecture definition, software design, requirement specification, and software implementation. |
| | • From the 22 articles in the final set, seven present some input to characterize when. |
| Why | • Adapt ambient to users' needs and behavior. |
| | • Provides comfort, quality of life, and benefit daily lives, accessibility, high productivity, reduce costs and effort, save time, use resources efficiently and give autonomy to users. |
| | • Benefit users on their activities by using cutting-edge technologies. |
| | • Helps on: health diseases, pollution management, traffic efficiency, deterioration and management of infrastructure, criminality, climate change, cyber-security, and economic development. |
| | • Provides natural and sustainable user-centric quality services. |
| | • From the 22 articles in the final set, 22 present some input to characterize why. |

The complete protocols can be found in a Technical Report[8] and the summary in Appendix A – Evidence Briefings.

## 4.4 Challenges in Engineering CSS

In the activities and studies carried out so far, we have extracted from the various sources **Concerns** (Section 3.6) and **Facets** (Section 4.2 and 4.3) and other information from the studies conducted so far. Our idea of the **Challenges** to be addressed by CSS deals with both aspects together. Therefore, a challenge in our perspective refers to addressing the concerns in each facet according to their specificities. This proposition is because a solution to concern is materialized in different ways in the proposed facets. Table 6 presents the high-level challenges collected throughout this work and shows how each facet see the concerns proposed.

---

[8] https://drive.google.com/file/d/1PyUC4U0p-AjGxC0pmtqrhP0_TNv6SOJX/view?usp=sharing

Table 6. Some challenges in Engineering CSS.

| Concern | Facets | | | | | |
|---|---|---|---|---|---|---|
| | **Connectivity** | **Things** | **Behavior** | **Smartness** | **Interactivity** | **Environment** |
| **Management** | One of the concerns for connectivity is traffic management and control to deal with the enormous data generated by these devices and guarantee the quality of service (Bera, Misra, and Vasilakos, 2017; Li, Xu, and Zhao, 2018). | This scenario involves distributed systems consisting of hundreds to thousands of devices, involving the coordination of their activities, requiring a high-level ability of reasoning and management (Patel and Cassou, 2015). | In the literature, we have behavior patterns (Haynes et al., 2017), separation of concerns (Ruppel, no date), state machines (de Lemos et al., 2013) and other solutions to manage behaviors. However, many authors argue that it is still an open issue. | Management issues and smartness are intimately connected. One example is the need to provide power consumption management with analysis and establishment of rules for optimization (Oliveira et al., 2017). | A goal is to allow systems to manage themselves so that human intervention could be minimized. For this, is necessary to automate management functions, according to the behavior of the components defined by a management interface (Dai et al., 2017). | It is necessary solutions to manage functionalities personalization and to interpret complex user needs in smart environments (Pons, Catala, and Jaen, 2015; Desolda, Ardito and Matera, 2017). |
| **Architecture** | SDN is an emerging network architecture, where network control can be decoupled from the traditional hardware. This change and researches in network architecture are crucial to connectivity (Bera, Misra and Vasilakos, 2017). | With the increase in complexity and the number of devices in CSS, new architectural styles are necessary to deal with their needs for scalability, fault isolation and flexibility for example (Herrera-Quintero et al., 2018) | When dealing with behavior, the architecture should encompass the system need to visualize/represent behaviors and interactions. System dynamics, agent-based modeling, and Monterey Phoenix are some commonly used behavioral modeling to describe the CSS architecture (Giammarco, 2017; Oquendo, 2017). | What makes a system smart in many cases is not only the devices that are used and the decision-making process but the whole solution architecture as well (Atabekov et al., 2015). | For interactivity many discussions are related to architecture especially focusing on decentralized solutions, supporting and monitoring assisted livings in heterogeneous contexts, integrating existing platforms (Giaffreda, Capra and Antonelli, 2016; Pace et al., 2017), | Activities like system architecture definition are important in Smart Environments and relate to designing and implementing it providing the reactivity, scalability, extensibility necessary for the environment (Cicirelli et al., 2016). |
| **Requirements** | A set of requirements could be captured that are intrinsic connected to the devices' nature | Regarding things, to deal with **heterogeneity** and **scale** (Rojas et al., | An emerging behavior arises from a lack of understanding of the system. For this | Different devices can capture data from the environment. Thus the systems in the future | The wide range of heterogeneity issues introduced by the among of different IoT | The increasing use of software in the embedded devices allows smart spaces |
| | but they directly influence connectivity such as **efficiency** - issues like low power capacity, low memory capacity, low processing (Murakami *et al.*, 2018) and **extended coverage** - to attend a large number of devices distributed, an extended coverage area is needed no matter the technology chosen (Chen, Tang, and Coon, 2018). | 2017), **distribution** - geographically distributed and sometimes, in inaccessible and critical regions (Chen, Tang and Coon, 2018) as well as **mobility** – IoT devices are not static they tend to move between different coverage areas (Bera, Misra and Vasilakos, 2017), are issues related to requirements to be covered in CSS. | reason, the initial phases of the project are very relevant, and in CSS one of the primary emphasis is attributed to the initial phase of requirements engineering (Rainey, Mittal and Rainey, 2015). | can make decisions and act. It should be planned, and it composes one of the parts of smartness in the systems (Medina *et al.*, 2018). | devices. Standardization, therefore, is a must but is not enough as no single standard can cover everything, as well as some organizations (manufacturers, software companies), would like to follow different standards or even proprietary protocols (Dalli and Bri, 2016). | development. The use of standard software engineering technologies needs some modification and defining a systematic process focusing in for smart space development (Aziz, Sheikh, and Felemban, 2016). |
| **Professional** | Many nodes in IoT undergo constant movement that may result in intermittent interconnectivity between the devices which may encounter frequent topology changes. Due to these frequent topological changes and limited resources available in the IoT devices, now a day's routing of the data has become a significant challenge requiring the proper skills and technologies to be overcome (Dhumane, Prasad and Prasad, 2016). | IoT application development is a multi-disciplined process where knowledge from multiple concerns intersects. Traditional IoT application development assumes that the individuals involved in the application development have similar skills. This is in apparent conflict with the varied set of skills required during the overall process involving this engineering (Patel and Cassou, 2015) | Managing a CSS project requires different profiles involved, each with a different skill (Gabor *et al.*, 2016). | Technological solutions can be better achieved in smart cities by making different stakeholders work together (Neuhofer, Buhalis, and Ladkin, 2015). | Being a multidisciplinary ecosystem, it is a considerable challenge for the developers to engineer consumer applications. With no widely followed guidelines, the creation and use of best practices will ease the development time, maintenance, and update cycle (Datta *et al.*, 2017). | To exploit the abundance of the related resources users could compose the different "behaviors" exposed by the surrounding environment, becoming an active part of the systems and adding a new perspective in development (Desolda, Ardito and Matera, 2017). |
| **Things** | Objects will be the primary users of the | Decentralized approaches, | When dealing with things, one should | One of the reasons for this smartness concern | New challenges | Things can be created, adapted, personalized, |
| | internet and will communicate with each other for gathering, sharing and forwarding the information' about the environment. Intermittent connectivity, mobility and the death of nodes are essential issues to address (Dhumane, Prasad and Prasad, 2016). | robustness, timely control, and independent decision-making, automatically adaptation are still issues in this context (Witthaut *et al.*, 2017) | functional attributes, individual parameters and the environment they are inserted are the factors that play can strong influence behaviors in different contexts (Roca *et al.*, 2016). | in CSSs may be the lack of standardization or understanding of what a smart thing is. In our perspective, it should have some level of autonomy and capable of sensing, data collection and data processing, decision-making and acting to orchestrate things in the environment that is immersed (Chen *et al.*, 2017). | arise when various technologies combine in a common architecture design where "things," "people," "places." and "data" should coherently communicate with each other (Davoudpour, Sadeghian, and Rahnama, 2015). | and rely on contextual data. The integration of things and social networks can contribute to improving this contextual data and is a research opportunity (Davoudpour, Sadeghian, and Rahnama, 2015). |
| **Network** | Applications in IoT domain require extensive connectivity, security, trustworthy, the ultra-reliable connection among other requirements for a large number of devices and, though used in IoT scenarios, 2G, 3G, and 4G technologies are not fully optimized for IoT applications (Li, Xu, and Zhao, 2018). | Although connectivity is the core of this new technology, the traditional network infrastructure is not prepared to support IoT requirements. Traditional devices, such as switches and routers, are usually preprogrammed to do particular tasks and follow particular rules. This does not meet the IoT application-specific requirements since can be necessary that the devices should adapt themselves to multiple different rules (Bera, Misra and Vasilakos, 2017). | Some behavior emerges that cannot be attributed to a single system but results from the interplay of some or all CPS in the network. Therefore, each system involved must adjust its behavior according to the common goal of the network (Brings, 2017). | The solution encompasses shock sensor, GPS, NFC reader, and cellular IoT. Those combined spontaneously notify the rescue team whenever an accident takes place. As for the higher layers in the IoT protocol stack, the emerged protocols, the Constrained Application Protocol over User Datagram Protocol, Datagram Transport Layer Security can be used to overcome the limitations of the IoT devices' constraints (Nasr, Kfoury and Khoury, 2016). | Current vehicular networks mostly utilize IPv6, which 1) does not support mobility natively and 2) is host centric, not data centric. We need a datacentric and network-independent approach to IoT mobility (Datta *et al.*, 2017). | Wireless Sensors Network, Vehicular Ad-hoc Network, and new network topology and strategies can contribute to achieving a sustainable smart city (Faria *et al.*, 2017). |
| **Security** | Some protocols guarantee essential | The paper (Dalli and Bri, 2016) highlights | There are issues | There are at least two different opportunities | Although there is many challenges in the | The Information about the environment is |

| | | | | | |
|---|---|---|---|---|---|
| data confidentiality and integrity, securing communication channels using cryptography, but there are still critical challenges related to network control (Beltrán, 2018). | some security challenges about things: 1) IoT devices spend most of their time unattended, thus can be easily physically attacked; 2) Wireless communication between Things is vulnerable to eavesdropping; 3) Complex and resource-demanding security mechanisms are not suitable to be implemented on resource-constrained IoT devices. | that the IoT community needs to address in order to prevent privacy violation, which includes **self-aware behavior of interconnected devices**, data integrity, authentication, heterogeneity tolerance, efficient encryption techniques, secure cloud computing, data ownership, and governance, as well as policy implementation and management (Mendez Mena, Papapanagiotou and Yang, 2018). | to obtain access in a smart home to control functions: network attacks and device attacks. In network attacks, an adversary may try to intercept, manipulate, fabricate, or interrupt the transmitted data. Device attacks can be classified into software attacks, physical or invasive attacks, and side-channel attacks (Ali and Awad, 2018). | design and implementation of an effective ambient assisted living (AAL) system such as information architecture, interaction design, human-computer interaction, ergonomics, usability and accessibility, there are also social and ethical problems like the acceptance by the older adults and the privacy and confidentiality that should be a requirement of all AAL devices (Marques, Roque Ferreira, and Pitarma, 2018). | autonomously and continually collected by IoT devices without human awareness (ex. smart home applications recording inhabitants' living habits) that represent a security issue (Dalli and Bri, 2016). |

The challenges presented represent the high-level facets' visions of the concerns collected in the studies, not an exhaustively detailed list. Instead, this view shows the need for differentiated targeting for each facet and the many open questions the subject still has. Before presenting solutions or alternatives to the challenges, we propose to list them and bring them to the attention. Being one of the initial contributions of the work to point out them as research gaps and argue that is necessary a multifaceted view when dealing with CSS.

## 4.5 Validity Threats

The same validity threats presented for **literature review** and **qualitative analysis** presented in the previous chapter can be applied at this point.

Regarding the RRs, we followed a research protocol and reviews guidelines, but the entire review procedure was conducted by students accompanied by more experienced researches, to reduce the selection bias. Also, we did not conduct a quality assessment.

## 4.6 Conclusion

The emergence of CSS brings new challenges in software engineering. To address these challenges, we should change our way of developing a software system from a monolithic structure to a broader multidisciplinary approach. So far in this work, we have presented the results obtained by analyzing data acquired through different strategies, which identified challenges in engineering CSS.

First, we identified concerns from the technical literature, practitioners, and a government report. Next, we presented the facets that compose IoT software systems, derived from a qualitative study and RRs. These results can support practitioners in evaluating risks to construct such systems and highlight some research opportunities for researchers. One of the contributions of this work is to explore a set of facets that need to be considered in CSS, showing that it is necessary to distinguish this contemporary software system from traditional ones.

This scenario represents excellent opportunities, especially in the context of Software Engineering, which can be glimpsed in the characterization carried out in this chapter, for example:

- The presentation of a baseline of concepts related to CSS projects presented in the technical literature of each facet. Define an initial understanding and what needs to be developed gives us a direction for the actions to be taken in the nearest future;

- The definition of research protocols based on RR for filling CSS gaps. It is one of the most significant recent innovations, which has attracted extensive attention from both industry and academia due to its potential impact on various fields. Theory and practice must go hand in hand, with research delivering usable technologies, and industry proposing problems. For us, it is an opportunity to conduct evidence-based research.

For this, the framework in the proposal, presented in the next chapter, was organized, structured and defined, aiming to fill in the gaps observed in the conducted studies.

# 5 A Framework to Support the Decision-Making on Engineering CSS

*This chapter presents a proposal for a framework to support the decision-making on engineering CSS, using IoT as the surrogate. The proposal was derived from the results obtained in the studies and based on research in the area.*

## 5.1 Introduction

The domain knowledge can reveal concepts, descriptions, and relationships that could be organized to evidence at each stage of development what needs to be investigated (Bjarnason and Sharp, 2017). In aligning these concepts and the problem understanding across stakeholders, such knowledge could be used to highlight relevant information to be considered to support the engineering of CSS.

The discussions in the conception phase of a project aims to deliberate the work involved in the software product being specified, designed, built, and, afterward, evolved. This initial step clarifies the overall scope and establishes a basic understanding of the problem, the people who seek to solve the problem, the type of solution desired, the collaboration with other stakeholders and the team that will oversee the solution (Pfleeger and Atlee, 1998).

A contemporary software system's problem domain is inherently multidisciplinary (Motta, de Oliveira, and Travassos, 2018), therefore it is necessary a way of characterizing and describing it across different areas and disciplines (facets). However, there is no well-established base for this characterization yet (Ncube, 2018). By addressing this issue, we aim to contribute with the clarification of what is involved in the engineering of CSS. We believe that this strategy is necessary since *incomplete knowledge* and *communication flaws* constitute the most frequently stated problems in the project conception phase (Bjarnason and Sharp, 2017; Fernandez, 2018). The scope of the proposal is limited to the initial problem understanding, seeking to reduce uncertainties and risks by promoting shared knowledge leading to directions based on the context and fitted for the project in question.

Hence, this chapter presents a proposal of strategy based on the results of studies previously described (Chapters 3 and 4) to support the decision-making while engineering CSS (Figure 13).

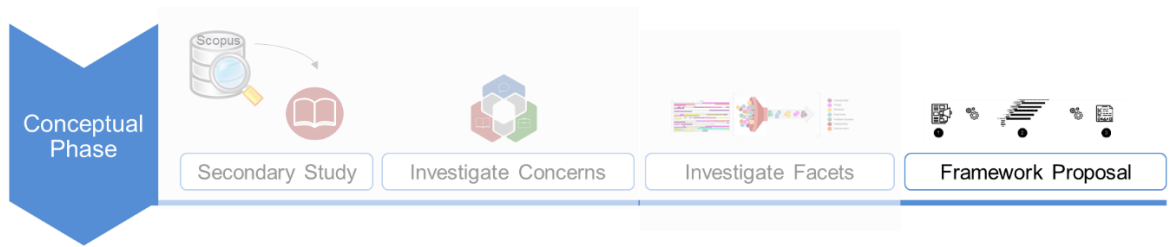

Figure 13. Propose Framework - methodology step

The previous findings allowed us in identifying research gaps and opportunities to support the proposition of this research. Based on the issues pointed out previously, the following elements were identified to compose a framework when considering the scope of engineering CSS:

- To provide engineering support by considering the area multidisciplinarity;
- To provide decision-making support considering the challenges and technologies retrieved from the studies, and;
- To support the problem domain clarification and act as a bridge between the problem and solution conception phase.

## 5.2 Framework Proposal

As observed in our investigations, the CSS scenario is covered by concerns that are seen and treated according to the facets involved, which leads to challenges for its development. In this context, the CSS conception phase is an essential development step and needs to handle all factors involved in CSS as well as the traditional issues regarding current software projects. For that, the initial alignment should be conducted prospectively to minimize the uncertainty and overcome the challenges.

Figure 14 presents an overview of the proposed framework with its respective steps aiming to define a strategy to support the decision-making on engineering satisfactory solutions for CSS problem domains.

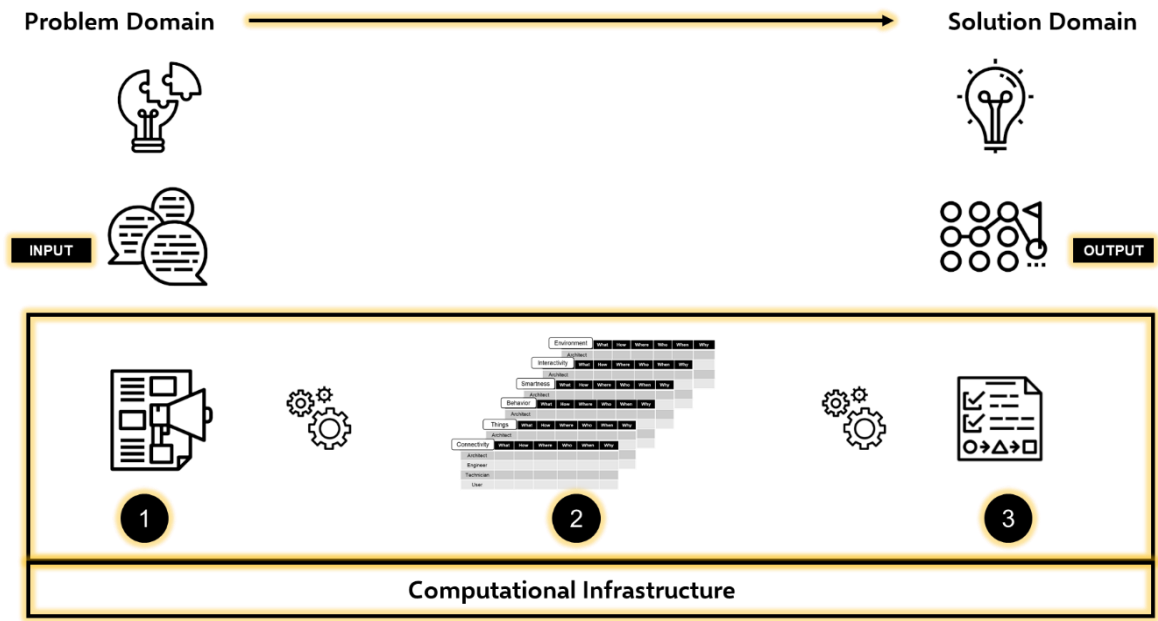

Figure 14. The framework proposed.

In this sense, we would like to contribute to the conception phase of CSS using a framework to support the decision-making regarding the engineering of CSS (focusing on IoT applications). By aligning different stakeholders' control perspectives, we want to characterize the problem domain. From the problem domain characterization, we aim to provide a general decision-making strategy. The strategy relies on the organization and analyzes of a CSS Body of Knowledge.

Our idea is based on the Zachman framework adaptation to organize a body of knowledge related to CSS engineering, that will be filled for the different kinds of CSS (in our case, focused on IoT). Once the body of knowledge is filled, we aim to analyze it to retrieve relevant information to support the decisions that allow the solution engineering according to problem characterization.

In the next sections, we will first present the organization of the body of knowledge, since it should be available to use the framework, and then describe the steps that compose the proposed strategy – Steps 1, 2 and 3 numbered in Figure 14 focused on IoT as surrogates of CSS. Further steps include its development and empirical evaluation.

### 5.2.1 Body of Knowlegde organization

Our aim regarding the conceptual organization of the recovered data is that it should take into account the requirements of different stakeholders and the activities in the various facets. Having such a conceptual structure, we do not aim to guide the software development directly but rather to support the organization the concepts more explicitly and the decision-making on CSS engineering. In this step, we want to organize our findings in a body of knowledge that adequately represents the concepts and in a structure that contains relevant information in a way that can support CSS engineering.

As previously described the Zachman Framework (Section 2.4) inspired the framework's organization to encompass the facets proposed. This multi-faceted view seeks to show that each facet must be treated according to its particularities and perspectives in CSS. The desired solution is more significant than the sum of its parts, according to the holistic view of systems engineering Figure 15.

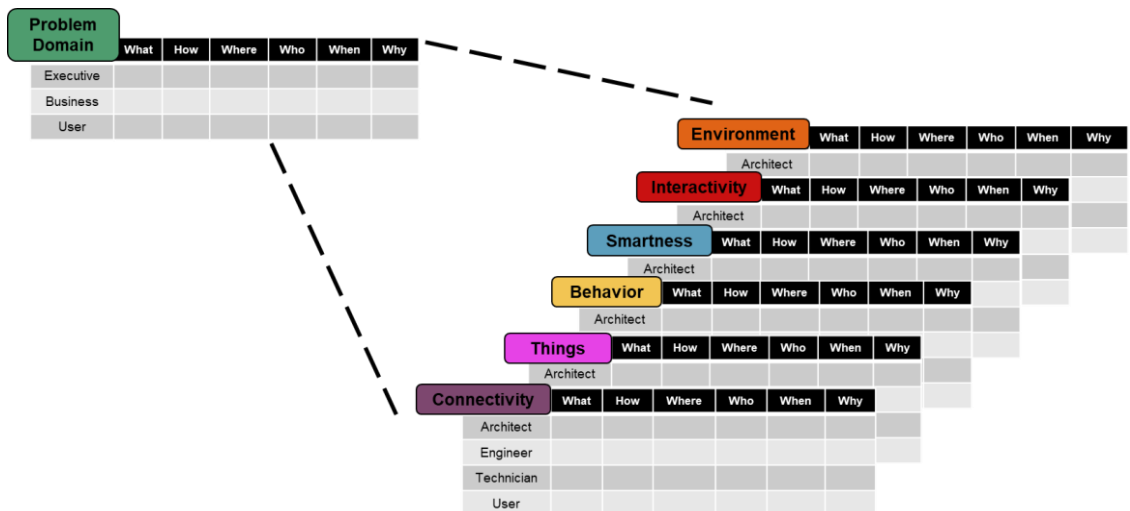

Figure 15. CSS Body of Knowledge Representation.

The idea is that the Problem Domain will direct and contextualize how the other facets will be derived, implemented and managed to achieve CSS solutions. To go from the problem to a software solution is the primary challenge in developing and this is especially difficult in contemporary systems since some of the facets should be part of the same solution, one related to the other aiming at the solution completion. Therefore, during the conception of any of the facets the integrity of the others could be impacted, and in turn the overall solution. To visualize CSS in this multi-perspective way can help the understanding of this relationship.

Alongside with the facets we have: **Perspectives** and **Communication Interrogatives** evolved both from the Zachman Framework (Sowa and Zachman, 1992). The perspectives were divided as **control** (Business, Executive and User), who support the definition of the problem domain, and **construction** (Architect, Engineer, Technician, and User) parts, that will specialize the facets to solve the problem. We are considering the user perspective as a hybrid because the future vision is that the user has active participation in the construction of CSS (Singh and Kapoor, 2017).

**Perspectives:** The framework considers all the perspectives involved in planning, conception, building, usage and maintaining activities of software systems:

- **Executive Perspective** - It focuses on the system scope and management plans, and how it would relate to a particular context.
- **Business Perspective** - It is concerned with the business models, which constitute business design, how they relate and how the system will be used.
- **Architect Perspective** - This perspective translates the system model designed and determines the logic behind a system considering data elements, process flows, and functions representing the business entities and processes.
- **Engineer Perspective** - It corresponds to the technology models, which must tailor the model to the programming languages details, devices, or other required supporting technology.
- **Technician Perspective** - The developer follows detailed specifications to build modules, sometimes without being concerned with the overall context or structure of the system.
- **User Perspective** - It concerns the functioning system in use.

From the guidelines provided in the Zackman's framework, we consider the questions as communication interrogatives for our context (CSS engineering) since the answer to each question in each perspective, and each facet will give us more direct information leading an engineer closer to the solution specification. These are fundamental questions to outline each perspective:

- **What** - Referring to the information required for the understanding and management of a system. It begins at a high level, and as it advances in the perspectives, the description of the data becomes more detailed;
- **How** - It relates to translating abstract goals into the definition of its operations using software technologies (techniques, technologies, methods, and solutions);
- **Where** - It concerns the activities location; it can be a geographical distribution or something external to the system;
- **Who** - It describes the roles involved with the systems to deal with the facet development, detailing the representation of each one as it advances;
- **When** - It concerns the effects of time over the system, such as the life cycles, describing the transformations and states of the systems;
- **Why** - It concerns to translate the motivation, goals, and strategies going to what is implemented in the facet.

We plan to follow a bottom-up approach and conduct studies on the technical literature, practitioners and real cases to fill the body of knowledge. We aim to achieve a complete solution on a small scale, to be evolved incrementally. If any adjustment is necessary, we sought to make available the protocols available at Delfos[9] (The Observatory of Contemporary Software Systems Engineering) to facilitate access, dissemination, re-execution and evolution of the findings in order to keep the body of knowledge updated.

### 5.2.2 Describing the Framework steps

**STEP 1 - Define Project Characterization**

It is necessary some kind of discussion among stakeholders from the control perspective, in a high level of abstraction to identify projects goals and align with the problem domain. To support it, an artifact reflecting each perspective and characterizing the problem domain in the light of the facets, including the project context will be prepared to materialize the discussion. From the problem definition, stakeholders from the control perspective will fill in the Problem Characterization Template.

- Input: *A problem that requires a CSS solution*
- Perspectives: Executive, Business and User
- Required Artifact: Problem Characterization Template
- Artifact Produced: Problem Characterization
- Post-activity: Analyzing the Body of Knowledge for decision making

**STEP 2 - Analyzing the Body of Knowledge for decision making**

Performing this step implies the body of knowledge regarding CSS engineering is available (in a first moment IoT body of knowledge). This step represents the rationale involve for the use of the body of knowledge. Once we have a filled the body of knowledge with information regarding practices and technologies, we will use it as the basis to drive the strategy. The information in the problem characterization artifact will be used to direct and specialize the body of knowledge according to the facets described, focusing on the construction part, to present the concerns that should be taken into account and be used in the strategy to decision making for the specific project.

- Pre-activity: Define Project Characterization
- Perspectives: Architect, Engineer, Technician, and User
- Required Artifact: Problem Characterization and CSS Body of Knowledge
- Artifact Produced: Relevant Specialized Knowledge for the Specific Project

---

9 http://146.164.35.157/

- Post-activity: Generate decision-making Strategy

**STEP 3 - Generate decision-making Strategy**

The Relevant Specialized knowledge for the specific project defined in the previous step will support the strategy definition, as a guide, for decision-making in CSS engineering. This support decision-making regarding the development directions and an action plan for the development, translating the problem to the solution domain. In this activity, the construction perspective align and decide based on the general strategy, from the project characterization to the definition of requirements so that the project needs can be addressed.

- Pre-activity: Analyzing the Body of Knowledge for decision making
- Perspectives: Architect, Engineer, Technician, and User
- Required Artifact: General Decision-Making Strategy Template and CSS Body of Knowledge specialization
- Artifact Produced: Decision-Making Strategy
- Output: *The use of a strategy to support decision-making in engineering CSS.*

Although a computational infrastructure is not entirely necessary for achieving the objective to support the clarification of the problem domain, we consider that it may facilitate the use of the proposed strategy in the context of software projects and can engage practitioners to use it. Therefore, we intend to develop a computational infrastructure to support the steps 1, 2 and 3.

This proposal does not aim to replace defined activities that are common in the development of traditional software projects. In this work, we hope to address the particularities of CSS projects since they present different and additional characteristics that can bring challenges to its engineering. Thus, the contribution of this proposal is to minimize the uncertainty of software technologies and engineering perspectives based on the problem domain and supported by the application of a body of knowledge defined.

## 5.3 An Example of Use

This section aims to exemplify the use of the proposed framework. For this, we rely on the results of a project carried out in the context of an undergraduate discipline of the Computer and Information Engineering course at the Federal University of Rio de Janeiro. Five last periods bachelors' students with previous knowledge in engineering conventional software systems construct the software system. Members of the Experimental Software Engineering Group and the professor mentored the project. The project was executed in the first semester of 2018 and the product (Camarão IoT) deployed in July of 2018.

The goal of the discipline was to discuss software engineering concepts in the light of IoT software systems engineering. Considering an infrastructure where "things" can be easily made available and a multitude of software applications can be engineered taking these resources into account. Practicing and observing issues related to engineering IoT systems revealed challenges regarding: a) Ideation and experimentation of new software applications from specific resources (things); b) Analysis of application requirements based on IoT and inspection of artifacts; c) Viability of the proposed technology platform (or others) for composing IoT software systems, d) Testing of the software system solution.

As a case for practice, the students` team should develop a solution to support the creation of shrimp in farms. A SEBRAE's claim motivated it[10] (Brazilian Service to Support Micro and Small Enterprises): "*Due to the complexity of the production process, and a large number of variables that must be constantly monitored, we suggest the acquisition of software of management, which was not found on the market with enough to be indicated here. Most companies that produce software can provide such a solution, provided that there is a customization of the software.*"

From a demand requiring a software-based solution, it is possible to foresee distinct technologies that can meet it. However, this scenario presents several challenges to the engineers regarding the different decisions that should be taken into account when engineering such a solution. One of the significant challenges then is the steps before the decisions that will guide the engineering of the solution. Some examples of decisions are: Why choose Wi-Fi or Bluetooth? Is an actuator required or is human intervention enough? What kind of resources and skills are required to implement it? Therefore, our expectation is that the proposed framework can support decision-making in order to answer such questions. For it, we want to fill in the body of knowledge with the necessary information for each facet in the different perspectives proposed in order to support decision-making, reducing the concerns and risks involved in CSS (starting with IoT) engineering.

Based on the motivation above, we present the toy example of a solution organized in the structure of the proposed framework. This example enables us to show different facets' arrangements in a basic solution. Because the software system had been implemented and design decisions were taken, we mapped the results to exemplify the use of the framework.

---

[10] https://goo.gl/qA3udh

**STEP 1 - Define Problem Characterization**

Given the lack of software solutions and the market opportunity for this product, the proposal was to idealize and to build an IoT software system to support freshwater carniciculture in Brazil. Our intention with this exemplification was to take what they accomplished and translate it into the proposed framework.

In this characterization step, from the project context, the Executive, Business and User perspectives proposed in the framework are used to support the identification of different concerns and relevant information that must be considered in the solution to be developed. This information is mapped below in the proposed structure:

**Executive Perspective:**
- **What**: A solution to support Malaysia shrimp farming in freshwater that will enable remote and real-time business management. Also, to be able to monitor the overall state of production.
- **How**: Receive notifications of critical conditions and current status, receive periodic reports and estimations.
- **Where**: Anywhere.
- **Who**: The professor performed the Executive perspective, playing the owner role. He gave orientations to be followed and had a significant influence on the overall decisions in the project.
- **When**: Real-time.
- **Why**: Due to the complexity of the production, and a large number of variables that must be continuously monitored in the shrimp farming process, and because the manager is not always present in the production site. Make management decisions about the business.

**Business Perspective:**
- **What**: Receive quick and easy information at any time through the used technologies. Modernize production and have greater control to meet the foreign market.
- **How**: Same of the owner but also define deadlines and demands, receive information about the water tank, consult stock and production, notifications of critical conditions and current status, receive periodic reports and estimations.
- **Where**: Anywhere.

- **Who**: A persona was established for the Business perspective, representing the manager role. S/He oversees the activities that are being carried out and help in decision-making for the business.
- **When**: Real-time.
- **Why**: More accuracy in the received information and increase production reducing costs. S/He does not have a way of observing the state of the system in real time from anywhere. S/He wants to know what business decisions he can make to optimize the production process.

**User Perspective** – Different personas were established for the user perspective, representing the following roles:

a) **Who - Installation oversee:** S/He takes care of the installation and stock, reporting back to the manager when it is necessary.
- **What:** S/He wants to offer his best to continue with the job. He needs something that can help the work with clear and direct visual information of when and what actions he should take.
- **How:** Check stock status, receive notification of demands, and notify manager about the need for purchases.
- **Where**: Production site.
- **When**: Anytime.
- **Why**: S/He needs to check inventory frequently because there is no communication and he does not know if some change occurred since the last visit. There is no precise counting in the stock, which can end up causing an early or late purchase, in incorrect amounts.

b) **Who - Shrimp keeper:** S/He is responsible for preparing the ration and feeding the shrimps.
- **What:** Make the tanks documentation and their characteristics simpler and easier to understand, that would make the job less stressful. Another point that would help in the day-to-day professional life would be to facilitate the feeding process to avoid repetitive strain injuries.
- **How:** Receive feeding schedule, notify biologist shrimp status, visualize tank and shrimp status.
- **Where**: Production site.
- **When**: Feeding schedule.
- **Why**: S/He needs to read the confusing documents that contain the characteristics of the appropriate mixture for ration in a specific shrimp tank. He needs to prepare the mixture and needs to do all the manual and repetitive work that is the act of feeding the shrimps. The throw of the food can lead to muscle injuries and RSIs something widespread in this profession.

c) **Who - Tank keeper:** S/He monitors the tank status, perform measures and adjust tank conditions.

- **What:** S/He would like to control the tanks more accurately and with a better frequency, without the need to always be running between different tanks. He wants the peace of mind that work it is according to the need of the business.
- **How:** Monitor tank status, generate reports, notify critical conditions, secure tank to return to normal conditions, biologist shrimp status, check environmental conditions that can affect the tank and visualize tank and shrimp status.
- **Where**: Production site.
- **When**: Check-up schedule.
- **Why**: S/He needs to communicate frequently with other employees to see if the tank parameters are as required. S/He thinks it is repetitive to move between tanks to get the status manually. S/He is concerned if she is maintaining the best conditions in the controlled tank, but he does not know with precision how well he can control the tanks.

d) **Who - Biologist:** S/He sets the conditions and is responsible for the production health.

- **What:** S/He would like to have past information to be able to perform more precise analyzes and to minimize the error of his estimates, besides being able to compare the evolution of the production in addition to obtaining information about the shrimps in a more accurate and faster way.
- **How:** Update production demand required, update tank conditions to achieve the production demand, define and monitor shrimp health parameters, define and monitor feeding schedule, visualize tank and shrimp status and generate reports.
- **Where**: Production site.
- **When**: Anytime.
- **Why**: S/He needs to perform analyzes and estimates on shrimp production working with many variables. Since he does not understand the theory so much, he is sometimes in doubt about the overall health in the tanks. S/He cannot quickly check specific variables and perform sampling evenly.

As described above, from this step with framework structure, is possible to contemplate the different goals for the same solution, thus enriching the initial characterization of the project. Figure 16 shows the representation of problem domain for the project context previously presented.

| Problem Domain | What | How | Where | Who | When | Why |
|---|---|---|---|---|---|---|
| Executive | Scope definition | List of expectations | Anywhere | Owner | Realtime | Motivation for the solution |
| Business | Detailed scope | List of expectations | Anywhere | Manager | Realtime | Motivation for the solution |
| User | Detailed scope | List of expectations | List of locations | List of roles | List of times | Goals for the user |

Figure 16. Example of representation of problem domain.

Due to the full range of perspectives and goals, the team organizes and prioritizes the primary needs. From this initial part, we defined the main needs of a system that **(1)** allows the clear visualization of information regarding the whole process in real time; **(2)** support the feeding of shrimp; **(3)** assist in estimating production and **(4)** monitor the tank status (Figure 17).

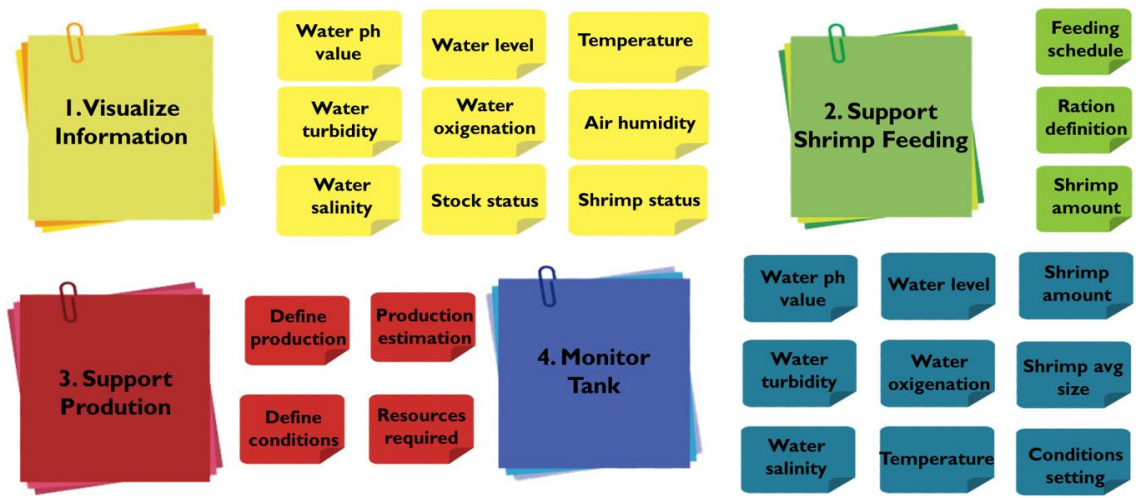

Figure 17. System's needs.

Alongside with the needs presented by the control perspectives, it is required to identify which information can indicate a match with facets, which will support the analysis of the body of knowledge to identify relevant knowledge to engineer a solution to that context.

The Problem Characterization template (used in step 1) will be defined to map the identified system needs with each facet of the body of knowledge in a way that could support the identification of the relevant knowledge. The next activity of this research in this step is the design of this template. It will comprise the investigation of the concerns defining the facets. A preliminary example of the Problem Characterization artifact for this case study is presented in Figure 18. The idea is to capture the need using questions and perspectives; then we want to map them throughout the artifact, enlightening which concerns should be considered in a given context. It aims to bridge the problem to the facets. The results of Rapid Reviews (Section 4.3) assists in the artifact's design, however, more investigation is required.

| Facet | Concerns | Status |
|---|---|---|
| Connectivity | Mobility required | ✓ |
| | Short-range communication only | ✗ |
| | Limited bandwidth | ✗ |
| Things | Hetereogeneous devices | ✓ |
| | Limited batery and memory resources | ✓ |
| | Remote control required | ✗ |
| Behavior | Traceability of the device is required | ✗ |
| | Capture context information | ✓ |
| | Behavior composition | ✗ |
| Smartness | Self-adaptation property is required | ✗ |
| | Decision-making trigger an action | ✓ |
| | Evaluate relevant contexto information | ✓ |
| Interactivity | Semantics influences the interpretation of information | ✗ |
| | Cooperation is required | ✗ |
| | Seamlessly interaction | ✗ |
| Environment | Context influences the solution | ✓ |
| | Critical environmental conditions | ✗ |
| | Limited infraestructure | ✗ |

Figure 18. A preliminary view of the Problem Characterization for this project..

**STEP 2 - Analyzing the Body of Knowledge for decision making**

After characterizing the problem domain with the characterization artifact, the next step is to analyze the body of knowledge.

In the context of this proposal, the body of knowledge will contemplate all the investigations conducted in order to fill in the body of knowledge. As presented in Section 5.2.1, the organization will be based on the filling of facets (by answering the questions for each perspective) representing the CSS multidisciplinary view, according to the problem domain needs. The body of knowledge has been proposed at a conceptual level but it has not yet been completely populated for the Iot systems. Hence, it is one of the next proposed activities, in which we plan to conduct studies to provide evidence-based findings to fill in the Body of Knowledge. Once Body of Knowledge is organized, it can be specialized to the problem context.

For instance, as presented in the problem (Figure 16) one of the needs refers to **(4) monitor the tank status.** This feature represents goals from the owner (Executive perspective), manager (Business Perspective), tank keeper and biologist (User Perspective) and can be developed by different solutions (Table 7). The body of knowledge specialization should assist in the decision-making to implement the desired solution considering this feature's properties..

Table 7. Possible solutions to monitor the tank status.

| Solution Option | Description |
|---|---|
| Manually | The manager defines the required shrimp production and requests production report, he communicates verbally to the biologist. The biologist sets new parameters for the tank and goes to the tank keeper to inform him. The tank keeper manually adjusts tank conditions to meet demand. He also manually collects information for the production report and deliver the report to the manager. There is no technical support in the process. |
| Communication Support | The manager defines the required shrimp production and requests production report; he uses a communication system to inform the biologist. The biologist defines new parameters for the tank and uses the communication system to inform the tank keeper. The tank keeper manually adjusts tank conditions to meet demand. He also manually collects information for the production report and deliver the report to the manager. There is technical support for communication in the process. |
| Control Support | The manager defines the required shrimp production and requests a production report; he uses a control system. The system notifies the biologist that defines new parameters for the tank. The system notifies the tank keeper that manually adjusts tank conditions to meet demand. He also manually collects information for the production report and make the report available in the system. There is technical support for control in the process. |
| Sensing Support | The manager defines the required shrimp production and requests a production report using the system. The system notifies the biologist that defines new parameters for the tank. The system notifies the tank keeper that manually adjusts tank conditions to meet demand. He automatically collects information from the sensors for the production report and makes the report available in the system. There is technical support for sensing in the process. |
| Actuation Support | The manager defines the required shrimp production and requests a production report using the system. The system notifies the biologist that defines new parameters for the tank. The system notifies the tank keeper that uses the system actuators to adjust the tank conditions to meet demand. He automatically collects information from the sensors for the production report and makes the report available in the system. There is technical support for actuation in the process. |

The solutions presented are simplified in high-level, but are only to exemplify the variety of options dependent on technology, to a greater or lesser degree. For example, if we choose the Sensor Support solution, exemplified in Table 7, it can analyze which relevant knowledge from the body of knowledge should be taken into account, as shown in Table 8. In order to support decision-making to guide the choice and development of the proposed solution, the body of knowledge aims to present the practices and technologies that allow engineers to develop the chosen solution.

Table 8. Some examples of possible practices and technologies from the body of knowledge.

| Sensing Support | Connectivity | Bluetooth Low Energy, ZigBee, Z-Wave, NFC (Near Field Communication), RFID (Radio-Frequency IDentification), Wi-Fi as enabling technologies, Low-Power Wide-Area technologies, SigFox, Ingenu-RPMA (Random Phase Multiple Access), 2G, 3G, 4G, Software-Defined Network (SDN) and Network Function Virtualization (NFV) and others. |
|---|---|---|
| | Things | Temperature sensor TTC104, temperature sensor DS18B20, luminosity sensor LDR 5mm, rain sensor FC37, rain sensor GROVE, humidity and temperature sensor RHT03, Gravity pH Sensor and others. |

| | **Behavior** | Collect water ph value, water level, water turbidity, water oxygenation, water salinity, water temperature. |

## STEP 3 - Generate decision-making Strategy

The output from the previous step (Table 8) should be presented as a set of software practices and technologies, with options of the body of knowledge specialization, and will compose the strategy to support decision-making to drive the solution.

From the problem domain established (the context), the team started the solution engineering. The project was conducted with the team working together. Therefore, there was no formal division of work for construction roles such as Architect, Engineer, and Technician perspectives. They worked as a group to achieve the expected results. For this reason, in this exemplification, we cannot represent the different perspectives. It is for illustration purposes and does not allow to demonstrate the full framework potential yet, which is an essential issue in the continuity of this research.

The solution implemented for the problem **(4)** monitor the status of the tank is presented in Figure 19 and was implemented in a floater. The floater collects data from the water tank where it was deployed, and it works at each determined interval of time. An operator can adjust the frequency in which the dashboard will update the information received from the sensors, implemented in the floater. A dashboard panel was implemented to enable the visualization of the data collected by the floater and attends the problem stated in **(1)** allows the clear visualization of information regarding the whole process in real time. In this context, it is a technological arrangement for data exhibition, where the data producers are the sensors in the floater, which through the connectivity with Wi-Fi can share data with the dashboard to exhibit the data. The overall Floater solution encompasses (Figure 19):

**Behaviors**: sensing and data collection to collect water level, water turbidity, and water temperature as well as processing, to provide data for the dashboard.

**Things**: the water level sensor, water turbidity sensor, water temperature sensor, and water salinity were implemented in an Arduino board that worked as the processing unit.

**Interactivity**: interacts with the dashboard to provide data.

**Connectivity**: used to provide data for the dashboard, implemented by a Wi-fi module in the Arduino.

**Environment**: the water tank was the environment settled for the sensors collect data, and the network layer used for connectivity.

| FLOATER | What | How | Where | Who | When | Why |
|---|---|---|---|---|---|---|
| | Collect water level | Water level sensor | Water tank | Arduino | Schedule time | Monitor tank status |
| | Collect water turbidity | Turbity sensor | Water tank | Arduino | Schedule time | Monitor tank status |
| | Collect water temperature | Temperature sensor | Water tank | Arduino | Schedule time | Monitor tank status |
| | Collect water salinity | Salinity sensor | Water tank | Arduino | Schedule time | Monitor tank status |
| | Provide data | Dashboard / WIFI | Network Layer | Arduino | Anytime | Enable status visualization |

Figure 19. Floater solution implemented for the need (4).

One solution envisioned was to use an Actuation Support, presented in Table 7. The solution can be exemplified in a technological arrangement where from a demand to fill the tank, a human user can trigger an action of an actuator to do it automatically. It would require a Filler solution, where from the data collected from the Floater or a request from the system dashboard, it would trigger the action of releasing the water to meet the required water level. The envisioned Filler solution encompasses (Figure 20):

**Behaviors**: decision-making to receive trigger and release water, data collection to collect water level as well as processing, to provide data for the dashboard.

**Smartness**: the behavior of release water requires a level of smartness since it should take into consideration the water level and amount required in the tank before releasing it.

**Things**: the actuator implemented in a Raspberry Pi board that worked as the processing unit and the water level sensor, used from the Floater.

**Interactivity**: interacts with the Floater to receive water level and the dashboard to receive trigger and to provide data.

**Connectivity**: used to receive information from the Floater through Bluetooth and to receive and send data to the dashboard through Wi-Fi, both implemented with a Raspberry Pi module.

**Environment**: the water tank was the environment settled for Floater and Filler, and the network layer used for connectivity with the dashboard.

| FILLER | What | How | Where | Who | When | Why |
|---|---|---|---|---|---|---|
| | Recieve trigger to fill the tank | Dashboard / WIFI | Network Layer | Raspberry | On demand | Automaticly adjust tank conditions |
| | Release water until meet the parameters | Actuator | Water tank | Raspberry | On demand | Automaticly adjust tank conditions |
| | Collect water level | Floater / Bluetooth | Water tank | Raspberry | Anytime | Automaticly adjust tank conditions |
| | Provide data | Dashboard / WIFI | Network Layer | Raspberry | Anytime | Enable status visualization |

Figure 20. Filler solution envisioned for the need (4).

The solution presented was performed by undergraduate students and has the typical limitations of this type of project but is enough to illustrate a scenario of our proposal. For more information, there is a video available[11] of the released version of this application.

It is necessary to emphasize that the previously described context is only a simplified example of using the framework, and does not represent the complete use of the framework. It was used for illustration purposes only. We understand that more research is needed to address the open points and to evolve the proposal in general. It represents future tasks to be conducted throughout the continuity of this research.

In this sense, we foresee some scenarios of utilization of the proposed framework. As envisioned contributions of its use, initially, we expect the production of scientific research which considers the knowledge that is important to practitioners concerning the problem domain. This knowledge will compose the body of knowledge proposed, which can be useful for both researchers and practitioners sharing and exchanging knowledge. We consider that the facets and perspectives have the potential to support the collection of various practices and technologies that can be used in CSS engineering. We expect that the more a facet is filled in a given perspective in response to a question (for example, response to the how, in the engineering perspective, in the behavior facet) more evidence of information we have about it, which aids decision-making in practice. In turn, the lack of answers (for example, an empty cell in the body of knowledge) may represent a research opportunity for the academy. In this sense, opportunities and risks are opposites, since an opportunity for researchers is a risk for practitioners.

## 5.4 Computational Infrastructure

A computational infrastructure is not entirely necessary for achieving the objective of supporting decision-making in engineering IoT, as proposed in this doctoral research. However, we believe it may facilitate the conduction of the activities proposed in the framework context, as well as engage and motivate its use.

Hence, we intend to build a computational infrastructure to provide support to some activities of the framework proposed, primarily the following macro-requirements:

1. Characterization of the Project;
2. Specializing the Body of Knowledge;
3. Alignment of Engineering Strategy.

---

[11] https://goo.gl/eXWCKW

## 5.5 Framework Evaluation

After the conceptual and development phases, we intend to perform empirical studies to assess the framework and its computational infrastructure, considering the previously adopted research methodology (Spínola, Dias-neto and Travassos, 2008). This assessment can be done in different stages, each one with specific purposes. Considering the nature of this doctoral proposal, we intend to evaluate the proposed framework in the following steps:

- Feasibility Study: the objective is to evaluate if the proposed framework meets the proposed objectives. We believe that this study can be conducted in the context of research projects and we believe that the software practitioners' opinion can also contribute to the feasibility study. We plan to perform initial studies in the context of undergraduate and postgraduate courses of UFRJ. Leaving to evaluate only the complete proposal would bring risks. Therefore, we intend that studies can observe the three steps proposed in the framework individually before a complete evaluation. It contributes to more direct assessments besides leading to targeted adjustments and improvements.

- Observational study: The results of the feasibility study can contribute to improving the framework. After the adjustments in the framework, we intend to perform an observational study in an IoT real project. This observational study aims to identify if the practical application of the framework makes sense. For evaluations on concrete projects, we plan to conduct the study in partnership with the *Université Polytechnique Hauts-de-France*. Currently, the university has several multidisciplinary projects[12] characterizing a suitable scenario for observing CSS engineering. Among the projects, there are solutions for automation and logistics for the industry, applications for autonomy and smart mobility besides innovation in robotics and investigation of human aspects in different types of systems.

- Case study: The results of the observational studies can contribute to improving the framework. After adjustments, we intend to perform a case study in a software engineering organization. This case study aims to observe if the proposed framework fits industrial settings and observing real cases regarding the construction of IoT solutions.

---

[12] http://www.uphf.fr/LAMIH/en/projets/87

## 5.6 Conclusion

The previous sections presented the preliminary version of the proposed framework, which aims to support decision-making in engineering IoT solutions, considering the problem domain. With the proposal, we seek to capture the problem domain with the goals to be met and characterize it in an artifact that maps these goals in the facets. The artifact presents a mapping of facets to concerns that should be considered for the problem in question. This mapping will specialize a body of knowledge in CSS with practices and solutions that can support decision-making to the solution engineering. Thus, we go from of the problem domain to the solution with a framework proposed with evidence-based based research, designed for CSS in the context of IoT applications.

Despite the possible contributions, the current proposal has some limitations that should be stated. To consider the Zachman's framework as the basis for the introduced conceptual body of knowledge was due to its broad use in other areas and applicability in our context. So far, its use has demonstrated adequate to represent our perspective on the challenges regarding CSS engineering. Further evaluation based on the cells filling is going to be performed, but it can represent a threat that we need to mitigate. It is configured as future work to investigate whether the proposition of the perspectives for each facet and whether this organization by control and construction is adequate to this context. Also, the communication interrogatives were adapted for this research, and it is configured as future works the retrieving of answers for each one.

We aim to fill in the body of knowledge regarding the facets with data from the technical literature, practitioners and real cases by conducting a family of experimental studies focusing in the IoT context. Up to the present time, this proposal is given in a conceptual structure representing what we are calling the body of knowledge. Its completion with required knowledge will be performed in the next phases of this research.

The use of the proposed framework was exemplified in the context of an IoT software solution to support carniciculture in Brazil, which it may be a useful and straightforward strategy. We envision this framework may support project decisions to perceive and handle needs, demands and risks associated in engineering an IoT solution. Finally, the expectations for evaluating the proposed framework were also described.

# 6 Final Considerations

*In this chapter, the final considerations are presented up to the current moment of the research, as well as the activities in progress and the future activities.*

## 6.1 Current Results

The research has evolved to a preliminary answer the questions defined in Chapter 1. The current chapter presents considerations about the current state of the research and which questions can already be answered and the way forward to answer the others. Thus, taking up the issues previously defined above, we have:

**RQ1.1) What characterizes IoT applications?**

Based on the findings of the studies performed, we can identify a set of characteristics that define IoT, first as an independent paradigm with the secondary study, then in the context of CSS with the rapid reviews. Some of the characteristics are discussed in Section 3.2 and others are observed throughout the technical report and presented the light of the six defined facets. We can observe that some of the characteristics are fundamental to an application in order to fulfill the IoT definition: "a paradigm that allows composing systems from uniquely addressable objects equipped with identifying, sensing or acting behaviors and processing capabilities that can communicate and cooperate to reach a goal." Addressability, Unique ID, Heterogeneity, and Mobility are essential for IoT applications. In turn, some other characteristics are necessary for specific behaviors, which are governed by a greater or lesser degree of smartness. For instance, Context-awareness is required to enable sensing behavior, and Autonomy is needed in the acting behavior. Also, the characteristics of Interoperability and Security are significant challenges to IoT (Section 3.6). However, we understand that evaluations and further investigations are necessary for this response, even though it is based on the technical literature, we need more evidence that it is adequate to the characterization that was proposed.

**RQ1.2) What are the challenges for IoT applications development?**

For this question, as presented in Chapter 3, we extracted IoT concerns from the different points of views: literature, practitioners and government. Together, the three sources provided 14 different concerns, which must be met in favor of a higher quality IoT software systems. Also, in Section 4.4, we mapped the 14 Concerns in the six Facets and presented high-level challenges to be addressed by CSS. Thus, based on these premises, we identified a set of challenges that should be considered in the framework to support the development to overcome them or at least reduce the risks.

**RQ1.3) What areas are involved in IoT applications development?**

From the secondary study and supported by qualitative analysis we identified six facets - Connectivity, Things, Behavior, Smartness, Interactivity, and Environment - that represent areas, disciplines, and fields involved in IoT development. Then we conducted rapid reviews to characterize each facet regarding what, how, where, when and why is a facet used.

**RQ1.4) What software technologies support the engineering of IoT applications?**

The Rapid Reviews' results allowed to extract, in a general form, techniques, technologies, methods, and solutions able to operationalize each facet by answering the question of *how*. It was an initial investigation that needs to be deepened and enriched considering the facets' perspectives and other research strategies.

Up to the present time, the activities of the conceptual phase defined in the methodology adopted have been performed (Section 1.4). From these stages the following results were obtained:
- Is Software Engineering ready for the Internet of Things? (I Workshop de Engenharia e Qualidade do Software para Internet das Coisas - QualityIoT);
- Towards a more in-depth understanding of IoT (Journal of Software Engineering Research and Development – *to appear*);
- On Challenges in Engineering IoT Software Systems (XXXII Brazilian Symposium on Software Engineering).

## 6.2 Next Activities

The next phase is the development of the framework itself: filling the body of knowledge, definition of templates and the computational infrastructure (Figure 21). Many decisions still need to be made towards the proposal, and Table 9 presents a summary of the research plan to be used as a script (and be updated, if necessary) script for the remaining activities of this doctorate.

The last step in the proposed methodology is the evaluation phase (also in the research plan). A detailed plan for each study will be prepared according to the results and limitations found after the development phase, and we aim that the results of these evaluations will also serve as input for improvements in the framework.

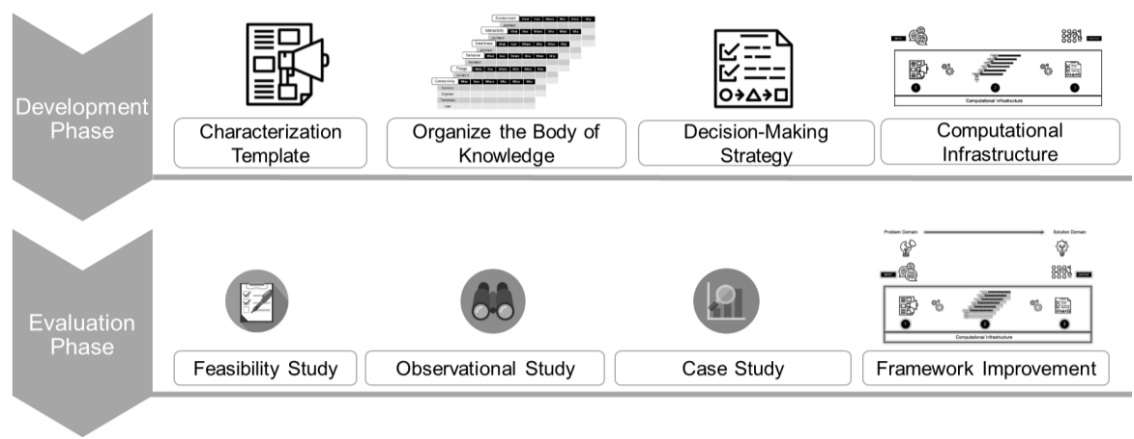

Figure 21. Next phases in the research methodology.

Table 9. Research Plan.

| |
|---|
| **1. Define the Problem Characterization Template** |
|     1.1. Identify the IoT problem domain definition from technical literature; |
|     1.2. Identify the IoT problem definition domain from real projects; |
|     1.3. Propose initial Characterization Template based on the findings; |
|     1.4. Conduct an Observation Study with the objective of evaluating the Characterization Template regarding its adequacy to define an IoT problem domain; |
|     1.5. Evolve the Characterization Template based on the study results. |
| **2. Organize the Body of Knowledge** |
|     2.1. Answer each question for each facet with inputs from technical literature; |
|     2.2. Answer each question for each facet with inputs from real projects; |
|     2.2. Answer each question for each facet with inputs from interviews with practitioners; |
|     2.3. Fill the body of knowledge with the findings. |
|     2.4. Conduct an Observation Study with the objective of evaluating the body of knowledge regarding the adequacy of its content; |
|     2.5. Evolve the body of knowledge based on the study results; |
|     2.6. Make protocols available for the body of constant knowledge evolution. |
| **3. Define the Decision-Making Strategy** |
|     3.1. Propose a decision-making strategy based on the problem domain characterization and the content of the body of knowledge; |
|     3.2. Conduct an Observation Study with the objective of evaluating the Development Guide regarding its adequacy to provide information to support the development; |
|     3.3. Evolve the decision-making strategy based on the study results. |

**4. Develop the Computational Infrastructure**

    4.1. Implement a computational infrastructure to assist the framework usage;

    4.2. Determine usage scenarios for the framework;

**5. Evaluate the proposed Framework**

    5.1. Feasibility Study: to observe if the framework is capable of meeting the objectives for which it was proposed;

    5.2. Observation Study: to observe if the practical application of the framework is reasonable;

    5.3. Case Study: to observe if the practical application of the framework is fitted for the context;

    5.4. Final Report with the results of the findings and the research as a whole.

Table 10 presents a planned schedule for the activities proposed in the research plan.

Table 10. Schedule for the Research Plan.

| Activity | 01/2019 | 02/2019 | 03/2013 | 01/2020 | 02/2020 | 03/2020 |
|---|---|---|---|---|---|---|
| 1. Define the Characterization Template | X | | | | | |
| 2. Organize the Body of Knowledge | X | X | X | | | |
| 3. Define the Decision-Making Strategy | | | | X | | |
| 4. Develop Computational Infrastructure | | | | | X | |
| 5. Evaluate the proposed framework | | | | | | X | X |

## 6.3 Future Work

Given the time available for the development of this thesis, it will not be possible to observe the full potential of the proposal, regarding its scope and applicability to different CSS and contemporary applications. However, considering the novelty and comprehensiveness of the area, some findings represent open-ended issues that should be addressed in future research. To investigate in depth each proposed facet and the challenges briefly presented, although extensive for the planned scope, is necessary and more studies need to be carried out in future times to strengthen the evidence of feasibility and applicability of this process.

# Appendix A – Evidence Briefings

*This appendix presents a summary of the Rapid Reviews results, presented in chapter 4, in the format of Evidence Briefings.*

## Introduction

The information acquired in the Rapid Reviews executed was aggregated and summarized to be presented in the format of evidence briefings (EBs), as discussed by (Cartaxo *et al.*, 2016).

EBs are medium to transfer knowledge from researchers to the industry and is motivated by the fact that software practitioners tend to not use research papers as a source of new knowledge (Cartaxo *et al.*, 2016). Thus, the idea is to present a more concise instrument, which summarizes the ideas and main findings of a paper to a broader audience. Some advantages presented by the authors is that this medium increases the research visibility and is considered an excellent way to share research findings since it promotes a "clear and understandable information" (Cartaxo *et al.*, 2016), also it has been used by other works in the area (Silva, Jeronimo, and Travassos, 2018).

The original template, available to use under an open source license (CC-BY) in the link http://cin.ufpe.br/eseg/briefings, was adapted it to our context, with the main elements, as described below and represented in Figure 22:

- The title of the briefing (1), sometimes simplifying the paper title to make the briefing more appealing to the practitioners;
- The logos and identification of the research group and the university (2).
- A summary (3) to present the objective, motivation, facet definition and context of the briefing;
- Informative box (4), separated from the main text, to highlight the target audience and the purpose of the briefing and answer the research questions;
- The additional information (5), extracted from the original empirical study;
- The references to the original empirical study (6);

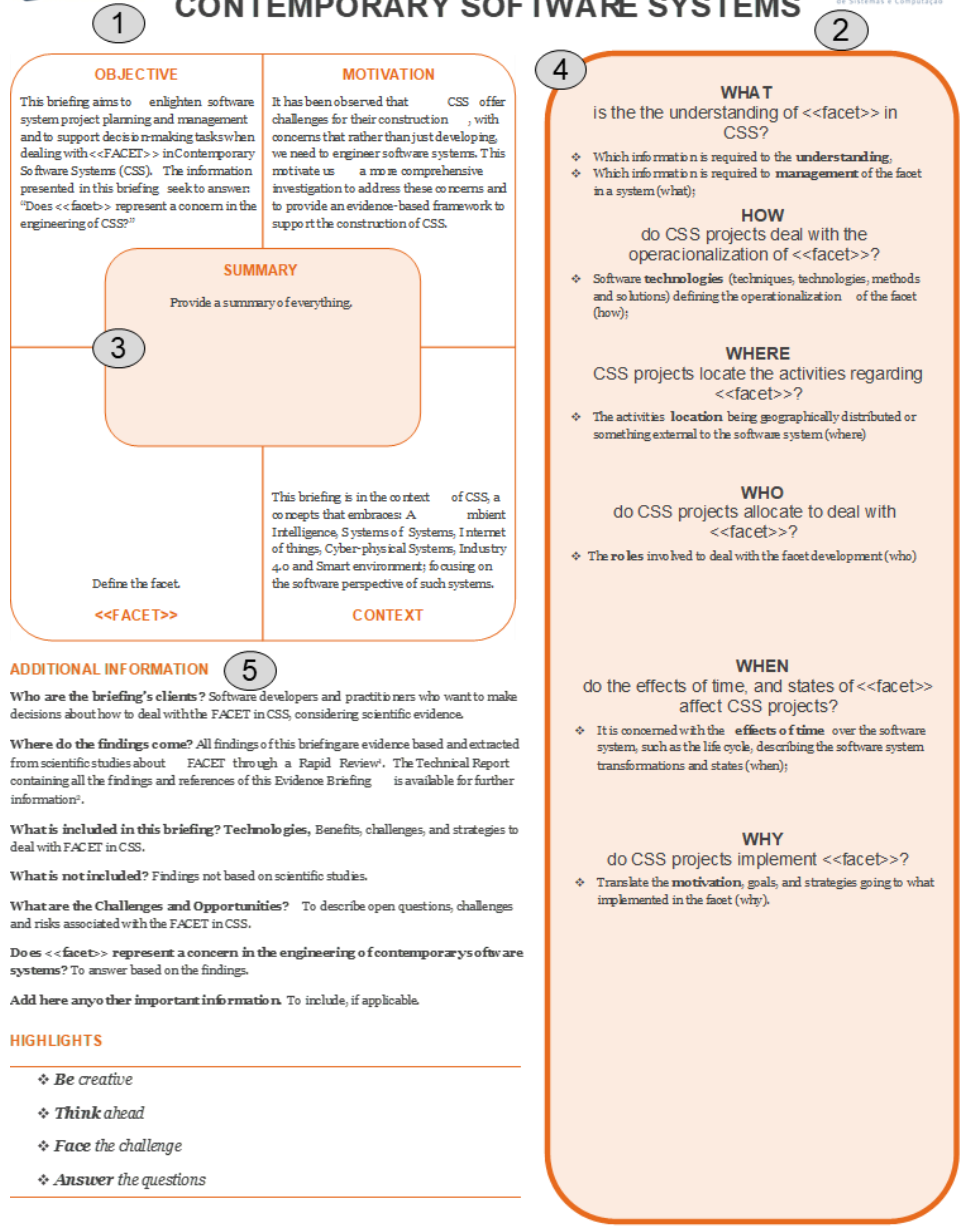

Figure 22. Overview of the elements of the EB template.

As we did on the protocol, this is a meta-template that should be instantiated in for each facet.

# A.1 Connectivity

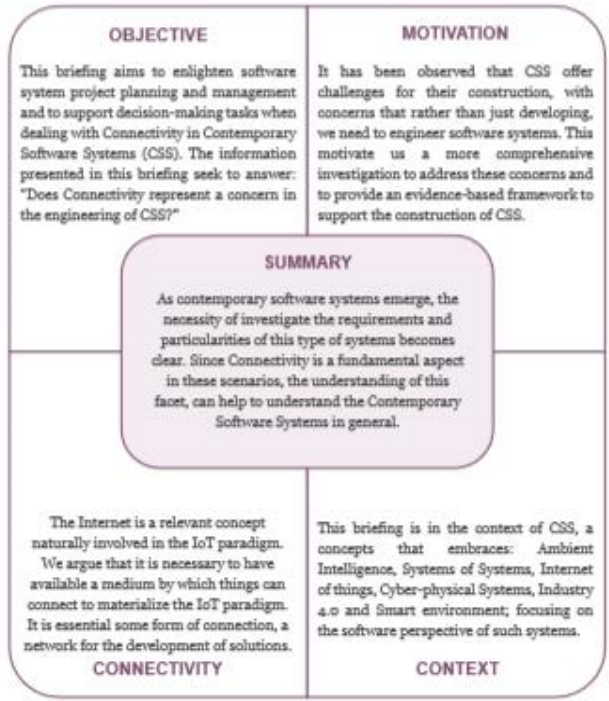
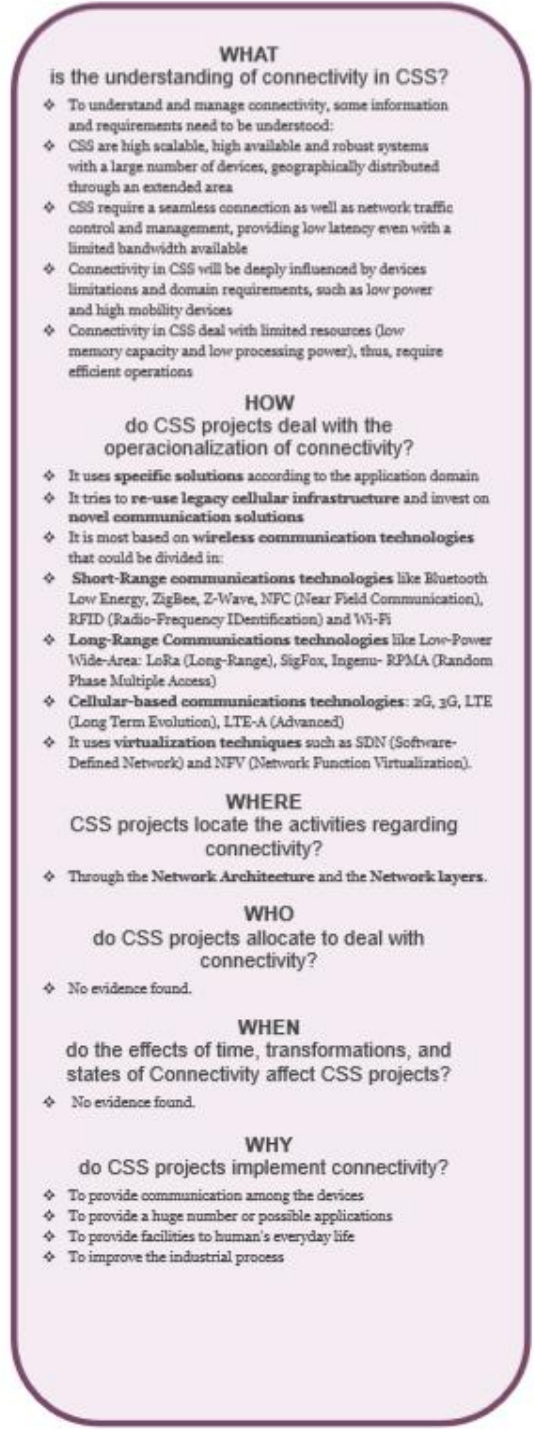

## A.2 Things

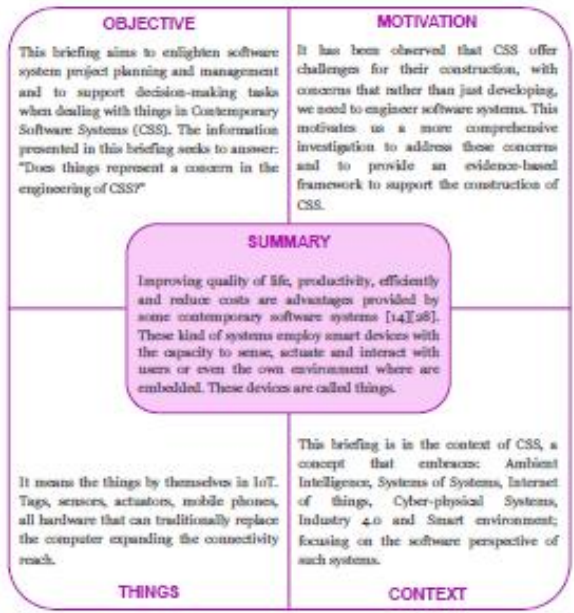

## A.3 Behavior

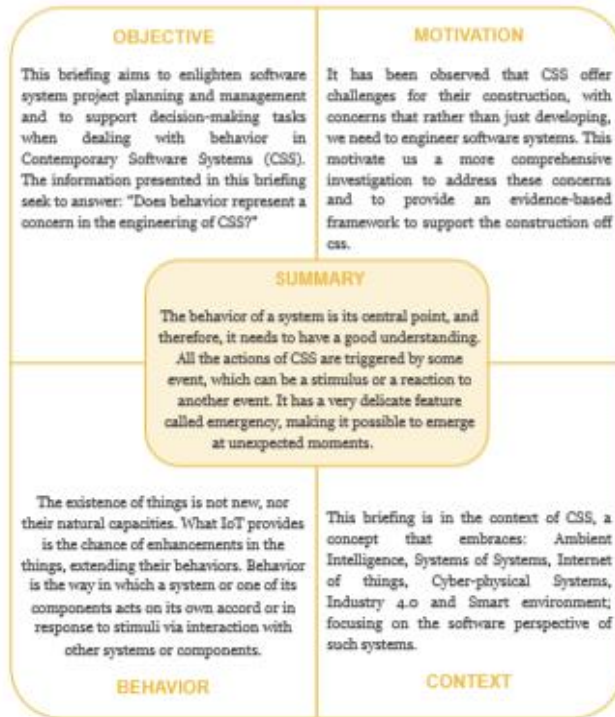
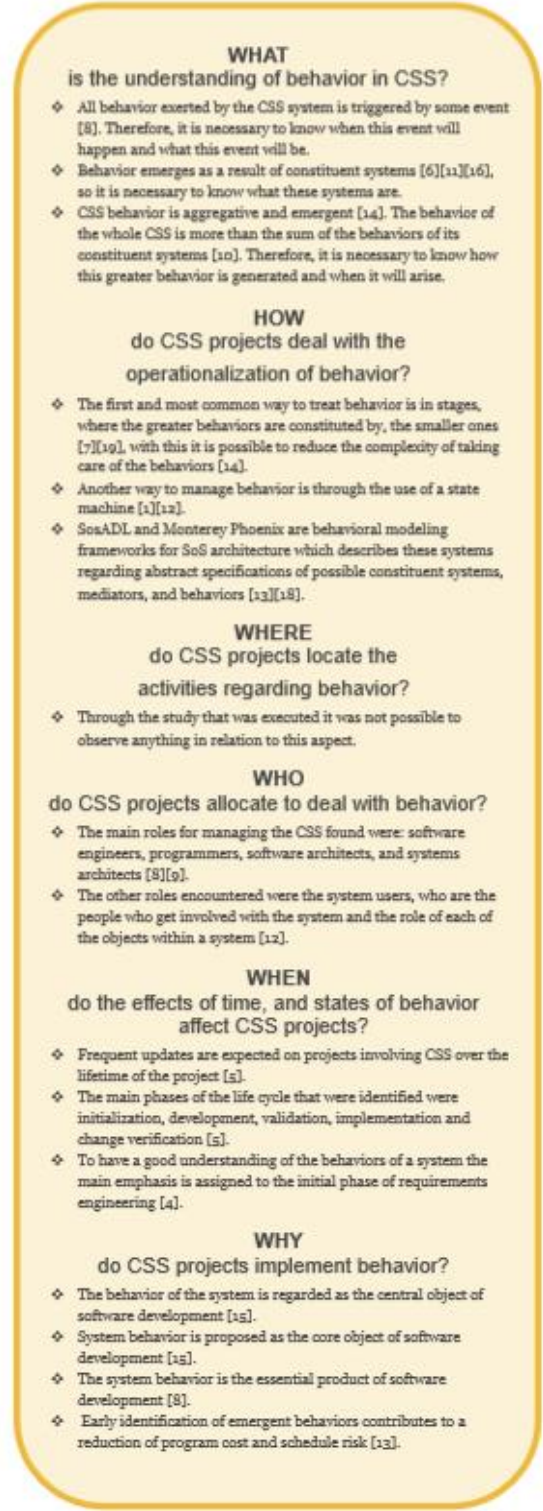

## A.4 Smartness

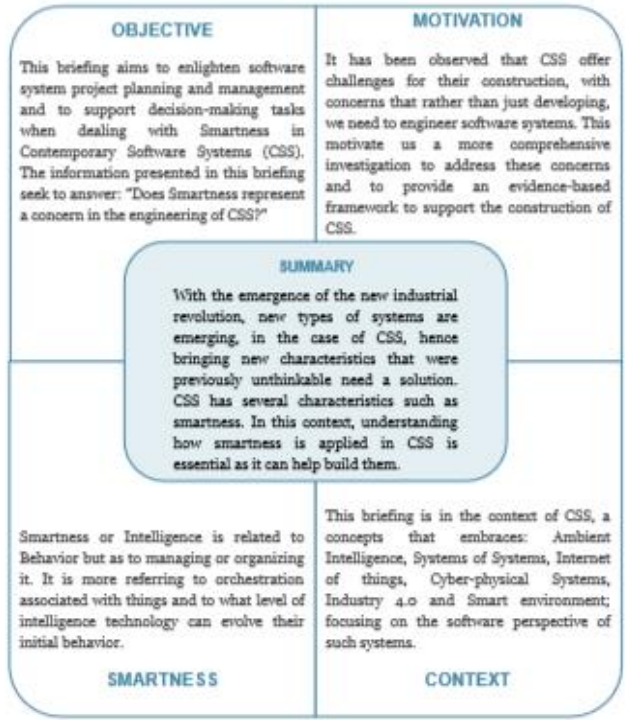

# A.5 Interactivity

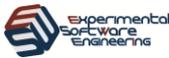 **DEVELOPING CONTEMPORARY SOFTWARE SYSTEMS WITH INTERACTIVITY** 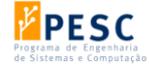

### OBJECTIVE
This briefing aims to enlighten software system project planning and management and to support decision-making tasks when dealing with Interactivity in Contemporary Software Systems (CSS). The information presented in this briefing seek to answer: "Does Interactivity represent a concern in the engineering of CSS?".

### MOTIVATION
It has been observed that CSS offer challenges for their construction, with concerns that rather than just developing, we need to engineer software systems. This motivates us to undergo a more comprehensive investigation to address these concerns and to provide an evidence-based framework to support the construction of CSS.

### SUMMARY
The challenge in making Contemporary Software Systems (CSSs) interoperable is the excessive heterogeneity present in an Internet of Things (IoT) system, due to several different disciplines, vendors, standards and protocols. All the articles found in this study try to fill this gap with the use of a solution based on devices, technologies, frameworks and methodologies.

### INTERACTIVITY
It refers to the involvement of actors in the exchange of information with things and the degree to which it happens. The actors engaged with IoT applications are not limited to humans. Therefore, beyond the sociotechnical concerns surrounding the human actors, we also have concerns with other actors like animals and the interactions thing-thing.

### CONTEXT
This briefing is in the context of CSS, a concept that embraces: Ambient Intelligence, Systems of Systems, Internet of things, Cyber-physical Systems, Industry 4.0 and Smart environment; focusing on the software perspective of such systems.

### WHAT is the the understanding of interactivity in CSS?
- In CSS projects, interactivity is characterized by the interaction involving things, systems, and humans where interaction is characterized by the ability to **communicate**, **exchange** information and control **actions**.
- Data must be collected (sensing the environment), processed (generally in some cloud), stored (using databases) and transmitted [9][19]. To transmit and receive the information, as well as interact with humans, they utilize networks a medium of communication [10][12].

### HOW do CSS projects deal with the operacionalization of interactivity?
- **To guarantee connectivity:** Zig-Bee, Bluetooth, Radio Frequency, RFID, 6LowPAN, WSN, WiFi, IPv6 and others.
- **To guarantee communication:** HTTP, XMPP, TCP, UDP, CoAP, MQTT and others.
- **To guarantee understanding:** JSON, XML, OWL, SSN Ontology, COCI and others.
- Also, real-world objects are virtualized and represented as Web Resources and accessed through Web Interfaces based on REST principles and Producer and Consumers methods.

### WHERE CSS projects locate the activities regarding interactivity?
- Most of the times, the activities regarding interactivity are locat in a system's architecture, on frameworks, middleware, and platform.

### WHO do CSS projects allocate to deal with interactivity?
- Designers, architects, developers, managers, and engineers deal with interactivity in different phases of CSS projects [3][5][7].
- **Changing the scenario:** "Engineering is no more a set of vertical activities developed by different engineers but a collaborative process in which people and technology is completely involved in the engineering process" [18].

### WHEN do the effects of time, and states of interactivity affect CSS projects?
- Activities regarding Interactivity are present all over a system or application **lifecycle**. Especially, they affect the design, development, integration, deployment, and operation phases [3][5][18].
- Interactivity is achieved by the exchange of information among "Things" and the orchestration of these operations occur during **runtime** [5].

### WHY do CSS projects implement interactivity?
- To **bridge the gap** between the massive heterogeneity present in CSS in order to create an interoperable systems, that can overcome different standards, protocols and technologies to perform more efficiently than isolated ones.
- **Interactivity** is one of the main characteristics of CSS projects, making new types of application possible (such as smart environments), facilitating everyday life, enhancing products competitivity, and sustainability.

### ADDITIONAL INFORMATION

**Who are the briefing's clients?** Software engineering practitioners who want to make decisions about how to improve customer collaboration based on scientific evidence.

**Where do the findings come?** All findings of this briefing were extracted from scientific studies about environment identified on a rapid review.

**What is included in this briefing?** Benefits, challenges, and strategies to deal with Interactivity in CSS.

**What is not included?** Findings not based on scientific studies.

**What are the Challenges and Opportunities?** Although this research was made specifically for the Interactivity facet, some important observations about Contemporary Software Systems could be noted. The most important of them is the fact that there is no more unique system, but an enormous quantity of systems, components, things, and applications, which has to work together.

**Does Interactivity represent a concern in engineering Contemporary Software Systems?** Yes, and to overcame heterogeneity to achieve interoperable and collaborative solutions is a major challenge for this facet.

### HIGHLIGHTS

- *Heterogeneity* is an intrinsic characteristic of CSS.

*It can affects the semantic (the meaning) and the syntax (the format) and harms interactivity because even if the systems can communicate, they cannot understand each other. That is a great challenge in Interactivity.*

# A.6 Environment

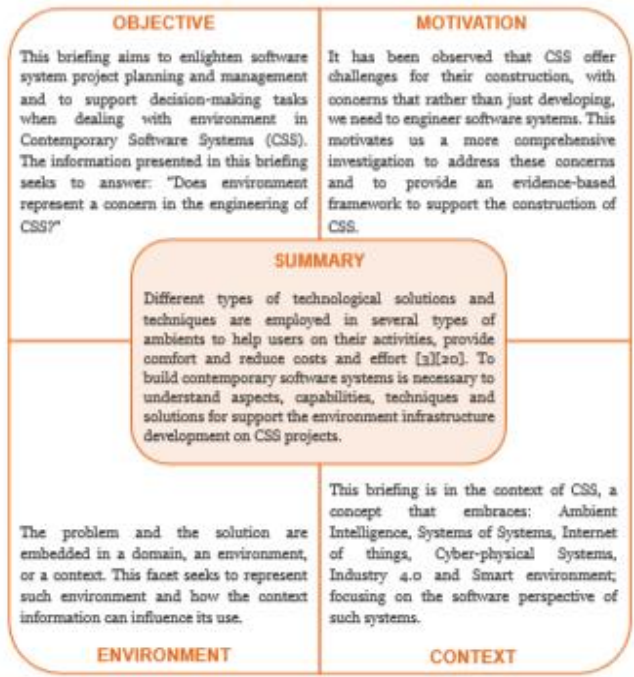